\documentclass[useAMS,usenatbib]{mn2e}
\usepackage{graphicx}
\usepackage{amsmath}
\usepackage{color}

\newcommand{\be}{\begin{equation}}
\newcommand{\ee}{\end{equation}}

\def\no{\noindent}

\newcommand{\ifm}[1]{\relax\ifmmode#1\else$\mathsurround=0pt #1$\fi}
\newcommand{\kms}{\ifmmode\,{\rm km}\,{\rm s}^{-1}\else km$\,$s$^{-1}$\fi}

\newcommand{\ltsima}{$\; \buildrel < \over \sim \;$}
\newcommand{\lsim}{\lower.5ex\hbox{\ltsima}}
\newcommand{\gtsima}{$\; \buildrel > \over \sim \;$}
\newcommand{\gsim}{\lower.5ex\hbox{\gtsima}}

\definecolor{green}{rgb}{0,0.5,0}
\definecolor{grey}{rgb}{0.4,0.5,0.7}

\def\M11{M_{11}}
\def\V100{V_{100}}
\def\R1{R_{Mpc}}
\def\T6{T_6}

\begin{document}

\title[Cold accretion, cooling and feedback]{GalICS 2.1: a new semianalytic model for cold accretion, cooling, feedback and their roles in galaxy formation}

\pagerange{\pageref{firstpage}--\pageref{lastpage}} \pubyear{2016}

\author[Cattaneo et al.]{A.~Cattaneo$^{1,2}$, 
I.~Koutsouridou$^1$,
E.~Tollet$^3$, 
J.~Devriendt$^4$, 
Y.~Dubois$^2$,
\\
\\
$^1$Observatoire de Paris, LERMA, PSL University, 61 avenue de l'Observatoire, 75014 Paris, France \\
$^2$Institut d'Astrophysique de Paris, CNRS, 98bis Boulevard Arago, 75014 Paris, France\\
$^3$Centre de Recherche Astrophysique de Lyon, 9 avenue Charles
Andr{\'e}, 69230 Saint-Genis-Laval, France \\
$^4$University of Oxford, Astrophysics, Keble Road, Oxford OX1 3RH, UK\\
}

\maketitle

\label{firstpage}


\begin{abstract}

{\citet{dekel_birnboim06} proposed that} the mass-scale that
  separates late-type and early-type galaxies is linked to the
  critical halo mass $M_{\rm vir}^{\rm crit}$ for the propagation of a
  stable shock
  and showed that they could reproduce the observed
  bimodality scale for plausible values of  the metallicity of the
  accreted gas $Z_{\rm accr}$ and the shock radius $r_{\rm s}$.
Here, we take their analysis one step further and present a new
semianalytic model that computes $r_{\rm s}$ from first principles. This advancement allows us to compute $M_{\rm vir}^{\rm crit}$ individually
for each halo.
{ Separating cold-mode and hot-mode accretion has little effect on
  the final galaxy masses if feedback does not preferentially couple
  to the hot gas.}
We also present an improved model for stellar feedback where
$\sim 70\%$ of the wind mass is in a cold galactic fountain with a
shorter reaccretion timescale at high masses.
The latter is the key mechanism that allows us to reproduce the low-mass end of the mass function of galaxies over the entire redshift range $0<z<2.5$.
Cooling must be mitigated to avoid overpredicting the number density of galaxies with 
stellar mass $M_{\rm stars}>10^{11}\,{\rm M}_\odot$
but is important to form intermediate-mass galaxies.
At $M_{\rm vir}>3\times 10^{11}\,{\rm M}_\odot$, 
cold accretion is more important at high $z$,
where gas is accreted from smaller solid angles,
but this is not true at lower masses because high-$z$ filaments have lower
metallicities. Our predictions are consistent with the observed metallicity evolution of the intergalactic medium at $0<z<5$.

\end{abstract} 


 
\section{Introduction}
 
\subsection{The classical semianalytic cooling model}

\citet{rees_ostriker77} and \citet{silk77} laid the foundations of the modern theory of galaxy formation by discovering that
the competition between the radiative cooling time $t_{\rm cool}$ and the gravitational freefall time $t_{\rm ff}$
determines the characteristic mass of galaxies.
For $t_{\rm cool}<t_{\rm ff}$, the gas cools, collapses in freefall and fragments into stars. 
For $t_{\rm cool}>t_{\rm ff}$, the gas contracts quasi-statically at the virial temperature.
In a baryonic Universe,
the condition $t_{\rm cool}\simeq t_{\rm ff}$ gives a maximum mass
$M_{\rm bar}^{\rm crit}\sim 10^{12}\,{\rm M}_\odot$ for cooling and star formation.

\citet{white_rees78} and \citet{blumenthal_etal84} extented this result to a cosmology with dark matter (DM). They found that 
$t_{\rm cool}$ is shorter than the age of the Universe for haloes with
virial mass $M_{\rm vir}<M_{\rm vir}^{\rm crit}\sim 10^{13}\,{\rm M}_\odot$.
The alternative condition $t_{\rm cool}\simeq t_{\rm ff}$ gives a critical halo mass about three times lower.

{ When \citet{white_frenk91} developed the first semianalytic model
(SAM) of galaxy formation in a cold DM cosmology, they were aware that
cold gas may survive shocks if the cooling rate is short
enough but they chose to make the simplifying assumption that all} the  gas initially relaxes to an isothermal distribution that exactly parallels that of the DM.
Cooling begins at the centre, where the gas is denser, and proceeds
inside out.
{ Gas cools if its cooling time is shorter than the time $t$
  available for cooling.

The choice of $t$ is the most problematic aspect of this model.}
\citet{white_frenk91} assumed that $t$ is the age of the Universe.
More realistically, the gas started cooling when the halo formed but halo formation is not a well defined concept in a hierarchical cosmology.
For \citet{cole_etal94}, a halo formed when it reached half its current mass.
For \citet{somerville_primack99}, cooling started after the last major merger because halo mergers can reheat and redistribute gas within haloes.
Most current SAMs follow \citet{springel_etal01} and take $t$ to be the dynamical time of the halo, 
which is directly related to the freefall time $t_{\rm ff}$.
In this article, we argue that the  gas began to cool when it was
shock-heated and present a physical model to compute that time.

\subsection{The cold-flow paradigm}

The SDSS has shown that the galaxy population is bimodal.
Star-forming (blue) and passive (red) galaxies  are separated at a
characteristic stellar mass $M_*\sim 3\times 10^{10}\,{\rm M}_\odot$ \citep{kauffmann_etal03}
and occupy different regions of the galaxy colour-magnitude diagram \citep{baldry_etal04}.
The SAMs of the time did reproduce the qualitative finding that the most massive galaxies are also the reddest
but they predicted that giant ellipticals lie on the extension of the colour-mass relation for blue galaxies. In the observations, they are shifted to redder colours.
The reason for this discrepancy is that,
in the classical semianalytic picture, the transition from rapid to
slow cooling is smooth.
Mergers can trigger starbursts and use up all the gas but they do
  not select any preferential mass-scale that may explain the galaxy
  bimodality\footnote{ The merging histories
    of DM are self-similar to the extent that the mass variance can be approximated with a power law
    \citep{lacey_cole93}.}. Feedback processes can cause the baryons to
    depart from self-similarity, so that the
    contribution of mergers to the stellar masses of galaxies becomes
    significant above $M_{\rm
      stars}\sim 10^{11}\,{\rm M}_\odot$ but,
without any process that reduces gas accretion at high masses, mergers
are never a major channel for mass growth 
    \citep{cattaneo_etal11}. Hence, in simulations
    without feedback from active galactic nuclei (AGN), the most massive galaxies are spiral rather than elliptical \citep{dubois_etal13,dubois_etal16}.

On the theoretical side,
cosmological hydrodynamic
simulations \citep{fardal_etal01,keres_etal05} showed that cold-mode accretion
(particles that radiate their energy at temperatures much lower than the virial temperature) accounts for a large fraction of the baryons that accrete onto galaxies.
At the same time,
one-dimensional simulations (\citealp{birnboim_dekel03}, BD03) highlighted a sharp transition from cold flows to shock heating above a critical halo mass.
Cosmological hydrodynamic simulations showed cold filamentary streams accrete onto galaxies, and their disruption and replacement
by hot atmospheres when the halo mass grew above $M_{\rm vir}^{\rm
  crit}\sim 10^{12}\,{\rm M}_\odot$  (\citealp{dekel_birnboim06}, DB06; \citealp{khalatyan_etal08}; \citealp{ocvirk_etal08}).
These developments lead DB06 and \citet{dekel_etal09} to champion a new paradigm, in which cold flows are the main mode of galaxy formation,
while hot-mode accretion (shock heating to $T\sim T_{\rm vir}$ followed by cooling) is heavily suppressed\footnote{In the most common version of this picture, the 
suppression of cooling is due to AGN feedback
(see \citealp{cattaneo_etal09} for a review), although this was not necessarily \citet{birnboim_etal07}'s viewpoint.}.

DB06 backed their conclusions with a shock-stability argument. 
A stable shock can propagate only if the cooling time of the compressed gas is longer than the compression time $t_{\rm comp}$.
The criterion $t_{\rm cool}<t_{\rm comp}$ derived by DB06
is almost identical to the classical criterion $t_{\rm cool}<t_{\rm dyn}$ of \citet{rees_ostriker77}, except that
DB06 evaluated 
$t_{\rm cool}$ and $t_{\rm comp}$ using post-shock quantities computed from the Rankine-Hugoniot jump conditions 
rather than using virial quantities. 
Therefore, the critical halo mass for shock-heating $M_{\rm vir}^{\rm crit}$ depends not only on the metallicity $Z_{\rm accr}$ of the accreted gas
but also on the shock radius $r_{\rm s}$.
DB06 showed that their model could reproduce a reasonable bimodality
mass-scale for plausible values of $Z_{\rm accr}$ and $r_{\rm s}$. They found 
$M_{\rm vir}^{\rm crit}\simeq 6\times 10^{11}\,{\rm M}_\odot$ for
$Z_{\rm accr}=0.1{\rm Z}_\odot$ and $r_{\rm s}=0.1r_{\rm vir}$
(a shock radius consistent with cosmological hydrodynamic simulations).

In \citet{cattaneo_etal06}, we put this picture to test by using the {\sc GalICS} SAM.  
Our SAM was based on three assumptions:
\begin{enumerate}
\item Below a critical halo mass $M_{\rm vir}^{\rm crit}$, the gas cools efficiently and accretes onto the central galaxy almost in freefall. 
\item Above $M_{\rm vir}^{\rm crit}$, all the gas is hot and unable to cool.
\item The critical shock-heating mass $M_{\rm vir}^{\rm crit}$ is the same for all haloes (at least at redshifts $z\ll 3$).
\end{enumerate}
Given the uncertainty on $Z_{\rm accr}$ and $r_{\rm s}/r_{\rm vir}$, we felt justified in treating  $M_{\rm vir}^{\rm crit}$ as a free parameter and we found an excellent 
agreement with the
colour-magnitude distribution of galaxies in the SDSS for $M_{\rm
  vir}^{\rm crit}=2\times 10^{12}\,{\rm M}_\odot$.

{ The interest of this work and, by extension, our article is that,
although cosmological hydrodynamic simulations with a volume comparable to ours are
now possible (e.g., \citealp{lebrun_etal14,schaye_etal15,dubois_etal16,nelson_etal18}), it is still not clear whether these simulations can
resolve the instabilities that are crucial for the survival of cold
filamentary flows \citep{berlok_etal19,han_etal19}
and their huge computational cost prevents a systematic exploration of
the impact of subgrid physics (e.g., feedback) on their results.}

\subsection{This work}

Despite the fit in \citet{cattaneo_etal06}'s being so good that even now we could hardly do any better,
assumptions (ii) and (iii) cannot both be correct because
\citet{mei_etal09} studied the colour-magnitude relation in eight clusters at $z\sim 1$
and found that the colours of the brightest galaxies could vary significantly from one cluster to another even when $M_{\rm vir}$ was comparable.

\citet{benson_bower11} improved previous work by applying DB06's shock-stability condition to individual haloes,
so that shock heating could occur at different halo masses. {
  Theirs has been the only attempt to date at separating cold- and
  hot-mode accretion in SAMs.}
However,  \citet{benson_bower11} assumed $r_{\rm s}=r_{\rm vir}$ for all haloes. 

Here, we present a new version of our SAM, {\sc GalICS~2.1}  in which {\it the parameters that determine} $M_{\rm vir}^{\rm crit}$ ($Z_{\rm accr}$, $r_{\rm s}/r_{\rm vir}$, $\Omega$)
{\it are computed from first principles on a halo-by-halo basis} ($\Omega$ is the solid angle from which gas is accreted).
By separating cold accretion and cooling, we are able to disentangle their contributions to the galaxy stellar mass function (GSMF).

The greatest obstacle to this research programme is that stellar feedback shapes the GSMF at low masses: predicted accretion rates are degenerate with respect to assumed ejection rates.
We overcome this difficulty through an improved model of stellar
feedback based on a Numerical Investigation of a Hundred Astrophysical
Objects (NIHAO; \citealp{tollet_etal19}, hereafter T19)
and through the comparison with cosmological hydrodynamic simulations
without feedback.

The structure of the article is as follows. 
Section~2 describes the SAM used  for this article, which we call {\sc GalICS~2.1} to distinguish it from the previous version {\sc GalICS~2.0} \citep[C17]{cattaneo_etal17}.
Section~2 contains our analysis of the shock-stability condition, explains how we compute the cooling rate of the hot gas, and presents our new model for stellar feedback, 
along with a few other improvements since C17.
Section~3 compares the predictions of models without feedback to cosmological hydrodynamic simulations without feedback and the predictions of models with feedback to observations 
of the GSMF and its evolution with redshift.
Section~4 discusses our results.
Section~5 summarises the conclusions of the article.

\section{Model}

Sections~2.1 and 2.2 present an overview of how {\sc GalICS~2.1} follows the DM and the baryons, respectively.
Sections~2.3 and 2.4 elaborate on the physics that are most relevant for this article: shock heating and cooling, respectively.
{ The presentation of the feedback model is split into two
  parts. Section~2.5 explains how we compute outflow
  rates. Section~2.6 discusses the fate of the ejected gas.}

\subsection{N-body simulation and orphan galaxies}

{\sc GalICS~2.1} is a SAM based on merger trees from cosmological N-body simulations
(in fact, {\sc GalICS} is an acronym for Galaxies In Cosmological Simulations).
The simulation used for this article assumes the same cosmology ($\Omega_{\rm M}=0.308$, $\Omega_\Lambda=0.692$, $\Omega_{\rm b} = 0.0481$,  $\sigma_8=0.807$; \citealp{planck14}, {\it Planck} + WP + BAO),
the same volume (a cube with side-length $L=100\,$Mpc) and the same initial conditions as in C17.
The only difference is that the number of particles has increased from $512^3$ to $1024^3$.
Due to improved resolution, we can now reproduce the halo mass
function down to $M_{\rm vir}\simeq 4\times 10^9\,{\rm M}_\odot$.

The completeness limit above refers to isolated haloes. A subhalo may
be ten times more massive than that and still be missed by the halo
finder\footnote{ Our halo finder ({\sc HaloMaker};
  \citealp{tweed_etal09})  is based on {\sc AdaptaHOP} \citep{aubert_etal04}.}
if it is at the centre of a massive structure, such as a group or cluster.
If we assume a one-to-one correspondence between galaxies and N-body haloes,
then the premature disappearance of subhaloes due to the  halo
finder's incapacity to detect them artificially  hastens mergers {\citep{springel_etal01}.}
Hence, the masses of central galaxies are overestimated and the number
of satellite galaxies is underpredicted {\citep{guo_etal10}}.

The situation is even worse if the subhalo is not detected only temporarily, e.g, while it crosses a massive system.
If the structure that reemerges to the other side is mistaken for a new halo, then a new galaxy may be created in it.
If the galaxy that reemerges falls back again and merges with the central galaxy of the massive system, as it often does, its contribution to the mass of the latter will be counted twice.

The problem of mergers' being counted more than once
is easily dealt with because the reemerging structure will be identified as a descendent of the massive system\footnote{A halo descends from a progenitor if most of its particles belonged to the progenitor at the previous timestep.}
but not as its main descendent. Hence, it  will be classified as a fragment or splinter halo, and we do not allow the formation of galaxies in such systems.
However, even if all mergers are counted only once, there can still be overmerging.

To solve the overmerging problem, we must allow satellite galaxies to survive for some time after the subhaloes they live in are no longer detected.
{ We refer to these galaxies as orphan galaxies and to their subhaloes, which are no longer detected, as ghost subhaloes.}

{ Orphan galaxies are a well established ingredient of SAMs
  (\citealp{somerville_primack99,hatton_etal03,cattaneo_etal06,cora06,delucia_blaizot07,benson12,lee_yi13,gonzalez_etal14,gargiulo_etal15};
also see \citealp{knebe_etal15} and references therein)}.
Our model for orphan galaxies is taken from \citet{lee_yi13} and \citet{tollet_etal17}, who treated ghost subhaloes as test particles and integrated their equations of
motion in the gravitational potential of the host halo starting from the position ${ r}$ and the velocity ${ v}$ 
in the reference frame of the host halo at the time of last detection.
This new approach was introduced in {\sc GalICS~2.0} by
\citet{koutsouridou_cattaneo19} but we describe it in detail so that
all the changes since C17 are fully documented.

The equation of motion for a ghost subhalo is:
\begin{equation}
\dot{ \bf v}=-{{\rm G}M_{\rm h}(r)\over r^3}{\bf r}-{4\pi{\rm G}^2\rho_{\rm h}M_{\rm s}{\rm\, ln}\Lambda
\over v^3}
f\left({v\over \sqrt{2}\sigma}\right){\bf v}
\label{eom}
\end{equation} 
{\citep{chandrasekhar43}}, where $M_{\rm h}(r)$ is the mass of the host halo within radius $r$, $\rho_{\rm h}$ is density of the host halo at radius $r$,
both computed assuming an NFW \citep{navarro_etal97} model, $M_{\rm s}$ is the mass of the ghost subhalo, ${\rm ln}\Lambda = {\rm ln}(1+M_{\rm h}/M_{\rm s})$ is the so-called Coulomb logarithm,
$\sigma$ is the radial velocity dispersion of the DM particles, and
\begin{equation}
f(x)= {\rm erf}(x)-{2\over\sqrt{\pi}}xe^{-x^2}.
\label{fofx_chandra}
\end{equation}
The first term on the right-hand side of Eq.~(\ref{eom}) corresponds to the gravitational attraction that the host halo exerts on the subhalo.
The second term is the deceleration due to dynamical friction, without which there would be no orbital decay and thus no merging.

The solution of Eq.~(\ref{eom}) is complicated by the dependence on $M_{\rm s}$ in the second term.
Assuming that $M_{\rm s}$ maintains the same value it had at the time of last detection overestimates the dynamical friction force 
and hastens orbital decay. We solve this problem by fitting an NFW profile to the subhalo at the time
of last detection and by truncating this profile at the tidal radius $R_{\rm t}$ computed with the formula:
\begin{equation}
{M_{\rm s}(R_{\rm t})\over R_{\rm t}^3}=3{M_{\rm h}(r)\over r^3},
\label{Rt}
\end{equation}
where $M_{\rm s}(R_{\rm t})$ is the subhalo mass within $R_{\rm t}$ from the centre of the subhalo
\citep{tollet_etal17}. { This procedure gives subhalo masses
  consistent with those that the halo finder returns for detected
  subhaloes. The only difference is that for ghost subhaloes $R_{\rm t}$ is not allowed to
  increase again after it has decreased.}

{ By construction, in our model mergers can occur only at pericentric passages}. To decide after how many pericentric passages a satellite merges with the central galaxy,
we use \citet{jiang_etal08}'s fit to the merging timescale $t_{\rm df}$ measured in cosmological hydrodynamic simulations:
\begin{equation}
t_{\rm df}={1.17(0.94\epsilon^{0.6}+0.6)\over{\rm log}\Lambda}{M_{\rm h}\over M_{\rm s}}{r_{\rm vir}\over v_{\rm vir}},
\label{tdf}
\end{equation}
where $t_{\rm df}$ is computed from the time the satellite enters the virial radius of the host and $\epsilon$ is the orbital circularity, that is,
the ratio of the actual angular momentum to that of a circular orbit with the same total energy
($\epsilon=1$ for circular orbits and $\epsilon=0$ for radial orbits).
The only difference between Eq.~(\ref{tdf}) and \citet{chandrasekhar43}'s dynamical friction formula is the term in brackets, which
contains the dependence on circularity.
 
The dynamical friction time $t_{\rm df}$ computed with Eq.~(\ref{tdf}) sets the initial value of a merging countdown timer, which begins to tick for 
all detected and ghost subhaloes at the time they first enter $r_{\rm vir}$. Satellites merge with the central
galaxy at the pericentric passage that is closest in time to when the merging countdown timer comes to zero (see \citealp{tollet_etal17} for further details).

{ Ghost subhaloes did not exist in the original (sub)halo catalogues
  generated by the halo finder. The process of completing the
  (sub)halo catalogue at each N-body output timestep 
and modifying the merger trees accordingly was done once, while
preparing \citet{tollet_etal17}, and is not repeated every time we run
our SAM.}

A tree contains all haloes that merge into one object by $z=0$. A subhalo that survives at $z=0$ is not part of its host's merger tree.
The notion of forest {\citep{behroozi_etal13,rodriguezpuebla_etal16}} extends the notion of tree by considering not only progenitor - descendent relationships but also host halo - subhalo relationships.
Two trees belong to the same forest if any of their haloes are linked by a host - subhalo relation or share a common host halo.
The notion of forest is important because  the evolution of baryons is computed in parallel forest by forest and this affects the way 
the chemical enrichment of the intergalactic  medium (IGM) is computed (Section~2.2).

\subsection{The modelling of baryons}

Our presentation of how {\sc GalICS~2.1} follows the evolution of
  baryons within haloes
 concentrates on those areas where the code has changed since C17.
 We refer to that article for a general overview of the structure of our SAM. {\sc GalICS~2.1} is identical  to {\sc GalICS~2.0} (C17) where no differences are explicitly mentioned.
 The current section refers to a number of processes without describing them because there are entire sections in which we elaborate on them in detail
  (shock heating: Section~2.3; cooling: Section~2.4; feedback: Sections~2.5 and 2.6; the Kelvin-Helmholtz instability: Appendix~B).

\subsubsection{Gas accretion onto haloes}

In the beginning all the baryons were in the IGM and there were no
metals. As soon as the first haloes appeared, they began to
accrete baryons from the IGM and to return mass and metals to the IGM
through stellar feedback (there is no AGN feedback in {\sc
  GalICS~2.1}).

To achieve greater computational performance, {\sc GalICS~2.1} computes the evolution of the baryons in several forests simultaneously.
As each forest is followed separately, each forest must have its own IGM, which we initialise
by giving each forest a share of the total baryon mass $\Omega_{\rm b}\rho_{\rm c}L^3$  in the computational volume
($\rho_{\rm c}$ is the critical density of the Universe). The mass of the IGM assigned to each forest is proportional to its maximum  DM mass 
throughout the lifetime of the Universe (the mass  of a forest is the sum of the masses of all haloes and subhaloes within the forest).

\begin{figure}
\includegraphics[width=1.0\hsize]{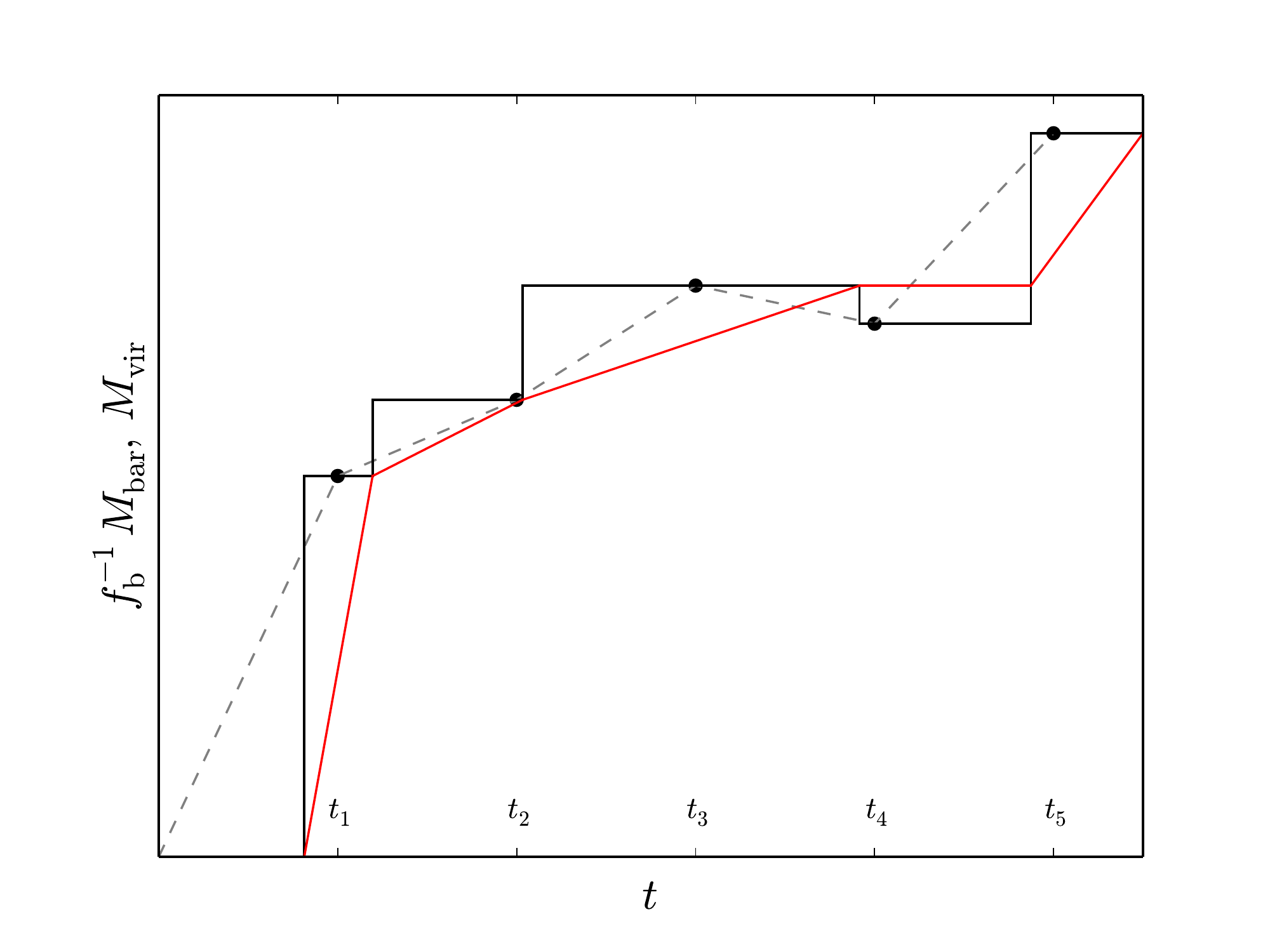} 
\caption{ Growth history of a DM halo and its baryonic content.
  The black symbols show the virial masses measured by the halo finder 
 at the times $t_1$, ..., $t_5$. The black line
 shows $M_{\rm vir}(t)$ in {\sc GalICS~2.1}.
 The transition between the mass at $t_i$ and the mass at $t_{i+1}$
 occurs instantaneously at a random time between $t_i$ and $t_{i+1}$.
 $M_{\rm bar}(t)$ is the halo baryon content in our SAM without feedback.
 The red solid line shows $M_{\rm bar}(t)$ rescaled by
 $f_{\rm b}^{-1}$ to make the comparison with the black line
 easier. 
$M_{\rm bar}/M_{\rm vir} < f_{\rm b}$ when the red curves
 are below the black histograms.
 The gray dashed
 line linearly
 interpolates the masses measured by the halo finder.
 ). 
}
\label{accretion}
\end{figure}

Introduced for greater computational speed, the parallel calculation forest by forest has the implication that the metals produced by a forest remain within the forest.
{ Cosmological hydrodynamic simulations by \citet{khalatyan_etal08}
  suggest that this is not an unreasonable assumption.
  \citet[SIMBA simulations]{borrow_etal19} have shown that  10 per cent of
  the cosmic baryons
  have moved by more than $5\,$Mpc, although $\lsim 20$ per cent of the baryons in a typical
  Milky Way-mass halo came from another halo.
  This cosmological baryon transfer is primarily due to entrainment of gas by jets from active galactic nuclei (AGN).
  To the best of our knowledge, {\sc GalICS~2.1} is the only
  SAM that follows the chemical
  enrichment of the IGM by forest or otherwise. In all other SAMs, there is no cosmological baryon transfer at all.

Our model for the accretion of baryons onto haloes is based on the
notion that, in {the} absence of feedback, all haloes should contain the 
universal baryon fraction. Hence, the baryonic accretion rate should
be $\dot{M}_{\rm accr}=f_{\rm b}\dot{M}_{\rm vir}$ with $f_{\rm b}=\Omega_{\rm b}/\Omega_{\rm M}$
The problem is that $M_{\rm vir}$ grows in
jumps and its growth is stochastic rather than monotonic
(Fig.~\ref{accretion}).

This stochasticity is largely due to our excellent time
resolution, the purpose of which is to follow the orbits of satellite
galaxies. {Let $t_1$
and $t_2$ be the ages of the Universe that correspond to two consecutive N-body snapshots.}
If $t_2-t_1$ is so small that the real mass growth between $t_1$
and $t_2$ is smaller than the accuracy of the halo finder, then
the evolution of $M_{\rm vir}$ on short timescales is dominated { by}
noise, in which case a continuos interpolation of the mass measurements
(Fig.~\ref{accretion}, gray dashes)
{would} not lead to more accurate results.

The simplest solution to the non-monotonic growth of halo masses would be to update the halo baryon content $M_{\rm bar}$ whenever $M_{\rm vir}$ grows and
to assume that $M_{\rm bar}(t)=f_{\rm b}M_{\rm max}(t)$, where $M_{\rm
  max}(t)$ is the maximum $M_{\rm vir}$ at all times $\le t$.
The main concern is that this could trigger spurious starbursts when
$M_{\rm bar}$ suddenly increases. 
This concern is probably exaggerated because there is a delay between
accretion onto haloes and accretion onto galaxies, but we have preferred to
be cautious and to spread the baryonic mass increase $f_{\rm b}(M_2-M_1)$
associated with the jump in halo mass at $t_{1,2}$ over
the entire time interval between $t_{1,2}$ and $t_{2,3}$, the latter being
the time when the halo mass jumps from $M_2$ to $M_3$.
This scheme gives:
\begin{equation}
\dot{M}_{\rm bar}(t_{1,2}<t<t_{2,3})=f_{\rm b}{M_2-M_1\over
  t_{23}-t_{12}}.
\label{dotMbar}
\end{equation}
for $M_2>M_1$, and $\dot{M}_{\rm bar}(t_{1,2}<t<t_{2,3})=0$ for
$M_2\le M_1$.
The  red line in Fig.~\ref{accretion} shows the $M_{\rm bar}(t)$ that corresponds to Eq.~(\ref{dotMbar}).

It is important to realise that the choice of spreading the accretion over the
time interval $(t_{1,2},t_{2,3})$ is a numerical scheme, not a
physical assumption. By construction, our scheme can underestimate
$M_{\rm bar}$ but cannot overestimate it if the
baryonic mass does not decrease when $M_{\rm vir}$
does\footnote{Bar numerical errors, tidal interactions are the only
  physical process that could cause a halo to lose mass.
Their effect is greater on the DM than on the more concentrated
baryonic component.}.

We can test the goodness of our scheme by comparing the baryon content of
our haloes with the universal baryon fraction $\Omega_{\rm b}/\Omega_{\rm M}=0.16$.
Without feedback, the {\it halo} baryon fraction  is $f_{\rm b}=0.166\pm 0.020$ for
haloes with $M_{\rm vir}>10^{11}\,{\rm M}_\odot$ at
$z=0$. A difference of less than 4 per cent with the universal baryon fraction is
entirely acceptable
if we consider that $M_{\rm vir}$ is measured with an accuracy of 
10--20 per cent (\citealp{knebe_etal13} on the accuracy of halo finders).

Two competing mechanisms make the scatter in $f_{\rm b}$ larger for subhaloes.
First, { in {\sc GalICS~2.1}, tidal stripping reduces $M_{\rm vir}$  but not the baryonic mass because
 tidal stripping of baryons is not included, while stripping of DM is calculated self-consistently.}
Second, we do not allow subhaloes to accrete
gas even if the DM mass increases. In practice, this rarely happens in our
merger trees. Most subhaloes lose mass through tidal stripping already before the
halo finder classifies them as subhaloes
\citep{behroozi_etal14,tollet_etal17,koutsouridou_cattaneo19}.
The two above mechanisms push the baryon fraction for subhaloes up and down,
respectively.
Subhaloes with $M_{\rm vir}>10^{11}\,{\rm M}_\odot$ have a baryon fraction
of $f_{\rm b}=0.161\pm 0.070$ at $z=0$. The variation in the mean value of $f_{\rm b}$ is not significant
but the standard deviation from the mean is three to four times larger for subhaloes.
}{ We note that the predicted $f_{\rm b}$ for satellites is almost certainly overestimated because
the SAM used for this article does not include ram-pressure stripping.}

{ 

Feedback suppresses the accretion of gas onto the haloes of
dwarf galaxies so that
$f_{\rm b}=M_{\rm bar}/M_{\rm vir}$
is lower than the universal baryon fraction $\Omega_{\rm b}/\Omega_{\rm M}$ \citep{brook_etal12,hopkins_etal14,wang_etal17,mitchell_etal18}.
We assume that $f_{\rm b}$ depends on $v_{\rm vir}$ through the
  functional form:
\begin{equation}
f_{\rm b}(v_{\rm vir})= {\Omega_{\rm b}\over\Omega_{\rm M}}\left[{\rm erf}\left({v_{\rm vir}\over \sigma}\right)-{2\over\sqrt{\pi}}{v_{\rm vir}\over\sigma} e^{-\left ( {v_{\rm vir}\over\sigma}\right)^2}\right].
\label{fofx}
\end{equation}
{In C17, we had presented a physical justification for Eq.~(\ref{fofx}) based on the notion that  the low baryon fractions of low-mass haloes are
due to heating of the IGM by a photoionising 
ultaviolet background  \citep{blanchard_etal92,efstathiou92,gnedin00,croton_etal06}.
Hence, the parameter $\sigma$ was called $\sigma_{\rm reio}$.
The problem with that interpretation, to which we no longer adhere, is that we require $\sigma> 25{\rm\,km\,s}^{-1}$ to match the 
stellar mass function of galaxies and
photoionisation feedback 
 is important only at $v_{\rm
  vir}<20{\rm\,km\,s}^{-1}$ \citep{okamoto_etal08}.
 
 Supernovae (SNe) can suppress gas accretion onto haloes with virial velocities as high as
 $v_{\rm
  vir}=30$--$40{\rm\,km\,s}^{-1}$ by disrupting the infall of gas on scales as large as $6\,r_{\rm vir}$
  (T19; also see
\citealp{mitchell_etal18} and preliminary work by T.~Gutcke in
\citealp{cattaneo19})\footnote{Haloes with $v_{\rm vir}\gg 40{\rm\,km\,s}^{-1}$  have accreted the universal baryon
fraction even though the final baryon fraction may be lower than the
universal one because gas has been expelled from the halo (T19).},
and are more important than cosmic reionisation in the mass range that is relevant for this article ($v_{\rm
  vir}\simeq 20{\rm\,km\,s}^{-1}$ is our resolution limit in circular velocity).
  
 Eq.~(\ref{fofx}) is still used for convenience rather than based on the physical considerations that motivated its introduction.
 We find that  Eq.~(\ref{fofx}) matches the results of T19 for $\sigma=34{\rm\,km\,s}^{-1}$ and that is the value of $\sigma$ that we use in {\sc GalICS~2.1}.}

\subsubsection{Baryonic components within a halo:  differential equations}

Gas that accretes onto a halo at the rate computed with Eq.~(\ref{dotMbar}) can either stream onto the central galaxy in cold filamentary
flows or be shock heated and contribute to the formation of a hot atmosphere within the halo.
We model these two possibilities by introducing a discrete parameter
$f_{\rm hot}$ that takes the value $f_{\rm hot}=0$ in the first
situation and $f_{\rm hot}=1$ in the second one.
{ We treat $f_{\rm hot}$ as a binary variable because our analysis
  of shock heating is based on a stability criterion. A shock is either stable or unstable.
The shock-stability criterion will be discussed at length 
in Section~2.3.}

Depending on whether $f_{\rm hot}=0$ or $f_{\rm hot}=1$, the gas that accretes onto the halo is added to the mass $M_{\rm fil}$ of the cold filaments  within the halo 
or the mass $M_{\rm hot}$ of the hot gas, respectively.

The gas in the cold filaments accretes onto the central galaxy on the freefall time: 
\begin{equation}
t_{\rm ff}=\int_0^{r_{\rm vir}}{{\rm d}r\over u_1(r)},
\label{tff}
\end{equation}
where $u_1=\sqrt{-2\phi}$ is the freefall speed from rest at infinity and $\phi(r)$ is the gravitational potential of the halo, computed assuming an NFW  profile.

In haloes with massive hot atmospheres, filaments may be disrupted by
the Kelvin-Helmholtz (KH) instability { on a timescale $t_{\rm KH}$
  that we compute in Appendix~B not to interrupt the main flow of the article.}

Not only do galaxies accrete gas from the IGM; they also
return gas to the IGM through feedback.
{ Re-accretion of ejected gas has been considered by
  SAMs for a long time (e.g., \citealp{font_etal08}).} The novelty of {\sc GalICS~2.1} is that it separates galactic winds into a cold component with outflow speed $v_{\rm w}^{\rm cold}\ll v_{\rm esc}$ 
and a hot component with outflow speed  $v_{\rm w}^{\rm hot}\gsim v_{\rm esc}$, where $v_{\rm esc}$ is the escape speed from the halo.

The cold component remains within the halo, forms a galactic fountain
with total mass $M_{\rm fount}$, and accretes back onto the galaxy
{ if the reaccretion timescale $t_{\rm reaccr}$ is shorter than the
  timescale $t_{\rm ev}$ for thermal evaporation
(the values of $t_{\rm reaccr}$ and  $t_{\rm ev}$
are discussed in Section~2.6 ``Stellar feedback: fate of the
ejected gas''). 

Hot winds escape from the halo unless a confining hot atmosphere is present, in which
case they mix with the hot atmosphere. Gas that escapes from
subhaloes mixes with the hot atmospheres of their hosts.}

The differential equations for the exchanges of matter between the IGM, filaments within the halo, the hot atmosphere, the galaxy 
and the galactic fountain are:
\begin{equation}
\dot{M}_{\rm fil}=(1-f_{\rm hot})\dot{M}_{\rm bar}-{M_{\rm fil}\over t_{\rm ff}}-{M_{\rm fil}\over t_{\rm KH}},
\label{dotMfil}
\end{equation}
\begin{equation}
\dot{M}_{\rm hot}=f_{\rm hot}\dot{M}_{\rm bar}-\dot{M}_{\rm cool}+(1-f_{\rm esc})\dot{M}_{\rm w}^{\rm hot}+{M_{\rm fil}\over t_{\rm KH}}+{M_{\rm fount}\over t_{\rm ev}},
\label{dotMhot}
\end{equation}
\begin{equation}
\dot{M}_{\rm gal}={M_{\rm fil}\over t_{\rm ff}}+\dot{M}_{\rm cool}+{M_{\rm fount}\over t_{\rm reaccr}}-\dot{M}_{\rm w}^{\rm cold}-\dot{M}_{\rm w}^{\rm hot},
\label{dotMgal}
\end{equation}
\begin{equation}
\dot{M}_{\rm fount}=-{M_{\rm fount}\over t_{\rm reaccr}}+\dot{M}_{\rm w}^{\rm cold}-{M_{\rm fount}\over t_{\rm ev}},
\label{dotMfount}
\end{equation}
where  $M_{\rm gal}$ is the baryonic mass of the central galaxy. In the equations above,
the terms proportional to $\dot{M}_{\rm bar}$ represent the accretion of gas from the IGM.
$M_{\rm fil}/t_{\rm ff}$ is the rate at which gas accretes from the filaments onto the galaxy (cold-mode accretion).
$M_{\rm fil}/t_{\rm KH}$ is the rate at which the filaments evaporate through the KH instability.
$M_{\rm fount}/t_{\rm ev}$ is the rate at which the fountain evaporates because of thermal conduction 
(whatever gas evaporates  from the filaments or the fountain is added to the hot atmosphere).
$\dot{M}_{\rm cool}$ is the rate at which the hot gas cools onto the galaxy (hot-mode accretion).
$\dot{M}_{\rm w}^{\rm cold}$ and $\dot{M}_{\rm w}^{\rm hot}$ are the outflow rates for the cold wind and the hot wind, respectively.
$M_{\rm fount}/t_{\rm reaccr}$ is the rate at which gas is reaccreted through the fountain.
Finally, $f_{\rm esc}$ is the fraction of the hot-wind material that
escapes from the halo instead of mixing with the hot atmosphere\footnote{
In our code, $f_{\rm esc}$ and $f_{\rm hot}$ are binary variables (0 or
1) rather than real variables that can take all intermediate values.}.
Gas that escapes from a halo is not preferentially reaccreted but enriches the IGM.
It therefore increases the metallicity of the filaments that accrete onto the halo and other haloes within the same forest.

The version of {\sc GalICS~2.1} used for this article assumes strangulation (there is no gas accretion onto subhaloes) but does not contain any stripping.
{ Gas that accreted onto subhaloes prior to their becoming subhaloes 
remains within subhaloes and
keeps accreting onto satellite galaxies.}

\subsection{Shock stability condition and radius}

Having introduced the general framework used to compute gas accretion, we now enter the key part of this work: the analysis of the shock-heating condition,
which determines the value of $f_{\rm hot}$ in Eqs.~(\ref{dotMfil}) and (\ref{dotMhot}).

BD03 demonstrated that a spherical shell of shock-heated gas in hydrostatic equilibrium is stable to compressive perturbations
if its effective polytropic index is larger than a critical value $\gamma_{\rm c}$:
\begin{equation}
\gamma \equiv {{\rm dlog\,}p\over{\rm dlog\,}\rho} =\gamma_{\rm ad}-{1\over 3}{r_{\rm s}\over u_2}{|\dot{q}|\over e}>\gamma_{\rm c}.
\label{gamma}
\end{equation}
Here $p$ and $\rho$ are the pressure and the density of the gas, respectively,
$\gamma_{\rm ad}=5/3$ is the adiabatic index for a monoatomic gas, $r_{\rm s}$ is the radius of the shell, $u_2>0$ is the speed at which the shell is compressed 
(the post-shock infall speed in the reference frame of the shock),
$\dot{q}<0$ is the heat radiated by the gas per unit mass and $e$ is the specific internal energy.
Throughout this article, the subscripts 1 and 2 refer to pre-shock and post-shock quantities, respectively.
To avoid heavy notations, we have not used the subscript 2 after quantities that describe the thermal properties of the gas (internal energy, heat loss, temperature), 
since these quantities always refer to the post-shock hot component.

In a polytropic gas, $\gamma$ is a constant and $\gamma_{\rm c}=4/3$. BD03 found $\gamma_{\rm c}=10/7$ because they
considered that the effective polytropic index $\gamma$ defined in Eq.~(\ref{gamma}) may depend on radius
and eliminated the freedom introduced by this possibility by assuming
that the temperature $T$ of the shock-heated gas is constant across
the shell.
{ We shall follow their approach although the difference between
  $10/7$ and $4/3$ is too small to produce any significant difference
  in our results.}

Let 
\begin{equation}
t_{\rm comp}={r_{\rm s}\over u_2}
\label{tcomp}
\end{equation}
be the compression time  and let
\begin{equation}
t_{\rm cool} \equiv {e\over |\dot{q}|} ={{3\over 2}\langle n\rangle
kT{\rm\,d}\Omega\over
\langle n_{\rm H}^2\rangle\Lambda(T,Z_{\rm accr}){\rm\,d}\Omega}
\label{tcool}
\end{equation}
be the cooling time, where $n$ and  $T$ are the post-shock number density and temperature, respectively,
$Z_{\rm accr}$ is the metallicity of the accreted gas,
$k$ is the Boltzmann constant, and
$\Lambda(T,Z_{\rm accr})$ is the cooling function of \citet{sutherland_dopita93} normalised to a number density of hydrogen atoms of 
$n_{\rm H}=1{\rm\,cm}^{-3}$. { For a plasma that is three quarters
  hydrogen and one quarter helium in mass,
  the particle number density is $n={27\over 12}n_{\rm H}$, since
  there are 27 particles (12 protons, 1 helium nucleus and 14 free
  electrons) for every 12 hydrogen atoms.

$\langle n\rangle$ and $\langle n_{\rm
  H}^2\rangle$ are functions of $r_{\rm s}$. Averages appear because
the density $n$ within a filament is not homogeneous even at a given radius.
The quantity $C=\langle n^2\rangle/\langle n\rangle^2$ is known as
the clumping factor.  
Its significance is that a clumpy medium radiates more efficiently.
In filaments, the Jeans length is usually large enough to prevent the
gas from fragmenting into clouds\footnote{ The Jeans length
  $\lambda_{\rm J}=c_{\rm s}\sqrt{\pi\over{\rm G}\rho_{\rm fil}}$ is
  determined by the sound speed $c_{\rm s}$ and the density $\rho_{\rm
    fil}$ of the gas in the filaments. After reionisation (the
  Universe was completely reionised by $z=6$) , the
  temperature of the gas in the filaments is $\gsim 10^4\,$K and thus
  $c_{\rm s}\sim 17{\rm\,km\,s}^{-1}$.
The mean density of the gas in the filaments is $\sim 40$ times the
cosmic baryon density (e.g., \citealp{mandelker_etal18}).
At $z=6$, this gives $\rho_{\rm fil}\sim 10^{-24}{\rm\,g\,cm}^{-3}$
and thus $\lambda_{\rm J}\sim 4\,$kpc for $c_{\rm s}\gsim 17{\rm\,km\,s}^{-1}$.
 $\lambda_{\rm
  J}$ grows as the density of the Universe decreases.
At $z=0$, $\lambda_{\rm J}\sim 50\,$kpc corresponds to a Jeans mass
of order $10^{10}\,{\rm M}_\odot$.
This calculation assumes that gaseous filaments are self-gravitating.
The presence of DM stabilises the gas even further.
Hence filaments are not expected to break into clumps unless the
clumps are galaxies with masses comparable to the Milky Way.
},
but the formation of clouds is not the only mechanism that can produce
$C>1$.
There is also the tendency of the cold gas to concentrate in the cores
of the filaments.
In this article, we assume $C=9$ based on a numerical study of the distribution of cold gas within filaments (Ramsoy et al., in prep.; also
see Appendix~A).}

\begin{figure}
\begin{center}$
\begin{array}{c}
\includegraphics[width=1.0\hsize]{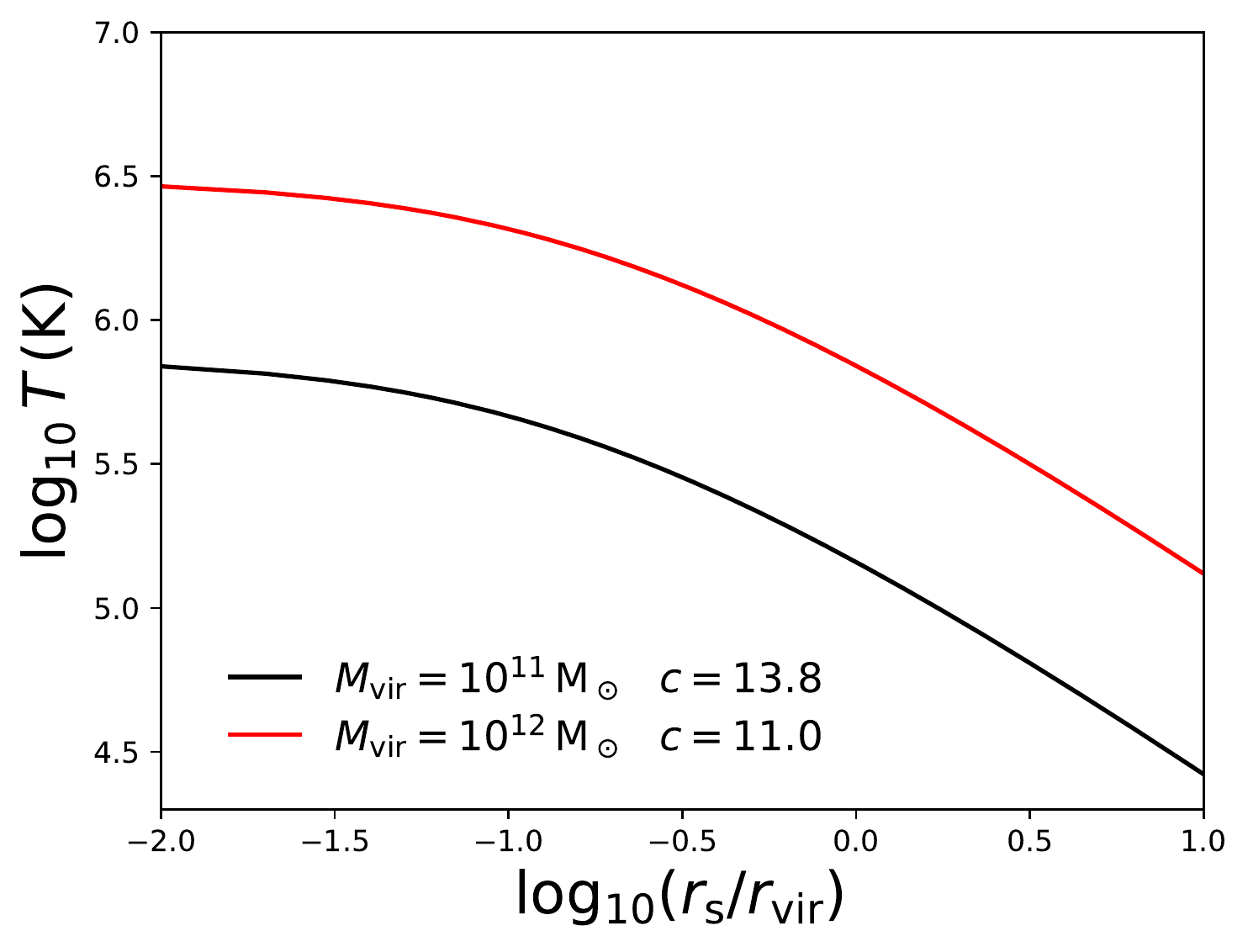} \\
\includegraphics[width=1.0\hsize]{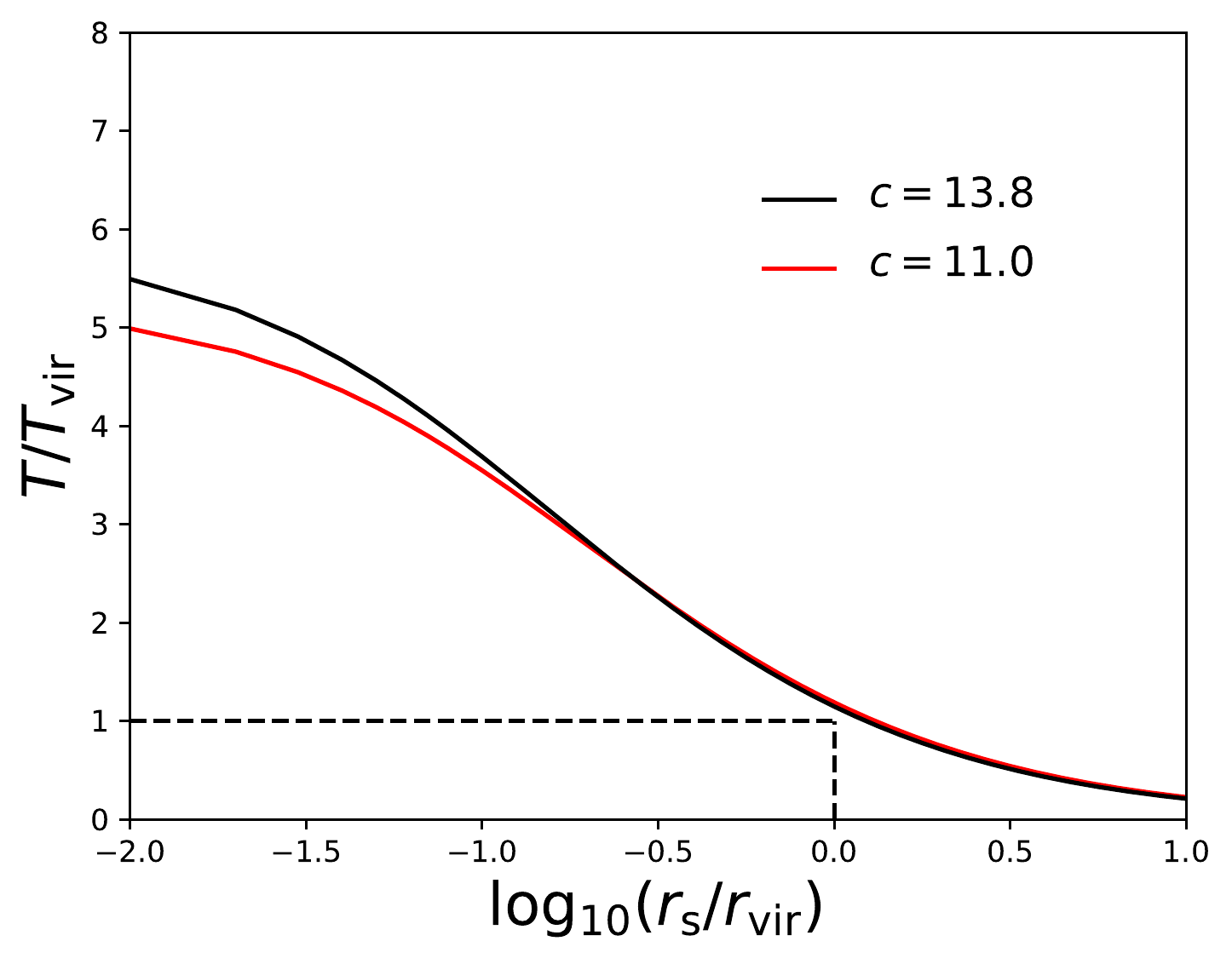} 
\end{array}$
\end{center}
\caption{Post-shock temperature $T$ as a function of the shock radius
  $r_{\rm s}$ for $M_{\rm vir}= 10^{11}\,{\rm M}_\odot$ (black curves) and
  $M_{\rm vir}= 10^{12}\,{\rm M}_\odot$ (red curves).
  { The figure is for $z=0$;
   $c=13.8$ and $c=11$ are the concentrations that correspond to the two halo masses
based on the fitting formulae by \citet{dutton_maccio14}.}
$T$ is shown in both kelvins (above) and virial
units (below).
The difference between the red curve and the black curve in the lower
panel is entirely due to the lower concentration of the halo with
$M_{\rm vir}= 10^{12}\,{\rm M}_\odot$
{ because there is no  dependence on mass  when the relation is shown in virial units, apart from the one through $c$.}
The dashed lines in the bottom panel show that $T\simeq T_{\rm vir}$ for $r=r_{\rm vir}$.}
\label{Trs}
\end{figure}

For a fully ionised plasma ($T>10^{4.5}-10^5\,$K) that is three
quarters hydrogen and one quarter helium in mass, 
the mean particle mass is  $\mu={16\over 27}m_{\rm p}$, where $m_{\rm p}$ is the proton mass.
With the substitutions $n_{\rm H}={12\over 27}n$ and $n={\rho_2\over \mu}$,
Eq.~(\ref{tcool}) becomes:
\begin{equation}
t_{\rm cool} ={3\over 2}\left({27\over 12}\right)^2{1\over C}{\mu\over\langle\rho_2\rangle}{kT\over \Lambda(T,Z_{\rm accr})}.
\label{tcool2}
\end{equation}

The condition $\gamma>\gamma_{\rm c}$ can be written as a condition for the ratio of the compression time to the cooling time:
\begin{equation}
{t_{\rm comp}\over t_{\rm cool}}<\Gamma_{\rm c}\equiv 3(\gamma_{\rm ad}-\gamma_{\rm c}).
\label{tcomp_over_tcool} 
\end{equation}
Using $\gamma_{\rm c}=10/7$ gives $\Gamma_{\rm c}=5/7$ (BD03 found a critical condition $t_{\rm comp}/t_{\rm cool}<1$ because they
reabsorbed $\Gamma_{\rm c}$ in their definition of $t_{\rm comp}$). 

The post-shock quantities $u_2$, $\rho_2$ and $T$  that enter Eqs.~(\ref{tcomp}) and (\ref{tcool2}) are computed from the pre-shock speed $u_1$ and the pre-shock density $\rho_1$ by
applying the Rankine-Hugoniot conditions for a strong shock. For a monoatomic gas ($\gamma_{\rm ad}=5/3$), the Rankine-Hugoniot conditions give $u_2=u_1/4$, $\rho_2=4\rho_1$ and
\begin{equation}
T={3\over 16}{\mu\over k}u_1^2.
\label{temperature}
\end{equation}
We remark that the Rankine-Hugoniot conditions assume conservation of energy and that this assumption is invalid when $t_{\rm cool}\ll t_{\rm comp}$.
In that case the post-shock density can jump to $\rho_2\gg\rho_1$ and the cooling time becomes even shorter, but
this has no practical implications for our results because all we care about is whether Eq.~(\ref{tcomp_over_tcool}) is satisfied or not.
Once we have established that $t_{\rm comp}/t_{\rm cool}>\Gamma_{\rm c}$ everywhere we do not care whether the real value of  $t_{\rm comp}/t_{\rm cool}$ is even higher than what we have calculated.

The calculations of $u_1$ and $\langle\rho_1\rangle$ are based on assuming that the pre-shock gas is in free fall from infinity and that it obeys the equation of continuity.
The approximation of free fall may appear extreme since the haloes into which filaments fall are not empty and since small shocks within filaments may slow down the cold gas.
However, hydrodynamic simulations by \citet{rosdahl_blaizot12} have shown that this is a reasonable assumption indeed.
If we further assume that the shock is static ($\dot{r}_{\rm s}=0$), 
so that the infall speed with respect to the shock is equal to the infall speed with respect to the halo, we find:
\begin{equation}
u_1=\sqrt{-2\phi(r_{\rm s})},
\label{u1}
\end{equation}
where $\phi(r_{\rm s},M_{\rm vir},c)$ is the gravitational potential of the DM halo, computed assuming an NFW profile.

The calculation of $\langle\rho_1\rangle$ is one of the main differences between this work and DB06.
A stationary accretion flow is a reasonable assumption as long as  $\dot{M}_{\rm bar}$
does not vary significantly on timescales comparable to the freefall time $t_{\rm ff}$ in Eq.~(\ref{tff}). For this assumption, the equation of continuity gives:
\begin{equation}
\langle\rho_1\rangle\Omega = {\dot{M}_{\rm bar}\over u_1r_{\rm s}^2},
\label{rho1}
\end{equation}
where $\dot{M}_{\rm bar}$ is the baryonic accretion rate onto the halo and $\Omega$ is the solid angle covered by the cold flows.
Eq.~(\ref{rho1}) gives average pre-shock densities $\sim 27$ times
lower than those  found by DB06\footnote{DB06 assumed $\rho_1\propto
  r^{-2}$ and normalised $\rho_1$ by requiring that
that $\rho_1(r_{\rm vir})$ equals the total density at the virial
radius times the universal baryon fraction.} but this
difference is largely compensated by taking $C=9$, so that the final difference in the density of the gas is only a factor of three.

\begin{figure}
\begin{center}$
\begin{array}{c}
\includegraphics[width=1.0\hsize]{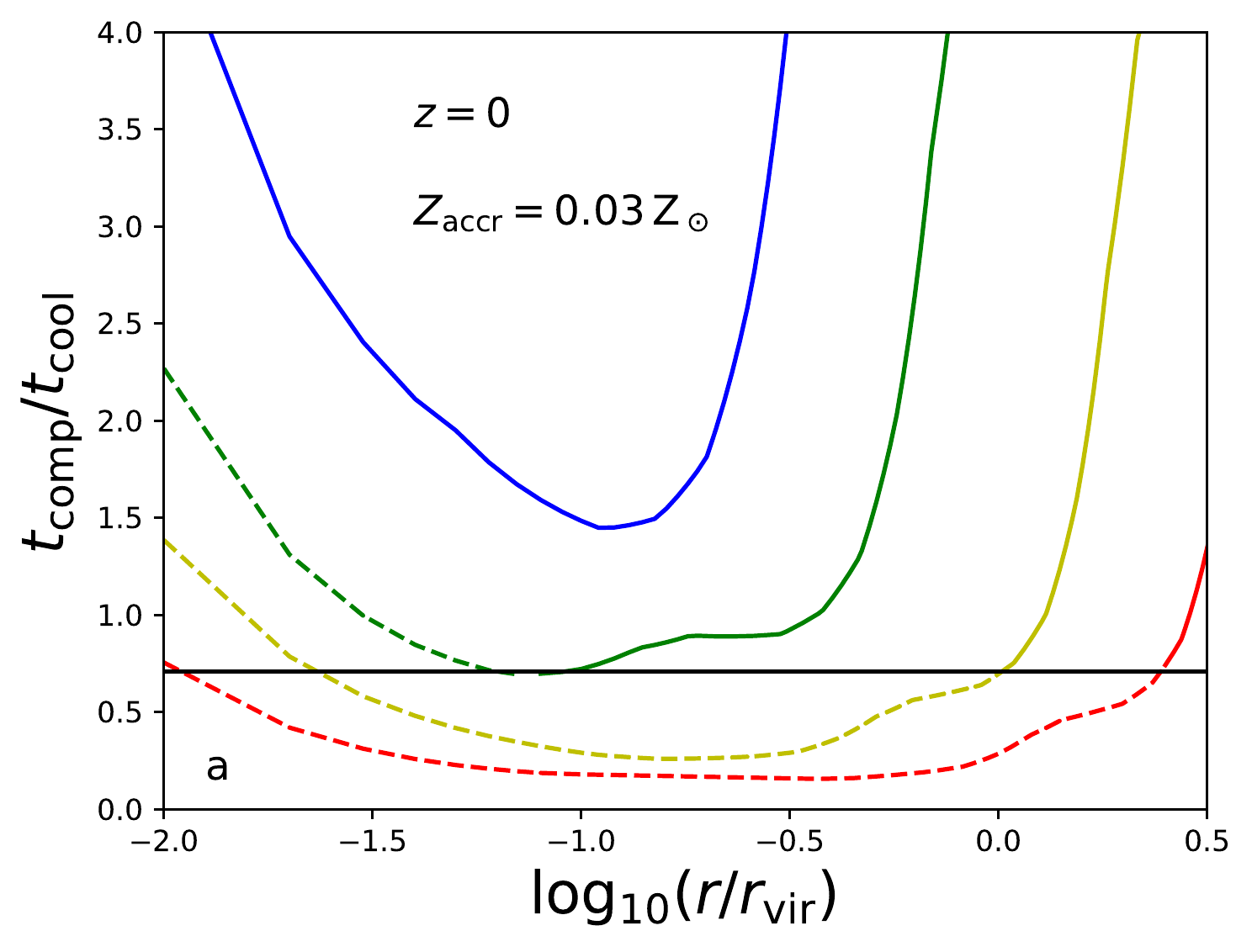}\\
\includegraphics[width=1.0\hsize]{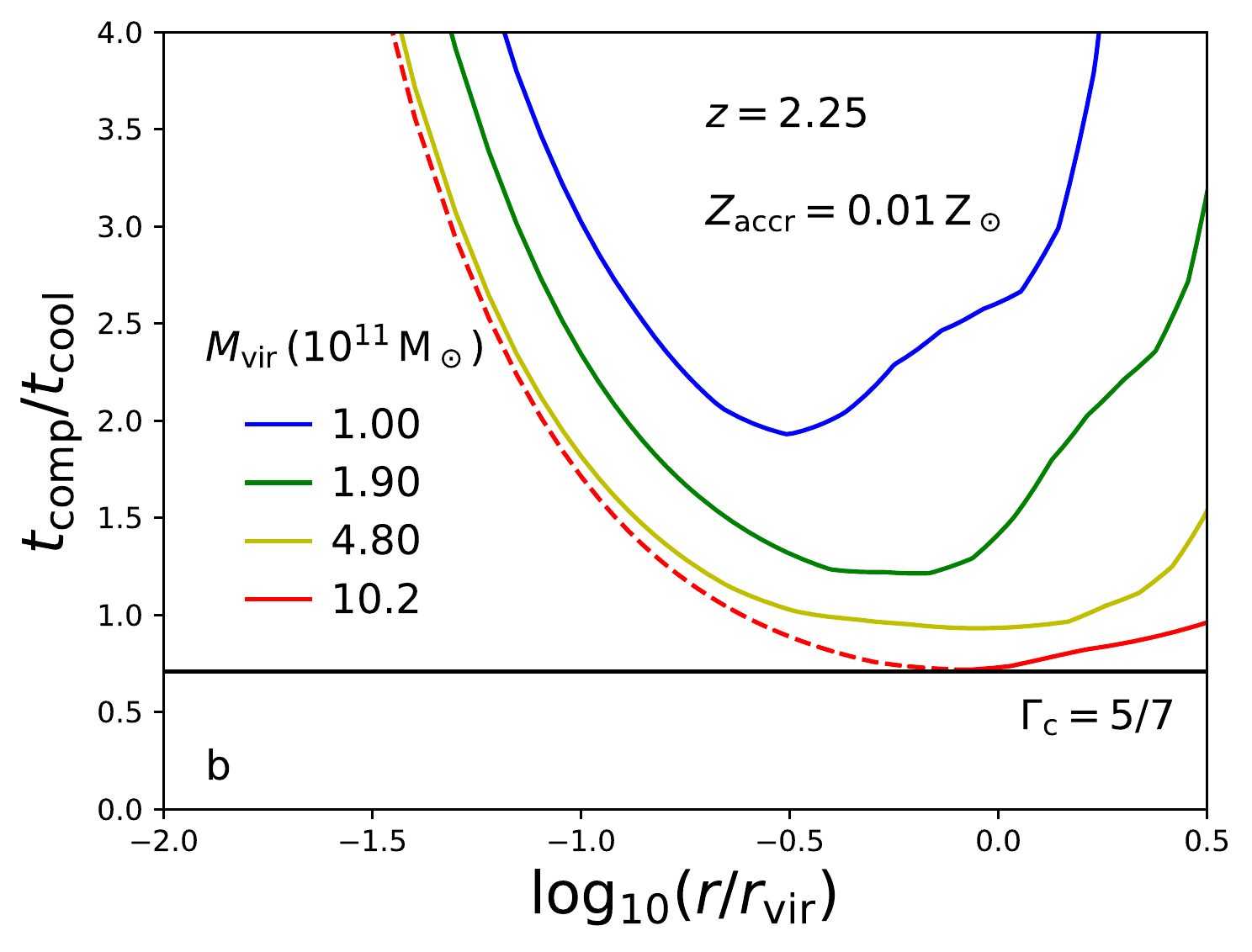}
\end{array}$
\end{center}
\caption{The ratio of the compression time $t_{\rm comp}$ to the
  cooling time $t_{\rm cool}$ as a function of $r_{\rm s}$ for
$M_{\rm vir}=10^{11}\,{\rm M}_\odot$ (blue curves),
 $M_{\rm vir}=1.9\times 10^{11}\,{\rm M}_\odot$ (green curves), $M_{\rm vir}=4.8\times 10^{11}\,{\rm M}_\odot$ (yellow curves) and
 $M_{\rm vir}=1.02\times 10^{12}\,{\rm M}_\odot$ (red curves). 
 The physical part of each curve ($r\ge r_{\rm s}$) is shown with a solid line. The part at $r< r_{\rm s}$ is shown with dashes.
 The two panels correspond to different combinations of redshift and the metallicity:
 a) $z=0$ and $Z_{\rm accr}=0.03\,{\rm Z}_\odot$, b) $z=2.25$ and
 $Z_{\rm accr}=0.01\,{\rm Z}_\odot$. 
 The horizontal black line $t_{\rm comp}/t_{\rm cool}=\Gamma_{\rm c}=5/7$ shows the critical ratio below which the infalling gas is shock heated.}
\label{tcomp_over_tcoolSD}
\end{figure}

By combining Eqs.~(\ref{tcool2}), (\ref{tcomp_over_tcool}), (\ref{temperature}) and (\ref{rho1}), we obtain our final shock-stability condition:
\begin{equation}
{t_{\rm comp}\over t_{\rm cool}}\simeq 0.5\,C{\dot{M}_{\rm bar}\over\Omega r_{\rm s}}{\Lambda(T,Z_{\rm accr})\over (kT)^2}<\Gamma_{\rm c}=0.71.
\label{tcomp_over_tcool2}
\end{equation}

Eq.~(\ref{tcomp_over_tcool2}) shows that the stability of the shock-heated gas depends on: i) 
the shock radius $r_{\rm s}$, ii) the virial mass $M_{\rm vir}$ and concentration $c$ (which enter Eq.~\ref{tcomp_over_tcool2} through $T$),
iii) the baryonic accretion rate $\dot{M}_{\rm bar}$,
iv) the metallicity $Z_{\rm accr}$ of the accreted gas, and 
v) the  solid angle $\Omega$ from which the gas is accreted.

In {\sc GalICS~2.1}, $M_{\rm vir}$, $c$ and $\dot{M}_{\rm bar}=f_{\rm b}\dot{M}_{\rm vir}$ are measured from the N-body simulation used to extract the merger trees.
$Z_{\rm accr}$ is the metallicity of the IGM in the forest to which the halo belongs.
The calculation of $\Omega$ is presented in Appendix~A.
Hence, the only quantity that remains to be determined to evaluate Eq.~(\ref{tcomp_over_tcool2}) is the shock radius itself.
DB06 treated $r_{\rm s}$ as a free parameter and argued for $r_{\rm s}=0.1r_{\rm vir}$ based on both observations and hydrodynamic simulations.
Here we make a step forward by computing $r_{\rm s}$ self-consistently from our model.

The shock radius $r_{\rm s}$ enters Eq.~(\ref{tcomp_over_tcool2}) both
directly, through the factor $r_{\rm s}^{-1}$, and indirectly, through
the factor
$\Lambda(T)/T^2$, since:
\begin{equation} 
T={3\over 4}{|\phi(r_{\rm s})|\over v_{\rm vir}^2}T_{\rm vir}
\label{T_over_Tvir}
\end{equation}
(Eqs.~\ref{temperature} and \ref{u1}), where $T_{\rm vir}=0.5\mu v_{\rm vir}^2/k$ is the temperature of a singular isothermal sphere with circular velocity $v_{\rm vir}$.
We have introduced $T_{\rm vir}$ because it provides a useful reference temperature although we do not assume a singular isothermal sphere anywhere in our SAM.

Eq.~(\ref{T_over_Tvir}) implies that $T$ is a decreasing function of $r_{\rm s}$.
Fig.~\ref{Trs} shows that, although the post-shock temperature increases with halo mass, its value and radial dependence in virial units are almost universal except for a small concentration effect
(massive haloes are less concentrated; hence, they have lower central temperatures in virial units).
It also shows that gas could be shock-heated to  several times $T_{\rm vir}$ if it is in free fall all the way down to the centre of the DM halo.
However, this is likely to happen only when the shock-heating process begins. 
Once a stable shock forms, $r_{\rm s}$ rapidly propagates outward. Hence, gas is rarely shock-heated to $T\gg T_{\rm vir}$.

In {\sc GalICS~2.1}, $\rho_{\rm vir}$ and $c$ are taken from the N-body merger trees, and 
$\dot{M}_{\rm bar}$ is computed for each individual halo.
However, to explore the mean behaviour, we have plotted $t_{\rm comp}/t_{\rm cool}$ as a function of radius 
when $\rho_{\rm vir}$, $c$  and $\dot{M}_{\rm bar}$ are computed with
the fitting formulae in  \citet{bryan_norman98}, \citet{dutton_maccio14} and \citet{dekel_etal09}, respectively
(Fig.~\ref{tcomp_over_tcoolSD}; different colours correspond to different halo masses).

To understand the curves' qualitative behaviour,
let $\nu(T)$ be the local exponent of the cooling function, so that we can write $\Lambda\propto T^\nu$ in a neighbourhood of $T$. Eq.~(\ref{tcomp_over_tcool2}) implies:
\begin{equation}
{t_{\rm comp}\over t_{\rm cool}}\propto r^{-1}T^{\nu-2}.
\end{equation}
 $T^{\nu-2}$ is a growing function of $r$ because $T$ decreases with $r$ (Fig.~\ref{Trs}) and $\nu<2$ for all $T>2\times 10^4\,$K.
The product of a growing function with a decreasing one ($r^{-1}$)
gives a function with a minimum.
Physically, cold gas comes in from large radii. The ratio $t_{\rm comp}/t_{\rm
  cool}$ decreases as the gas falls in.
The minimum occurs when the temperature approaches its maximum value (Fig.~\ref{Trs}). At smaller radii, $T^{\nu-2}$ is almost constant and the growth 
of $t_{\rm comp}/t_{\rm cool}$ is driven by the $r^{-1}$ term.
Shock heating occurs  when the curves drop below the black horizontal
line $t_{\rm comp}/t_{\rm cool}=\Gamma_{\rm c}$.  The shock-heating
condition is therefore:
\begin{equation}
{\rm min}\left({t_{\rm comp}\over t_{\rm cool}}\right)<\Gamma_{\rm c},
\label{tcomp_over_tcool3}
\end{equation}
where the minimum is over all radii with $r\le r_{\rm vir}$.
We set $f_{\rm hot}=1$ when Eq.~(\ref{tcomp_over_tcool3}) is satisfied and $f_{\rm hot}=0$ when it is not.
  
Fig.~\ref{tcomp_over_tcoolSD}a and Fig.~\ref{tcomp_over_tcoolSD}b are
representative of the Universe at $z=0$ and $z=2.25$, respectively.
The main differences are the metallicity of the accreted gas, the density of the Universe, and the solid angle of accretion.

In Section~3.2, we shall see that our SAM predicts $Z_{\rm  IGM}\sim 0.03\,{\rm Z}_\odot$ at $z=0$ and
$Z_{\rm IGM}\sim 0.01\,{\rm Z}_\odot$ at $z=2.25$,
and that these values are consistent with the observations. Hence, the justification for the metallicities assumed  in Fig.~\ref{tcomp_over_tcoolSD}a and Fig.~\ref{tcomp_over_tcoolSD}b, respectively.
We warn, however, that our predictions for $Z_{\rm IGM}$ 
are sensitive to our assumptions about feedback and that observational measurements of $Z_{\rm IGM}$ contain large uncertainties, too.
This cautionary note is important because using different metallicities can change the critical mass for shock heating by up to an order of magnitude (Fig.~4 of DB06).

In Fig.~\ref{tcomp_over_tcoolSD}a, the blue curve ($M_{\rm vir}= 10^{11}\,{\rm M}_\odot$) corresponds to a
low-mass halo where
$t_{\rm comp}/t_{\rm cool}$ 
never falls below $\Gamma_{\rm c}$ and gas is never shock heated.
A shock appears for the first time
at $r_{\rm s}\lsim 0.1r_{\rm vir}$
when the halo mass reaches  $M_{\rm vir}^{\rm crit}\simeq 2\times
10^{11}\,{\rm M}_\odot$ (green curve).
The shock propagates to $r_{\rm s}=r_{\rm vir}$ by the time  $M_{\rm
  vir}\simeq 5\times 10^{11}\,{\rm M}_\odot$ (yellow curve). Only the solid parts of the curves
are physical because the assumption of freefall, on which all curves are based, does not apply to the post-shock gas  with $r<r_{\rm s}$.

At $z=2.25$, the lower metallicity of the IGM prolongs the cooling time 
and
pushes the curves down, but this is only one of the effects at play.

The higher virial density has a more complex effect because it reduces both the cooling time and the compression time.
 The compression time decreases with $\rho_{\rm vir}$ as $t_{\rm comp}\propto\rho_{\rm vir}^{-1/2}$.
The cooling time scales as $t_{\rm cool}\propto\rho_{\rm vir}^{-1}T^{1-\nu}$  if the gas density is proportional to the virial density. This gives
$t_{\rm cool}\propto\rho_{\rm vir}^{(1-\nu)/3-1}$, since $T\propto\rho_{\rm vir}^{1/3}$ at fixed $M_{\rm vir}$.
$M_{\rm vir}^{\rm crit}\sim 10^{11}\,{\rm M}_\odot$ corresponds to a temperature range where $\nu\sim -3/4$, from which one finds $t_{\rm cool}\propto\rho_{\rm vir}^{-5/12}$.
Hence, $t_{\rm comp}$ and $t_{\rm cool}$ decrease with $\rho_{\rm vir}$ in a  similar manner.
This is why DB06 find that $M_{\rm vir}^{\rm crit}$ varies very little with $z$ at constant $Z_{\rm accr}$ and $\Omega$.

In our model, however, the post-shock density used to compute $t_{\rm cool}$ is proportional to $\dot{M}_{\rm bar}$ (Eq.~\ref{rho1}) and is not constrained to scale as $\rho_{\rm vir}$.
 \citet{dekel_etal09}'s fitting formula for $\dot{M}_{\rm bar}$,  used in Fig.~\ref{tcomp_over_tcoolSD} (but not in {\sc GalICS~2.1}),
  contains a strong dependence on redshift,
 $\dot{M}_{\rm bar}\propto M_{\rm vir}^{1.15}(1+z)^{2.25}$, which causes the  post-shock density to grow with $z$ faster than the virial density.
 This is why, in  Fig.~\ref{tcomp_over_tcoolSD}, the density effect dominates and  $t_{\rm cool}$ decreases with $z$ more rapidly than $t_{\rm comp}$ does.
 
Accretion become highly anisotropic for
 $M_{\rm vir}\gg M_{\rm nl}$,  where $ M_{\rm nl}$ is the non-linear mass (Appendix~A).
At $z=0$, $M_{\rm nl}\simeq 10^{13}\,{\rm M}_\odot$, so that 
$\Omega\simeq 4\pi$ over the  entire mass range in Fig.~\ref{tcomp_over_tcoolSD}.
At $z=2.25$, however, $M_{\rm  nl}\simeq 1.3 \times 10^{11}\,{\rm M}_\odot$. Hence,
anisotropic accretion boosts the green, yellow and red curves in Fig.~\ref{tcomp_over_tcoolSD}b by factors of $4\pi/\Omega=1.1,\,1.5$ and $1.9$, respectively.

The combined effect of metallicity, density and solid angle pushes the characteristic shock-heating mass up from $M_{\rm vir}^{\rm crit}\simeq 2\times 10^{11}\,{\rm M}_\odot$ at $z=0$ (green curve in Fig.~\ref{tcomp_over_tcoolSD}a)
to $M_{\rm vir}^{\rm crit}\simeq 10^{12}\,{\rm M}_\odot$ at $z=2.25$ (red curve 
in Fig.~\ref{tcomp_over_tcoolSD}b). 
The fivefold increase of $M_{\rm vir}^{\rm crit}$ between $z=0$ and  $z=2.25$ is largely driven by the lower $\Omega$ at $z=2.25$.
If we assume $\Omega=4\pi$ at all $z$, we find $M_{\rm vir}^{\rm crit}=3.9\times 10^{11}\,{\rm M}_\odot$ at $z=2.25$, in which case the increase is only twofold.

\subsection{Cooling}

{ Our model to compute the cooling rate $\dot{M}_{\rm cool}$ of the
  hot gas follows closely the one by \citet{white_frenk91}. The
  cooling radius $r_{\rm cool}$ is the maximum radius within which
  both the cooling time:
  \begin{equation}
t_{\rm cool}={3\over 2}\left({27\over 12}\right)^2\epsilon_{\rm
  cool}{\mu\over\rho}{kT\over\Lambda(T,Z_{\rm hot})}={3\over 2}\left({27\over 12}\right)^2\epsilon_{\rm
    cool}{K^{3\over 2}\over\Lambda\sqrt{kT}}
\label{ttcool}
\end{equation}
and the freefall time:
\begin{equation}
t_{\rm ff}=\int_0^r{{\rm d}r'\over\sqrt{2[\phi(r)-\phi(r')]}}
\label{tff_cf}
\end{equation}
are shorter than the time $t$ available for cooling (in {\sc
  GalICS~2.1}, the time since $M_{\rm hot}>0$). Here, $\rho$, $T$, $Z_{\rm hot}$ and $K$ are the density,
temperature, 
metallicity and entropy of the hot gas, respectively  (see \citealp{cattaneo_etal09} for the relation between the entropy
constant $K=kTn^{-{2/3}}$ and the real thermodynamic entropy), while
$\epsilon_{\rm cool}$ is a fudge-factor calibrated on hydrodynamic
simulations.
Our semianalytic calculations overestimate the cooling time because
they are based on a static density profile
(Eq.~\ref{rho_prof}). The real cooling time is shorter because the
gas condenses while it cools (e.g., \citealp{cattaneo_teyssier07}). The parameter
$\epsilon_{\rm cool}$ was introduced to correct this shortcoming.

The freefall time for cooling flows
  (Eq.~\ref{tff_cf}) differs from the freefall time for cold streams
  (Eq.~\ref{tff}) because cooling flows start with zero speed
  (they are assumed to develop from
  initial conditions of
  hydrostatic equilibrium), while cold
  streams enter the virial radius at speeds comparable to the virial velocity.

The mass of the hot gas that cools and
flows onto the central galaxy between $t$ and $t+\Delta t$ is the
mass of the hot gas within $r_{\rm cool}$ at $t+\Delta t$ minus the
mass $M_{\rm cooled}$ 
of the gas that had already cooled at time $t$.

Appendix~C shows that this scheme gives cooling rates in reasonably
good agreement with hydrodynamic simulations for $\epsilon_{\rm
  cool}=0.8$
when the initial conditions for $\rho(r)$, $T(r)$ and $Z_{\rm hot}$ are
the same.

The initial conditions are the true problem, especially since the notion of initial
conditions is intrinsically ill-defined in a hierarchical cosmology.
 We argue that the initial density profile that gives
the correct $\dot{M}_{\rm cool}$ is the one that the hot gas would have if it
had not cooled. Hence, the initial density profile is a theoretical construct rather than an observable quantity (\citealp{benson_bower11} referred to it as the ``notional" profile).
To establish it, we start from
 adiabatic cosmological simulations, which show us how the Universe would have evolved without cooling.

The gas density distribution in adiabatic cosmological hydrodynamic
simulations follows a profile of the form:
\begin{equation}
\rho = \rho_0\left(1+{r\over 0.59r_0}\right)^{-3},
\label{rho_prof} 
\end{equation}
\citep{faltenbacher_etal07}, where $r_0$ is the core radius of the DM halo (fitted with an NFW
profile) and the central density $\rho_0$ is determined by the
condition that the baryon fraction converges to the universal one at
large radii. 

\citet{faltenbacher_etal07} found smooth cores in the entropy distribution, too. Their results for the entropy distribution at large radii,
$K\propto r^{1.21}$,  imply  $\gamma=1.26$
for a polytropic equation of state $T\propto\rho^{\gamma-1}$.
This polytropic index  agrees well with
observations of galaxy clusters, which are approximately adiabatic
because of their long cooling times. \citet{komatsu_seljak01},
\citet{ascasibar_etal06} and \citet{ghirardini_etal17} find
$\gamma\simeq 1.2$ for concentrations $5\lsim c\lsim 7$.

 $T\propto\rho^{0.26}$ implies that the temperature decreases with
radius much less rapidly than the density does. Hence, $\rho$
(or $K$) determines the radial dependence of
$t_{\rm cool}$, while $T$ is a nearly uniform quantity that sets its normalisation. 

Most SAMs assume $T=T_{\rm vir}$.
We determine $T$ by substituting Eq.~(\ref{rho_prof}) and
\begin{equation}
T = T_0\left(1+{r\over 0.59r_0}\right)^{-3(\gamma-1)}
\label{T_prof}
\end{equation}
into the hydrostatic-equilibrium equation:
\begin{equation}
{1\over\rho}{{\rm d}\over{\rm d}r}{\rho kT\over\mu}=-\rho{{\rm d}\phi\over{\rm d}r}.
\label{hee}
\end{equation}
We cannot require that Eq.~(\ref{hee}) should hold everywhere because
Eq.~(\ref{rho_prof}) is not an exact solution of the
hydrostatic-equilibrium equation. However, by
taking the limits for $r\rightarrow 0$
of both the left-hand side and the right-hand side, Eq.~(\ref{hee})
becomes:
\begin{equation}
T_0={0.59\over 6\gamma}{c\over f_{\rm NFW}(c)}{\mu v_{\rm vir}^2\over k},
\end{equation}
where  $f_{\rm NFW}=\ln(1+c)-c/(1+c)$.

Adiabatic simulations are most relevant to systems with a long cooling time, such as galaxy clusters,
where most of the baryons are in the form of hot gas. Adapting their results to the case $M_{\rm hot}/M_{\rm vir}\ll \Omega_{\rm b}/\Omega_{\rm M}$ is not straightforward, but we assume
that Eq.~(\ref{rho_prof}) can be extended to the non-adiabatic case by replacing the condition $\rho/\rho_{\rm NFW}\rightarrow \Omega_{\rm b}/\Omega_{\rm M}$ for $r\rightarrow\infty$ with the more general one:
\begin{equation}
\lim_{r\rightarrow\infty}{\rho\over\rho_{\rm NFW}}={M_{\rm hot}\over M_{\rm vir}},
\label{rho0_norm}
\end{equation}
where $\rho_{\rm NFW}$ is the halo density profile obtained by fitting  an NFW model.

Admittedly arbitrary from a theoretical standpoint, this assumption is nevertheless consistent with observations spanning more than three orders of magnitude in halo mass (\citealp{voit_etal19}b), which include:
 galaxy clusters \citep{cavagnolo_etal09}, elliptical galaxies \citep{babyk_etal18,singh_etal18}, the Milky Way \citep{voit19}, and galaxies with $10^{11.5}{\rm\,M}_\odot\lsim M_{\rm vir}\lsim 10^{12.5}{\rm\,M}_\odot$,
 the hot circumgalactic media of which have been studied through absorption-line measurements with the Cosmic Origins Spectrograph (\citealp{voitg_etal19}a).
 
 In Section~3.1, we show that Eq.~(\ref{rho0_norm}) gives results in good agreement with cosmological hydrodynamic simulations with cooling when feedback is turned off in both {\sc GalICS~2.1} and the simulations
 for a cleaner comparison.

\subsection{Stellar feedback: outflow rates}

Supernovae (SNe) enrich the surrounding gas with metals and drive winds that can eject large masses of gas from galaxies (ejective feedback).
In haloes with $M_{\rm vir}\lsim 10^{10}\,{\rm M}_\odot$, SN-driven winds can also disrupt the filaments that supply dwarf galaxies with gas (pre-emptive feedback; T19).

We model chemical enrichment by assuming the instantaneous-recycling approximation with a returned fraction of $R=0.43$ and a yield of $y=0.03$, both defined as in C17\footnote{The
values assumed here are those appropriate for a \citet{chabrier03}
initial mass function if all stars more massive than $50\,{\rm M}_\odot$ implode into a black hole and do not release any metals into the environment at the end of their lives.}) 

Pre-emptive feedback is implicitly accounted for by assuming $\sigma=34{\rm\,km\,s}^{-1}$ (Section~2.2.1). %
If this value were interpreted as the thermal velocity dispersion of the IGM,
it would imply an unrealistically high reionisation temperature of $80,000\,$K.
In reality, this value mimics the effects of pre-emptive feedback,
which limits the baryonic mass that can accrete onto a halo with
$M_{\rm vir}=7\times 10^9\,{\rm M}_\odot$ to $\sim 20\%$ the
universal baryon fraction (T19).

The modelling of ejective feedback can be separated into two problems: the calculation of the outflow rate (how much gas is blown out of galaxies?) and the fate of the ejected gas
(where does it go?)
{ Here we deal with the former question and postpone consideration
  of the latter to Section~2.6.}

Our model for the  outflow rate  is similar to the one in C17:
\begin{equation}
\dot{M}_{\rm w}={2\epsilon_{\rm SN}E_{\rm SN}\Psi_{\rm SN}\over v_{\rm w}^2}{\rm SFR},
\label{dotMw}
\end{equation}
\citep{silk03}, where $\Psi_{\rm SN}=1/140$ is the number of SNe for
$1\,{\rm M}_\odot$ of star formation assuming a \citet{chabrier03} initial mass function,
$E_{\rm SN}=10^{51}\,$erg is the energy released by one SN, $\epsilon_{\rm SN}$ is the fraction of this energy used to power the wind, and
$v_{\rm w}$ is the wind speed. We apply this model with the same parameters to both quiescent and bursty star formation.

Eq.~(\ref{dotMw}) hides the complexity of SN feedback in the values of $v_{\rm w}$ and $\epsilon_{\rm SN}$, 
which are intrinsically ill-defined because
of the multiphase structure of galactic winds.
In M82, for example, the outflow speed is $v_{\rm w}\sim 1400$--$2400{\rm\,km\,s}^{-1}$ for the ionized X-ray-emitting gas \citep{strickland_heckman09},
$v_{\rm w}\sim 600{\rm\,km\,s}^{-1}$ for the H$\alpha$-emitting clumps \citep{shopbell_blandhawthorn98}, and
only $v_{\rm w}\sim 160{\rm\,km\,s}^{-1}$ for ${\rm H_2}$ \citep{beirao_etal15}. \citet{heckman_etal15} and \citet{chisholm_etal17} find typical outflows speed of 
 $v_{\rm w}\sim 300$--$500{\rm\,km\,s}^{-1}$ for warm gas.
Hydrodynamic simulations confirm a broad distribution of outflow speeds and find that the wind mass per dex of $v_{\rm w}$ decreases rapidly with $v_{\rm w}$ (T19), 
so that low-speed gas contains most of the mass and high-speed gas contains most of the energy.

In {\sc GalICS~2.1}, we separate winds into a hot component (with outflow rate $\dot{M}_{\rm w}^{\rm hot}$) and a cold component (with outflow rate $\dot{M}_{\rm w}^{\rm cold}$).
The entraining hot component has outflow speed $v_{\rm w}\sim v_{\rm esc}$, where $v_{\rm esc}$ is the escape speed from the virial radius\footnote{The minimum outflow speed to escape to $r>r_{\rm vir}$ starting from $r=0$ is: $$v_{\rm esc}=\sqrt{2[\phi(r_{\rm vir})-\phi(0)]},$$ where $\phi$ is the gravitational potential, which we compute assuming an NFW profile.
We assume that the hot wind has outflow speed $v_{\rm w}^{\rm hot}\sim v_{\rm esc}$, even though particles with speeds much larger than $v_{\rm esc}$ do exist, because the wind's velocity distribution is a strongly decreasing function
of outflow speed (T19). Hence, most of the wind mass with $v_{\rm w}\ge v_{\rm esc}$ is composed by particles with $v_{\rm w}\sim v_{\rm esc}$.},
and carries most of the energy ($\epsilon_{\rm SN}^{\rm hot}\gg\epsilon_{\rm SN}^{\rm cold}$). The entrained cold component has much lower outflow speeds $v_{\rm w}^{\rm cold}\ll v_{\rm esc}$ and carries most of the mass, so that
$\epsilon_{\rm SN}^{\rm cold}/(v_{\rm w}^{\rm cold})^2>\epsilon_{\rm SN}^{\rm hot}/(v_{\rm w}^{\rm hot})^2$.

If we introduce the mass-loading factors $\eta_{\rm hot}=\dot{M}_{\rm w}^{\rm hot}/{\rm SFR}$ and $\eta_{\rm cold}=\dot{M}_{\rm w}^{\rm cold}/{\rm SFR}$ for the hot wind and the cold wind, respectively, then the mass-loading factor for
the dominant cold component can be rewritten as:
\begin{equation}
\eta_{\rm cold}={\eta_{\rm cold}\over\eta_{\rm hot}}{2\epsilon_{\rm SN}^{\rm hot}E_{\rm SN}\Psi_{\rm SN}\over v_{\rm esc}^2}.
\label{eta}
\end{equation}
The usefulness of this algebraic manipulation is that we can use cosmological hydrodynamic simulations to constrain $\eta_{\rm cold}/\eta_{\rm hot}$.

Cold winds have low outflow speeds, they remain within the halo, and form a galactic fountain with the reaccretion time given by Eq.~(\ref{treaccr}),
while hot winds are seldom reaccreted. Hence, the wind mass fraction processed through the fountain,
\begin{equation}
\epsilon_{\rm fount}={\eta_{\rm cold}\over \eta_{\rm cold}+\eta_{\rm hot}},
\label{epsilon_fount}
\end{equation}
is a proxy for the reaccreted fraction. If we know the reaccreted fraction, we can invert Eq.~(\ref{epsilon_fount}) to compute $\eta_{\rm cold}/\eta_{\rm hot}$.

Typical reaccreted fractions in cosmological hydrodynamic simulations are 20--70 per cent \citep{christensen_etal16}, 50--80 per cent (T19) and 60--90 per cent \citep{angles_etal17}.
{\sc GalICS~2.1} contains a parameter  $\epsilon_{\rm fount}$ that gives the fraction of the ejected gas processed through the galactic fountain.
A fountain fraction $\epsilon_{\rm fount}=0.7$ sits reasonably well within the above range and gives  $\eta_{\rm cold}/\eta_{\rm hot}=7/3$. Once this ratio is fixed,
$\dot{M}_{\rm w}^{\rm cold}$ and $\dot{M}_{\rm w}^{\rm hot}$ are entirely determined by $\epsilon_{\rm SN}^{\rm hot}$.

The energetic efficiency of SNe $\epsilon_{\rm SN}^{\rm hot}$ cannot be larger than unity but it cannot be constant either.
If we neglect a weak dependence on concentration, $v_{\rm esc}\propto v_{\rm vir}$. Therefore, a constant energetic efficiency would imply a mass-loading factor 
$\eta_{\rm cold}\propto\eta_{\rm hot}\propto v_{\rm vir}^\xi$ with $\xi=-2$.
In contrast, most SAMs fit the GSMF assuming a relation of the form $\eta=\dot{M}_{\rm w}/{\rm SFR}\propto v_{\rm vir}^\xi$ where $\xi<-2$.
\citet{cole_etal94} find $\xi=-5.5$. \citet{guo_etal11} find $\xi=-3.5$. \citet{somerville_etal12} find $\xi=-2.5$. \citet{lacey_etal16} find $\xi=-3.2$. C17 find $\xi=-6$. Only \citet{henriques_etal13} find $\xi=-0.92$.
Exponents $\xi<-2$ imply that $\epsilon_{\rm SN}$ must decrease with $v_{\rm vir}$, since $\epsilon_{\rm SN}\propto v_{\rm vir}^{\xi+2}$.

Cosmological hydrodynamic simulations confirm that both $\eta_{\rm cold}$ and $\eta_{\rm hot}$
decrease with $v_{\rm vir}$ faster than an inverse square law.
T19 found:
\begin{equation}
\eta_{\rm cold}\simeq\eta_0\left({v_{\rm vir}\over 100{\rm\,km\,s}^{-1}}\right)^{-5.6}
\label{eta_cold}
\end{equation}
with $\eta_0\simeq 2$ for $v_{\rm vir}\gsim 67{\rm\,km\,s}^{-1}$.
The relation flattens to $\eta_{\rm cold}\propto v_{\rm vir}^{-2}$ at  $v_{\rm vir}< 67{\rm\,km\,s}^{-1}$ where $\epsilon_{\rm SN}^{\rm hot}$ has reached its maximum value $\epsilon_{\rm SN}^{\rm hot}=1$.

In {\sc GalICS~2.1}, we use Eq.~(\ref{eta_cold}) and $\eta_{\rm cold}/\eta_{\rm hot}=7/3$ to compute $\epsilon_{\rm SN}^{\rm hot}$ 
unless this calculation yields a value larger than unity, in which case we assume $\epsilon_{\rm SN}^{\rm hot}=1$.
We treat $\eta_0$ as a free parameter of the model and we fit the GSMF with $\eta_0\simeq 4.5$.
This value is somewhat larger than the one in T19 ($\eta_0\sim 2$) but is still consistent with their results.

Eq.~(\ref{eta_cold}) is an {\it ad hoc} functional form introduced first to reproduce the observations and later to fit the results of cosmological hydrodynamic simulations.
Yet, it has a simple physical justification if the distribution of outflow speeds is similar for all galaxies.
At low masses, $v_{\rm w}>v_{\rm esc}$ for all particles, so the entire SN energy goes with the wind (except for radiative losses).
At high masses, a lot of this energy is wasted accelerating particles that do not reach the escape speed. Therefore, only
a small fraction of the SN energy ends up in the component with $v_{\rm w}\gsim v_{\rm esc}$.

\subsection{Stellar feedback: fate of the ejected gas}

Cold winds are assumed to be part of a galactic fountain with
reaccretion time:
\begin{equation}
t_{\rm reaccr}=1.1\left({M_{\rm vir}\over 10^{11}\,{\rm M}_\odot}\right)^{-0.18}{\rm\,Gyr}
\label{treaccr}
\end{equation}
based on cosmological hydrodynamic simulations (T19).

In low-mass haloes that lack a hot atmosphere, all the gas in the
fountain is eventually reaccreted.
If the fountain is embedded in a massive hot atmosphere,
thermal conduction may cause the clumpy cold phase to evaporate 
(the gas clouds in the fountain have much smaller diameters than the
filaments; therefore thermal conduction is more relevant to them than
the KH instability).

Thermal conductivity is proportional to the temperature of the hot
gas to the power $5/2$  \citep{spitzer56}.
Therefore, it is reasonable to assume that the timescale for thermal evaporation should follow a relation of the form:
\begin{equation}
  t_{\rm ev}=t_{\rm ev}^0\left({T_0\over 10^6{\rm\,K}}\right)^{-2.5}{\rm\,Gyr}
\label{tev}
\end{equation}
\citep{cowie_mckee77}, where $T_0$ is the central temperature of the hot gas.
Hydrodynamic simulations by \citet{armillotta_etal17} find that the cold phase has a half-lifetime of $\sim 0.25\,$Gyr for $T_0=2\times 10^6\,$K.
{This half-lifetime corresponds to an $e$-folding time of  $\sim 0.35\,$Gyr for $T_0=2\times 10^6\,$K if $M_{\rm fount}$ declines exponentially,
from which we infer
 $t_{\rm ev}^0\simeq 2\,$Gyr since our evaporation times are normalised at $T_0=10^6\,$K. Notice, however,
 that this figure is based on assuming a cloud radius of $a=250\,$pc; $t_{\rm ev}\propto a^2$ \citep{cowie_mckee77}
 and smaller $a$ cannot be excluded.}

{ Hot winds have $v_{\rm w}=v_{\rm esc}$ by construction (we use this
assumption to compute $\eta_{\rm hot}$; Section~2.5)} but
starting with $v_{\rm w}\ge v_{\rm esc}$ is not a sufficient condition
to escape from the halo 
if a massive hot atmosphere stifles their expansion.
We therefore introduce the supplementary escape condition:
\begin{equation}
\dot{M}_{\rm w}v_{\rm w}>M_{\rm hot}g_{\rm h},
\label{blowout_condition}
\end{equation}
where $g_{\rm h}=-{\rm d}\phi/{\rm d}r$ is the gravitational acceleration at the centre of the DM halo.
When Eq.~(\ref{blowout_condition}) is satisfied, the pressure force of
the wind is larger than the gravitational force. Therefore,
all the wind material escapes  into the IGM ($f_{\rm esc}=1$ in Eq.~\ref{dotMhot}) and all the hot gas is blown out with it.
If the condition in Eq.~(\ref{blowout_condition}) is not satisfied,
we set $f_{\rm esc}=0$ and all the gas that escapes from the galaxy is added to the hot atmosphere.

\section{Results}

Observations are the ultimate test of any theory but testable predictions are often the end point of a long chain of assumptions.
SAMs are no longer as degenerate as they were twenty years ago because the data used to constrain them have expanded considerably but outflow rates are still considerably uncertain
(see, e.g., \citealp{bouche_etal12,cicone_etal14,kacprzak_etal14,schroetter_etal15,schroetter_etal16,falgarone_etal17}).
Hence, it is possible to overestimate gas accretion and still reproduce the galaxy stellar mass function (GSMF) by compensating over-accretion with over-ejection. 

Cosmological simulations describe the hydrodynamics of gas accretion
much more accurately but their subgrid modelling of stellar feedback
suffers from uncertainties similar to those encountered in SAMs.
We therefore run {\sc GalICS~2.1} both without and with feedback. 
The model without feedback is compared to cosmological hydrodynamic simulations without feedback to test that we model gas accretion correctly.
The model with feedback is the realistic one that we compare with the observed GSMF.
The comparison with observations is used to calibrate the feedback model but not the accretion model, which is pre-calibrated on cosmological  hydrodynamic simulations.

\subsection{Models without feedback}

We want to compare {\sc GalICS~2.1} with a simulation for which we
have GSMFs both with cooling and cold accretion only.
This requirement restricts the comparison to a fairly old simulation with only $5\times 10^7$ particles in a cubic volume with side-length $50h^{-1}\,$Mpc (\citealp{keres_etal09}; K09)
This simulation also neglects metal enrichment, but this is an advantage rather than a problem for our comparison, because metals affect shock heating and cooling only when they mix
with the IGM and the circumgalactic medium. Neither possibility applies to {\sc GalICS~2.1} when there are no winds because feedback has been turned off.
Hence, a simulation without enrichment enables a cleaner comparison.
The cosmology assumed by K09 is very similar to ours and the only potential difference is the star-formation law,
which cannot be replicated exactly in a hydrodynamic simulation and a SAM, but the difference cannot be large because both K09's star-formation law 
and ours are calibrated to be consistent with the Kennicutt law.

{ Fig.~\ref{modelswofeedback} compares the local ($z=0.1$) GSMF in {\sc
  GalICS~2.1} (red curve) with the results by K09 (red
squares). The {\sc
  GalICS~2.1} results are for a clumping factor of $C=9$.
Its value has been determined from cosmological hydrodynamic simulations
(Ramsoy et al., in prep.; Appendix~A) and has not been tuned
to improve the agreement with K09.}

{ The red square at $M_{\rm stars}\lsim 10^{10}\,{\rm M}_\odot$ falls below the red curve because of K09's limited resolution}.
{\sc GalICS~2.1} has a much lower resolution limit $M_{\rm stars}\sim
10^{9}\,{\rm M}_\odot$ (Fig.~\ref{modelswofeedback}, gray shaded area).
Therefore, pre-emptive feedback (the dependence of $f_{\rm b}$ on
$v_{\rm vir}$) and not
resolution is the reason why the red curve departs from
a model that converts all baryons into stars  (the gray curve) at
$M_{\rm stars}<10^{10}\,{\rm M}_\odot$.

The discrepancy at $M_{\rm stars}\gsim 10^{12}\,{\rm M}_\odot$ has two causes.
First, K09's  smaller volume limits their ability to find rare massive objects.
Second, our N-body simulation contains an excess of massive objects with $\phi\lsim 10^{-3.8}\,{\rm Mpc}^{-3}$ (discussed in C17).

To study to what extent cosmic variance causes the red curve and the red squares to differ at high masses, we have used {\sc ramses} \citep{teyssier02} 
to run a low-resolution hydrodynamic simulation (equivalent to using
$512^3$ gas particles) that has the same initial conditions as the N-body simulation used to generate the merger trees. 
Its results are shown by the red circles. 
The red curve remains consistent with the new simulation without  any need to readjust $C$. 
In fact, the agreement with the red circles is better than the
agreement with the red squares. { The red curve passes through all red
  circles at $\phi> 10^{-3.8}\,{\rm Mpc}^{-3}$ (those above the gray
  shaded area at the bottom of the diagram).}

\begin{figure}
\includegraphics[width=1.00\hsize]{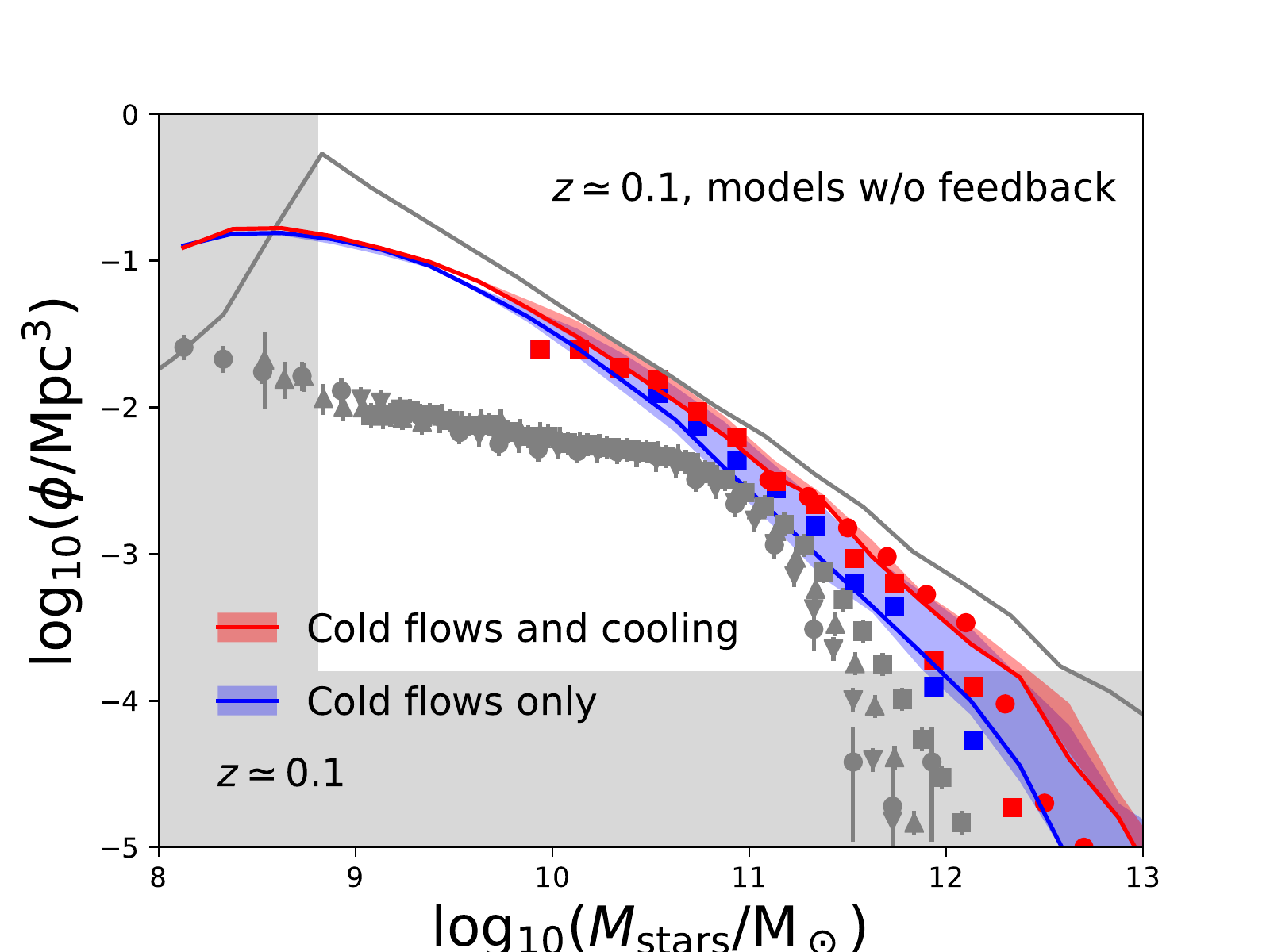} 
\caption{The local ($z=0.1$) GSMF in {\sc GalICS~2.1} (curves) and cosmological
  hydrodynamic simulations without feedback (symbols).
  Red curves and symbols refer to the case with cooling. Blue curves
  and symbols refer to the case with cold accretion only.
  Squares are for a simulation by \citet[ {\sc gadget-2}
  ]{keres_etal09}.
  Circles refer to a simulation that we have run
with {\sc ramses}.
The red and blue shaded areas show the uncertainties linked to the KH
timescale.
The upper envelope of the blue shaded area
corresponds to a model without KH instability.
The lower envelope of the blue shaded area corresponds to the extreme
KH model at the end of Appendix~B.
The gray curve is the GSMF in an extreme model where all baryons are converted into stars.
The gray symbols are the data points by Yang et al. (2009), Baldry et al. (2012), Bernardi et al. (2013) and Moustakas et al. (2013).
The gray shaded area shows our resolution limits in both stellar mass and number density.}
\label{modelswofeedback}
\end{figure}

The blue squares show the GSMF found by  K09 when they remove all particles that have passed through a hot phase. 
The blue curve is the prediction of {\sc GalICS~2.1} when galaxies grow through cold accretion only. The agreement is once again good. Expectedly, the blue curve runs slightly below the blue squares 
at all but the highest masses because K09 could not adequately resolve
the KH instability, { a description of which is included in our SAM.}
Varying { the clumping factor $C$ of the cold gas in the filaments shifts} the red curve and the blue curve up or down but
does not change the separation between them.
It is therefore significant that the difference between the red curve and
the blue curve 
is comparable to the one between the red symbols and the blue symbols.

{ The shaded areas around the red curve and the blue curve show the
  uncertainties linked to the KH timescale.
  Our predictions for the total amount of gas that accretes onto
  galaxies are robust (the red shaded area is very narrow).
  The KH instability can disrupt the filaments and mix their baryons
  with the hot gas but these low-entropy baryons have a short cooling
  time and find their way into the central galaxy anyway.
    The KH instability can, however, modify the relative importance of
  cold flows and cooling. In a model without KH instability (upper
  envelope of the blue shaded area), cold flows dominate all the way
  up to $M_{\rm stars}\sim 10^{12.5}\,{\rm M}_\odot$.

 Our model for the KH instability is robust in
  massive haloes ($M_{\rm vir}\gsim 10^{13}\,{\rm M}_\odot$ at $z=0$).
  At lower masses, the geometric approximation on which our model is
  based breaks down (Appendix~B) but this where the difference between
  the blue curve and the red curve is less significant.}

In conclusion, Fig.~\ref{modelswofeedback} demonstrates that:
\begin{enumerate}
\item Ejective feedback is necessary. Even a complete shutdown of cooling is not enough (the blue curve overestimates the observed GSMF, i.e., the gray data points with error bars, at all masses).
\item In a model without feedback, the increase in stellar mass due to
  cooling is small (less than { a factor of two} for a galaxy with
  $M_{\rm stars}= 10^{11}\,{\rm M}_\odot$).
\item In massive systems, { cooling is inefficient even if it is allowed}
(the red curve falls below the gray curve at high masses).
\end{enumerate}

\subsection{Models with feedback}

\begin{figure*}
\begin{center}$
\begin{array}{cc}
\includegraphics[width=0.5\hsize]{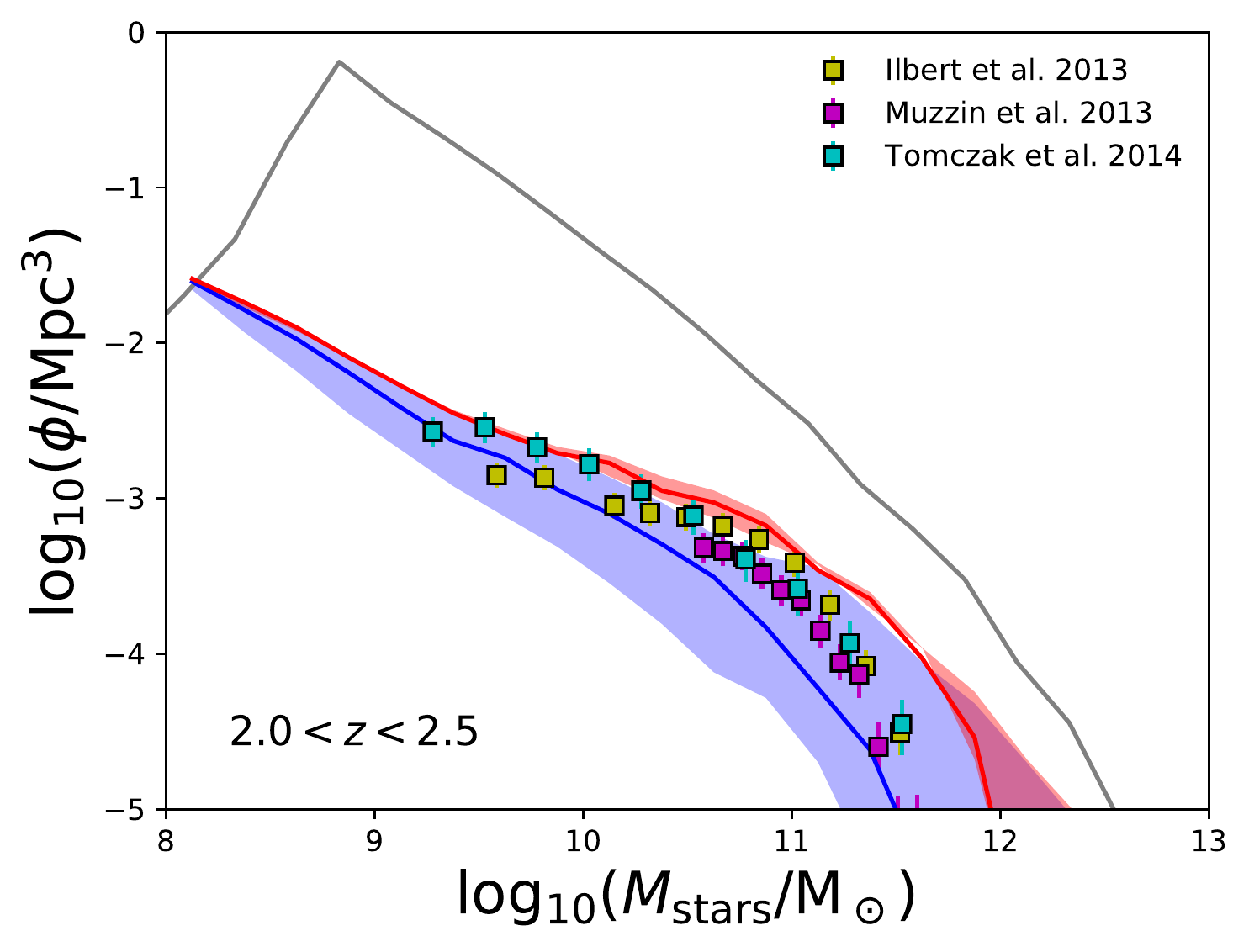}&
\includegraphics[width=0.5\hsize]{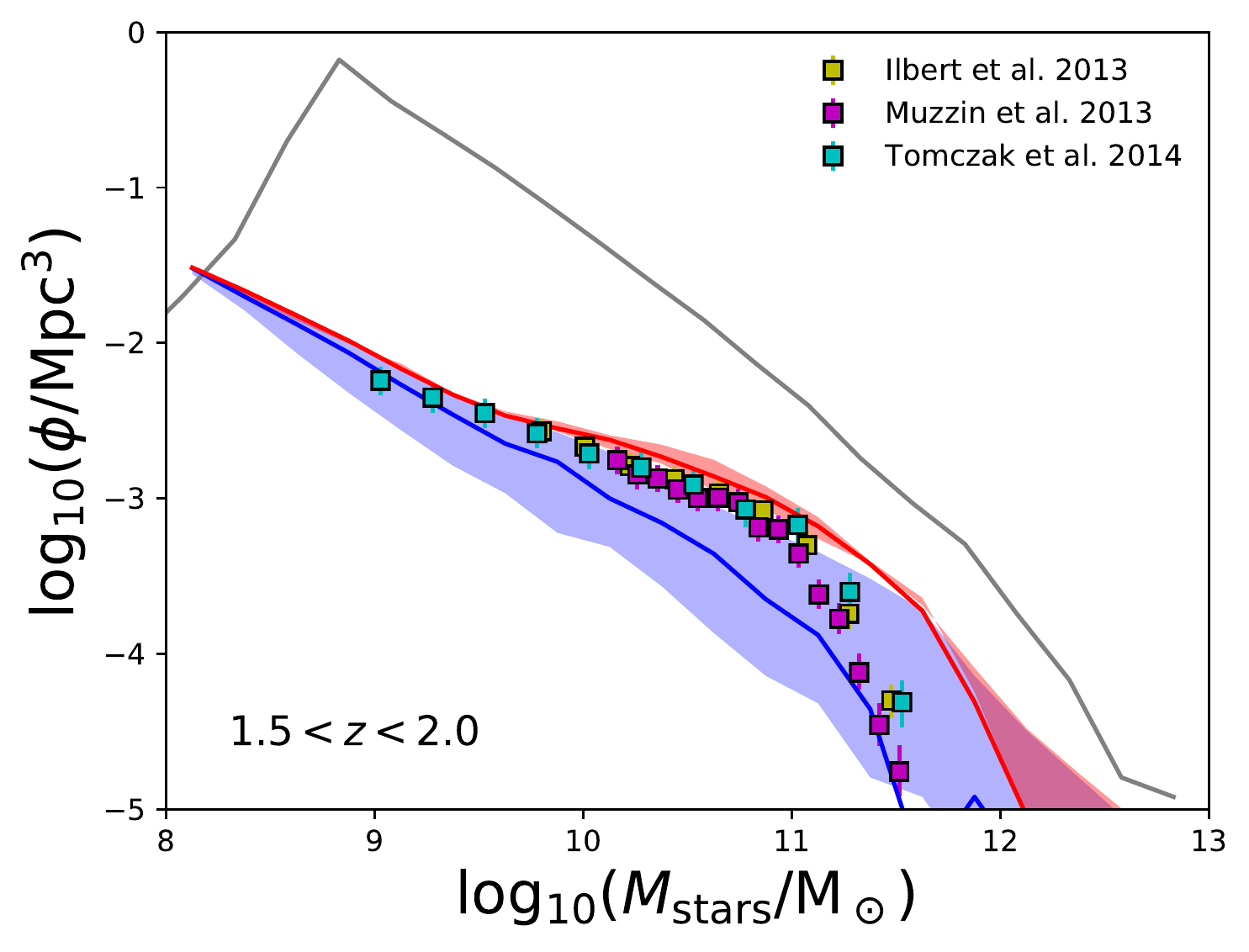}\\
\includegraphics[width=0.5\hsize]{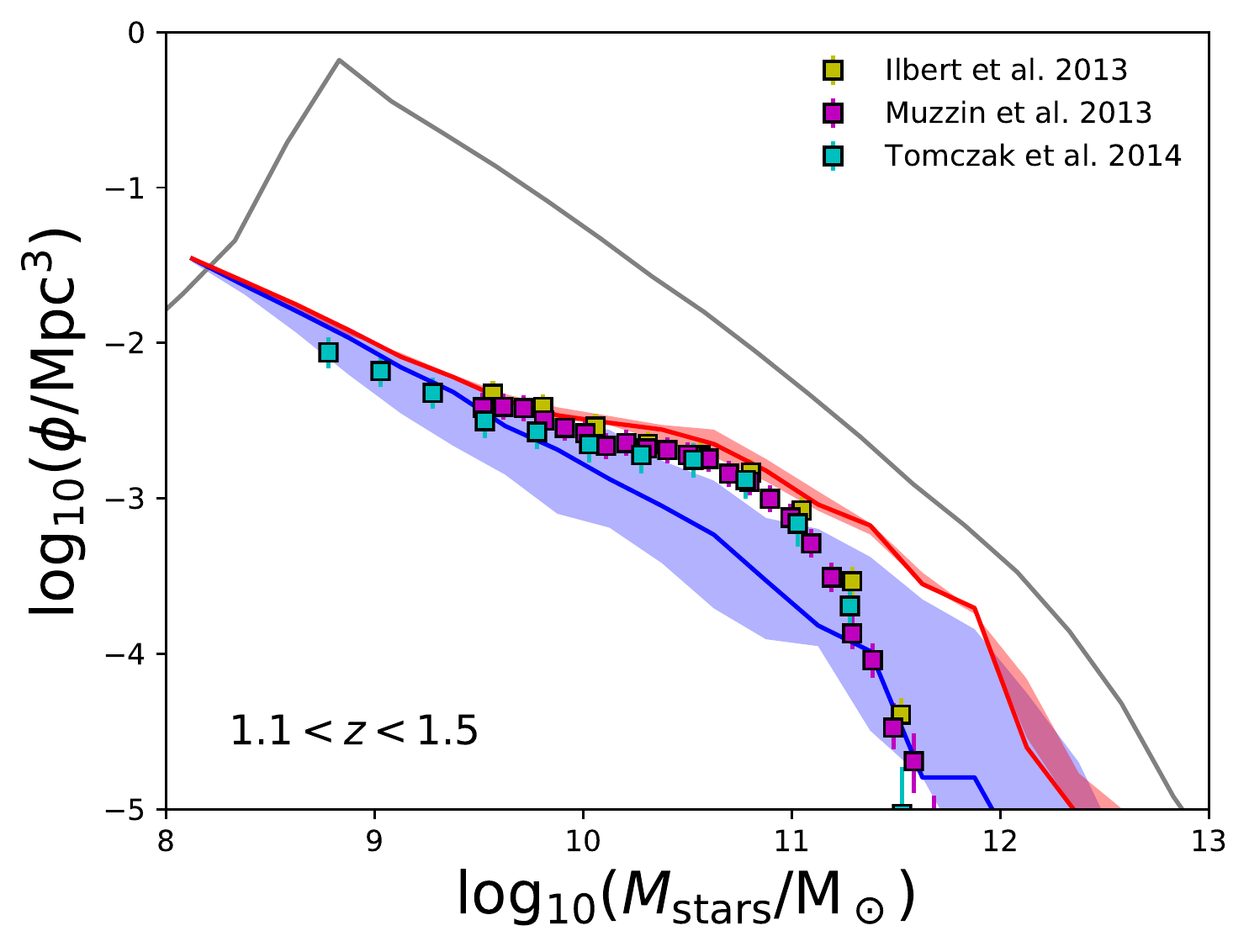}&
\includegraphics[width=0.5\hsize]{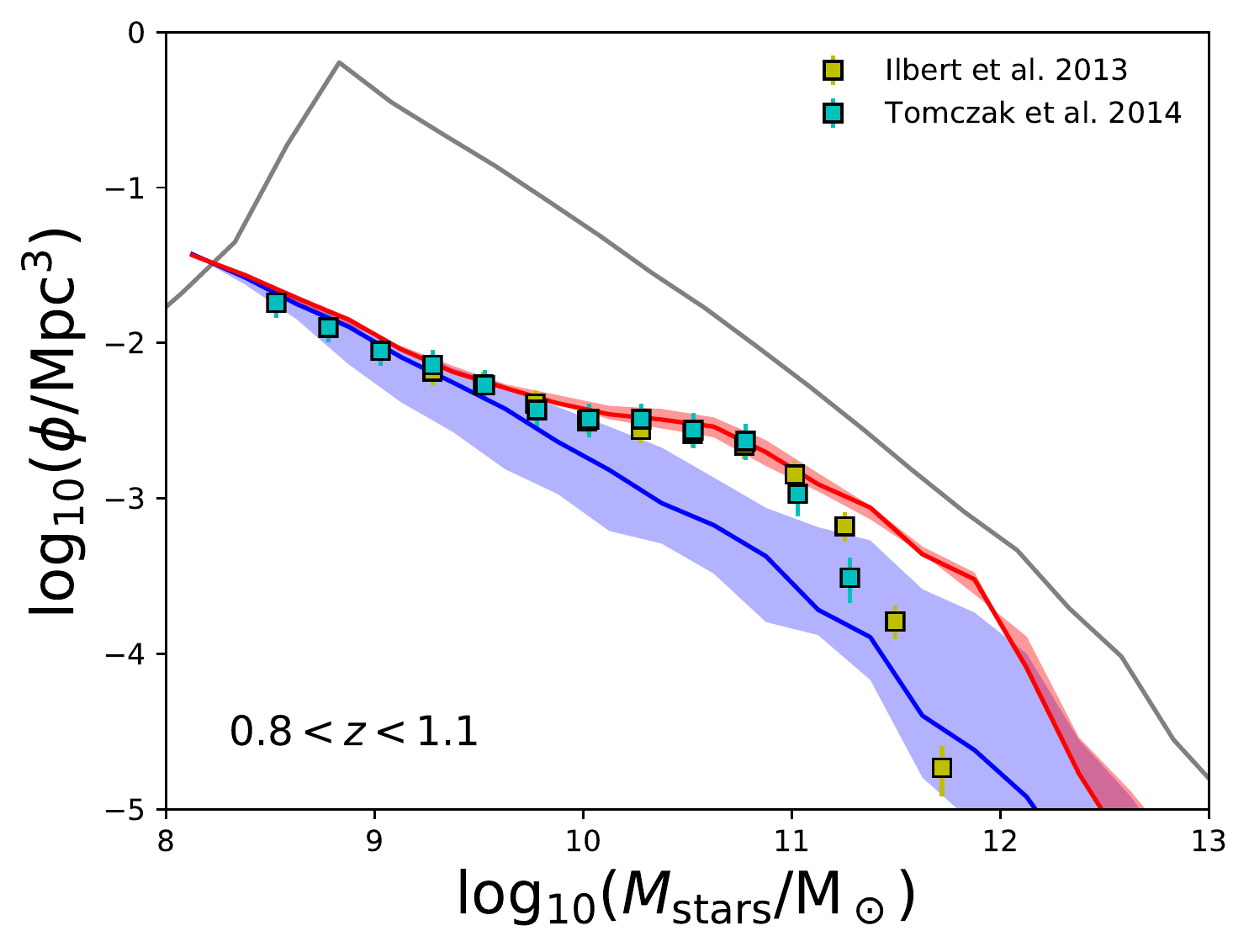}\\
\includegraphics[width=0.5\hsize]{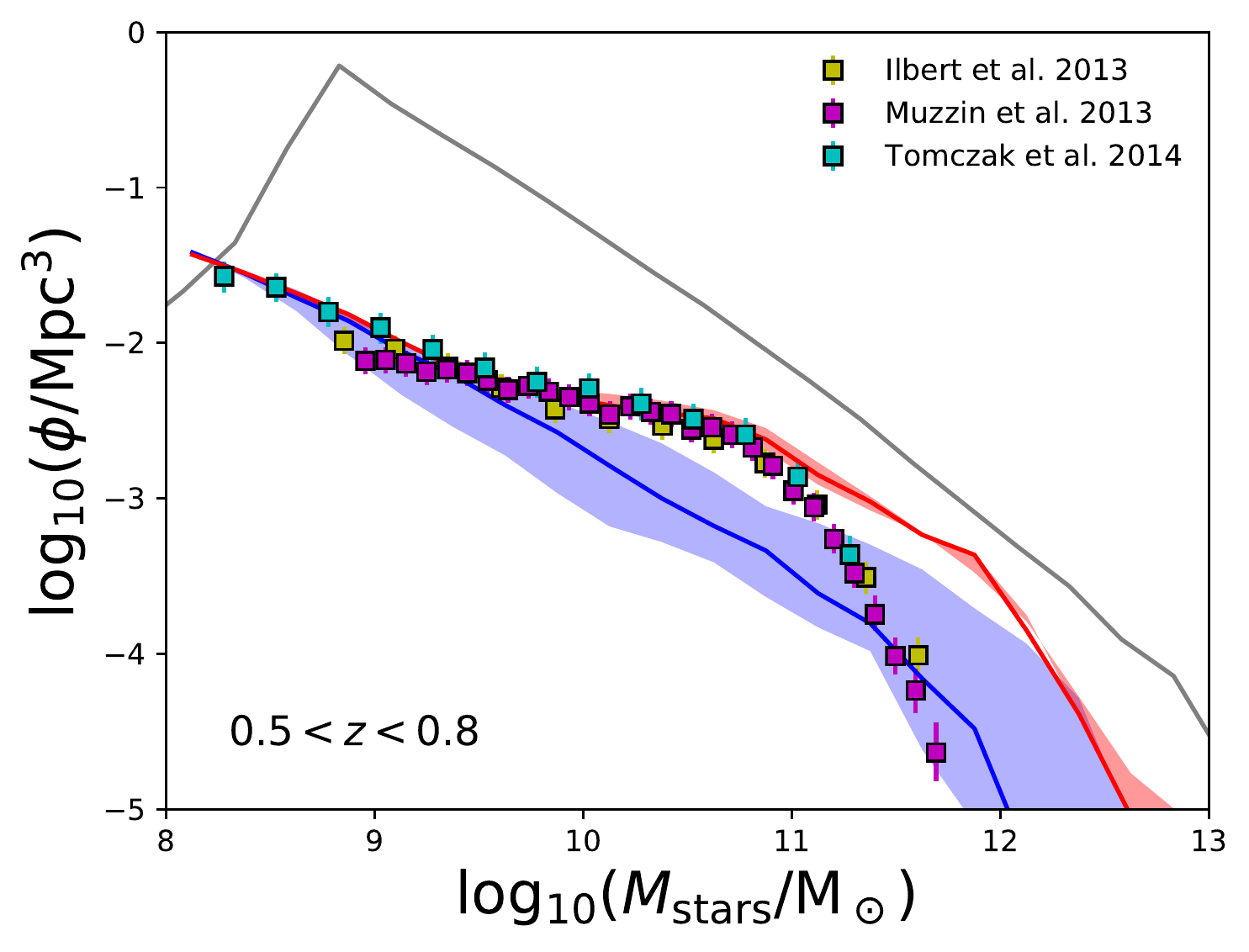}&
\includegraphics[width=0.5\hsize]{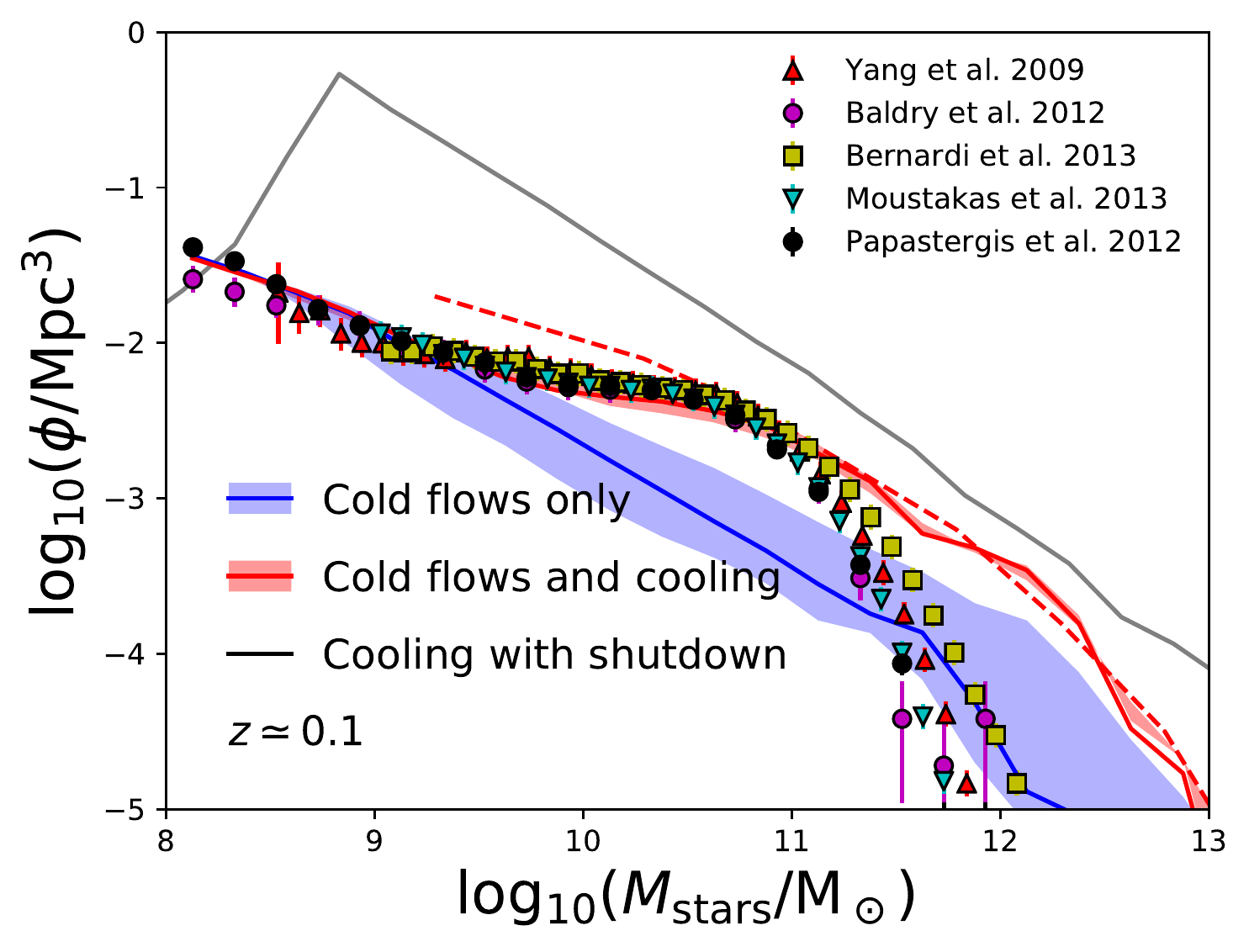}\\
\end{array}$
\end{center}
\caption{Evolution of the GSMF with redshift in the full {\sc GalICS~2.1} model (red curves) and the observations 
(\citealp{yang_etal09,baldry_etal12,papastergis_etal12,bernardi_etal13,ilbert_etal13,moustakas_etal13,muzzin_etal13,tomczak_etal14}; data points). 
The blue curves show how the predictions of {\sc GalICS~2.1} would
change if galaxies grew through cold accretion only { because the
  cooling of shock-heated gas (based on the shock-heating criterion in
  Eq.~\ref{tcomp_over_tcool3}) has been shut down}.
The red and blue shaded areas show the uncertainties that derive from
the KH timescale { (see Section~3.2 for the reason why the red
  curves can slip outside the red shaded areas).}
The red dashed curve at $z\simeq 0.1$ shows the GSMF in the Horizon-noAGN simulation  \citep{beckmann_etal17}.
The gray curves show the GSMF in an extreme model  where all baryons are converted into stars.}
\label{GSMFz}
\end{figure*}

\begin{table*}
\begin{center}
\caption{Model parameters in the full {\sc GalICS~2.1}}
\begin{tabular}{ l l l l }
\hline
\hline 
Parameter                                  & Symbol                & Value          & Units         \\
 \hline
{  Cosmology}                        &                              &                   &                       \\
Matter density                            & $\Omega_{\rm M}$        & $0.308$                 &           \\
Baryon density                           & $\Omega_{\rm b}$          & $0.0481$                 &         \\ 
Cosmological constant               & $\Omega_\Lambda$& $0.692$              &            \\
Hubble constant                         & $H_0$                    &$0.678$ & $100{\rm\,km\,s}^{-1}{\rm Mpc}^{-1}$   \\
\hline
{ N-body simulation}               &                       &                &                                      \\
Box size                                     & $L_{\rm box}$ & $100$ & Mpc                            \\
Resolution                                 & $N_{\rm part}$ &   $1024^3$              &                       \\
\hline
{ Dimensional parameters}    &                      &                 &                                  \\
Minimum $v_{\rm vir}$ for efficient baryonic accretion & $\sigma$                        &$34$& km$\,$s$^{-1}$ \\
SF threshold                            & $\Sigma_{\rm th}$  &$10$&
                                                                    ${\rm M}_\odot{\rm\,pc}^{-2}$\\                                          
  SN feedback saturation scale   & $v_{\rm SN}$       & 75&${\rm\,km\,s}^{-1}$\\
SN energy                                  & $E_{\rm SN}$       & $10^{51}$&erg\\
SN rate                                       & $\Psi_{\rm
                                                SN}$&$1/140$& ${\rm M}_\odot^{-1}$ \\
Fountain evaporation time       &$t_{\rm ev}^0$&$2$&Gyr \\
\hline
{ Adimensional efficiency factors} &                         &&\\
Star formation                              &$\epsilon_{\rm sf}$     &0.04&\\  
Filament clumping factor             &$C$                            &$9$             & \\
Efficiency of KH instability           &$\epsilon_{\rm KH}$&$1$         &\\
Critical $t_{\rm comp}/t_{\rm cool}$&$\Gamma_{\rm c}$& $5/7$      &\\
Cooling time fudge factor             &$\epsilon_{\rm cool}$&$0.8$    &\\
Mass ratio for major mergers   &$\epsilon_{\rm m}$ &$4$   &           \\
Maximum SN feedback efficiency  & $\epsilon_{\rm max}$ &$1$   & \\           
Returned fraction                      & $R$               &    $0.43$               &   \\
Metal yield                                & $y$               &     $0.03$              &   \\
Fountain fraction                       &$\epsilon_{\rm fount}$& $0.7$  &\\
\hline
{ SN feedback scaling exponents}  &          &                    &                      \\
$v_{\rm vir}$ scaling                 & $\alpha_v$        &    $-3.6$               &              \\
$z$ scaling                                & $\alpha_z$        &     $1.5$              &                     \\
\hline
{ Logical parameters}            &                          &                             &\\
Accretion onto subhaloes          & No                    &                             & \\
Compute cooling                       & Yes                   &                             &\\
Compute KH instability              & Yes                  &                              &\\
\hline
\hline
\end{tabular}
\end{center}
\label{model_parameters}
\end{table*}

Feedback introduces {five} additional parameters: $v_{\rm SN}$, $\alpha_v$, $\alpha_z$, $\epsilon_{\rm max}$ and $\epsilon_{\rm fount}$.
We do not treat the reaccretion time $t_{\rm reaccr}$ {and the
  evaporation times $t_{\rm ev}^0$} of the galactic fountain as free
parameters because
{their values are taken from hydrodynamic simulations: $t_{\rm reaccr}$ is computed with the fitting formula in T19 (Eq.~\ref{treaccr})
and $t_{\rm ev}^0=2\,$Gyr is taken from \citet{armillotta_etal17}.}
We also fix the maximum efficiency of SNe to $\epsilon_{\rm max}=1$ (a  large but not unreasonable assumption based on T19's analysis of the NIHAO simulations\footnote{Energetic efficiencies that approach unity are physically possible if the SN rate is so high
that SN bubbles overlap on a timescale shorter than the radiative time \citep{larson74,dekel_silk86}.}).

With these constraints,  the full {\sc GalICS~2.1} model with cooling, feedback and the KH instability
(Fig.~\ref{GSMFz}, red curves) fits the low-mass end of the GSMF at $0<z<2.5$ for $v_{\rm SN}=75{\rm\,km\,s}^{-1}$, $\alpha_v=-3.6$, $\alpha_z=1.5$ and $\epsilon_{\rm fount}=0.7$
(Table~1 presents a  full list of all parameter values).
{ All figures and results hereafter refer to this model
  unless the contrary is explicitly stated.}
For comparison, C17 found  $v_{\rm SN}=24{\rm\,km\,s}^{-1}$, $\alpha_v=-4$, $\alpha_z=2$ and $\epsilon_{\rm max}=1$ (default model;
$\epsilon_{\rm fount}$ did not exist in {\sc GalICS~2.0}); $v_{\rm SN}$ is much higher in {\sc GalICS~2.1} than in {\sc GalICS~2.0} because $v_{\rm w}$ is now the escape speed rather than the virial velocity.

{ As in C17, all GSMFs have been convolved with the observational
  error on $M_{\rm stars}$, which is $0.04(1+z)\,$dex according to
  \citet{ilbert_etal13}.
  The application of a Gaussian random error to each stellar mass
  explains why the red curves {(corresponding  to $\epsilon_{\rm
      KH}=1$)}
  can move out of  the red shaded areas.
  The boundaries of the red shaded areas correspond
  to a model without KH instability
  and an extreme model in which the KH instability is very efficient
  {($\epsilon_{\rm KH}=0$ and $\epsilon_{\rm KH}=26$, respectively).}
  The red curves slip out of the red shaded areas {because} 
 the effects of the KH instability on the model
  with cooling are smaller than the observational errors.}

{\sc GalICS~2.1} overpredicts the number densities of galaxies with
$M_{\rm stars}>10^{11}\,{\rm M}_\odot$ 
but this finding was totally expected, since we have  not included AGN feedback, which is known to be important to prevent overcooling in massive haloes
(\citealp{bower_etal06}, \citealp{cattaneo_etal06}, \citealp{croton_etal06}; also see \citealp{cattaneo_etal09} for a review).

For $M_{\rm stars}\lsim 10^{11}\,{\rm M}_\odot$, the red curves are in reasonably good
agreement with the observations over the entire redshift range $0<z<2.5$ \citep{baldry_etal12,bernardi_etal13,moustakas_etal13,papastergis_etal12,yang_etal09,ilbert_etal13,muzzin_etal13,tomczak_etal14}.
Differently from  {\sc GalICS~2.0}, {\sc GalICS~2.1} is not affected
by the common tendency of SAMs to overestimate the number densities of
low-mass galaxies at high-$z$ (\citealp{weinmann_etal12}; \citealp{knebe_etal18}; \citealp{asquith_etal18}; and references therein).
The discrepancy with observations at low masses disappeared after the introduction of a fountain component.
Fountain feedback reduces star-formation at high-redshift while producing lower net outflow rates at low $z$, since outflows are partly compensated by the reaccretion of previously ejected gas.

\begin{figure}
\includegraphics[width=1.00\hsize]{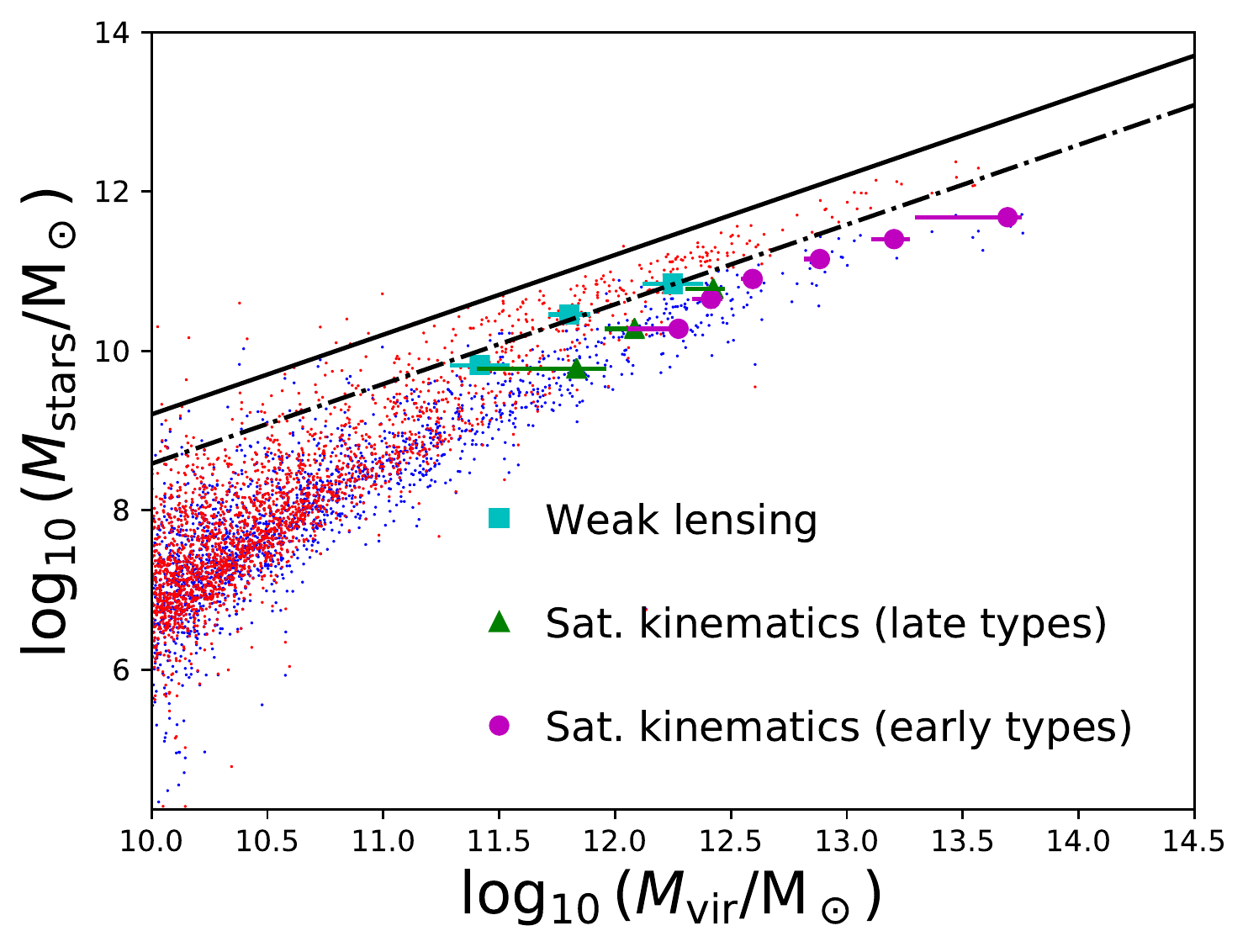} 
\caption{ Relation between stellar mass and halo mass in {\sc
    GalICS~2.1} for 10,000 randomly-selected galaxies at $z=0.1$ (point cloud). The red points show the full model. The
  blue points show the model with cold accretion only.
The black solid line corresponds to $M_{\rm
  stars}={\Omega_{\rm b}\over\Omega_{\rm M}}M_{\rm vir}$.
 {The black dottted-dashed line corresponds to the stellar-to-halo
   mass ratio of the Milky Way, which has $M_{\rm stars}=5\times
   10^{10}\,{\rm M}_\odot$ and $M_{\rm vir}=1.3\times 10^{12}\,{\rm M}_\odot$ \citep{bland_gerhard16}.}
  The cyan
  squares are based on combined weak-lensing and Tully-Fisher data for
  $\sim 10^5$ local disc galaxies \citep{reyes_etal12}. The {green}
  triangles and the {magenta} circles are based on satellite kinematics
  \citep{wojtak_mamon13}. They are for late-type and early-type
  galaxies, respectively.}
\label{Hod}
\end{figure}

The red dashed curve in the panel at $z\simeq 0.1$ shows the GSMF in the Horizon-noAGN simulation,
a version of the Horizon-AGN simulations \citep{dubois_etal16}
 without AGN feedback (see \citealp{beckmann_etal17} and
 \citealp{kaviraj_etal17} for a comparison of the Horizon-AGN and the
 Horizon-noAGN simulations). The red solid curve and the red dashed curve differ at low masses because stellar feedback in {\sc GalICS~2.1} is much stronger than stellar feedback in the Horizon-noAGN simulation,
which is purely thermal, but the agreement at high masses is rather
good.

The data points by \citet{baldry_etal12} and \citet{papastergis_etal12} in the local Universe and
\citet{tomczak_etal14} at higher redshifts show that the slope of the
GSMF increases below $M_{\rm stars}\sim 10^9\,{\rm M}_\odot$.
{\sc GalICS~2.1} reproduces this behaviour and proposes an explanation for its physical origin.
The red curves are shallow at intermediate masses compared to the slope of the halo mass function (gray curves) because the efficiency of SN feedback increases dramatically at low masses
($\epsilon_{\rm SN}\propto v_{\rm vir}^{-3.6}$). However,
the growth of $\epsilon_{\rm SN}$ at low $v_{\rm vir}$ cannot continue
indefinitely because the energetic efficiency of SN feedback cannot be
larger than unity.
The stellar mass at which the GSMF becomes steeper corresponds to
the virial velocity below which the efficiency of SN feedback saturates
(see C17 for actual proof that changing $\epsilon_{\rm max}$ modifies the slope of the GSMF at low masses).
T19 find that the mass-loading factor $\eta_{\rm cold}(v_{\rm vir})$ departs from  $\eta_{\rm cold}\propto v_{\rm vir}^{-5.6}$ for $v_{\rm vir}<70{\rm\,km\,s}^{-1}$. In {\sc GalICS~2.1}, we reproduce the GSMF for $v_{\rm SN}=75{\rm\,km\,s}^{-1}$
($v_{\rm SN}$ is the virial velocity for which $\epsilon_{\rm SN}=1$).

The blue curves in Fig.~\ref{GSMFz} correspond to the same model as
the red curves except that cooling has been shut down.
At $M_{\rm stars}\lsim 10^{9}\,{\rm M}_\odot$, the blue curves and the red
curves run very close to one another. 
At $M_{\rm stars}>10^{9}\,{\rm M}_\odot$, the contribution from cooling is no longer negligible { and}
accentuates the transition to a flatter slope at $M_{\rm
  stars}>10^{9}\,{\rm M}_\odot$.
  
{ The width of the blue shaded areas shows that the KH timescale is an
  important source of uncertainty when it comes to separating the
  contributions of cold accretion and cooling.
  A scenario in which the entire galaxy population is assembled through cold
accretion only would require a much longer $t_{\rm KH}$ than what we have estimated but is not ruled out at
$z>1.1$, where the upper envelopes of the blue shaded areas are in reasonably good agreement with the data points.
At low redshift, however, the need for cooling at intermediate
masses ($10^{10}\,{\rm M}_\odot\lsim M_{\rm stars}\lsim 10^{11}\,{\rm M}_\odot$)
is hardly avoidable.}
Cooling must be mitigated to avoid overproducing galaxies with $M_{\rm
  stars}>10^{11}\,{\rm M}_\odot$ but its shutdown cannot be complete.

{ The relation between stellar mass and halo mass (Fig.~\ref{Hod}) confirms some of
  the above points while illustrating them in a clearer manner. The cloud of
blue points corresponds to the model without cooling.
  The circles/squares/triangles are observational data.
The masses of early-type galaxies are consistent with
their formation at high-$z$ through cold accretion only ({the blue points and the magenta circles overlap}).
The discrepancy between the red points and the magenta circles
tallies with the well-known fact that the hot gas in massive
early-type galaxies cools very inefficiently {(e.g., \citealp{peterson_fabian06,mcnamara_nulsen07})}.

For late-type galaxies, there is a tension between the halo masses
measured from lensing and internal kinematics \citep[cyan
squares]{reyes_etal12},  and those measured from satellite kinematics
\citep[{green} triangles]{wojtak_mamon13}, which are much larger for a given $M_{\rm stars}$.
We put greater trust in the cyan squares because we have not managed
to find models that reproduce the GSMF and the {\rm green} triangles
simultaneously, while models that pass through the cyan squares reproduce the knee of the GSMF naturally.
{The cyan squares are also more consistent with the data for the Milky Way (\citealp{bland_gerhard16}; compare the cyan squares and the dotted-dashed line in Fig.~\ref{Hod}).

In the mass range that corresponds to the cyan squares (the intermediate-mass spiral galaxies of \citealp{reyes_etal12}), the red points are consistent with the cyan squares; the blue points are not.
This is another, more direct way to see that, in our SAM, cooling is important to explain the masses of normal spiral galaxies like the Milky Way.
}

To look at predictions beyond stellar and halo masses, we have plotted
the relation between SFR and stellar mass in {\sc GalICS~2.1}
and compared it to observational data by
\citet[Fig.~\ref{SFR_mass}]{wuyts_etal11}.
We have also added more recent data \citep{duarte_etal17} that extend
the main sequence of star-forming galaxies to lower masses.
The main sequence of star-forming galaxies is reproduced reasonably
well although it extends too much at high masses and the passive
population is almost inexistent.
Both are manifestations of overcooling in massive haloes.
}

\begin{figure}
\includegraphics[width=1.0\hsize]{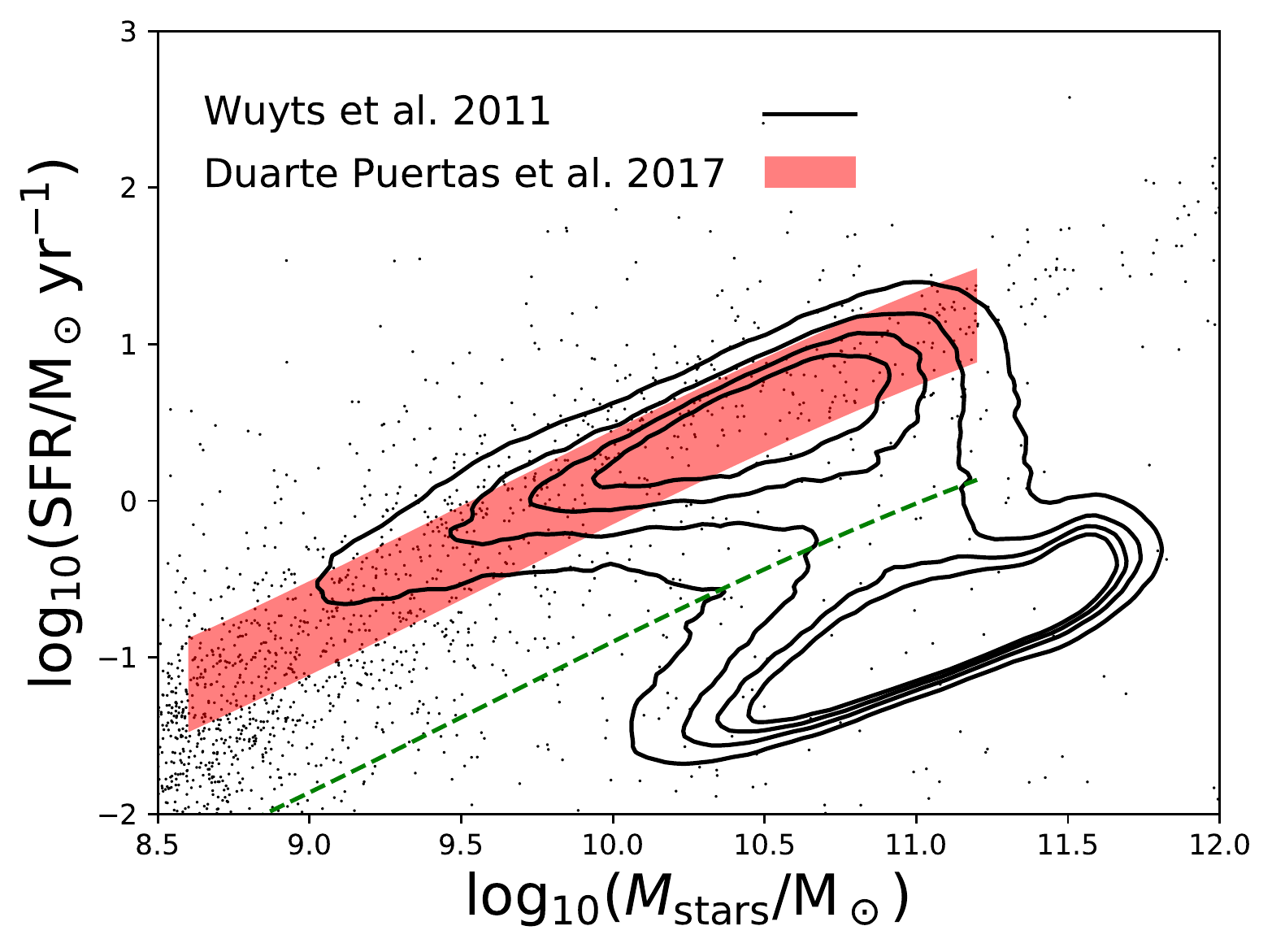} 
\caption{SFR - stellar mass relation  in the full {\sc GalICS~2.1}
  model (point cloud). The contours show the SFR - $M_{\rm stars}$
  distribution in the SDSS \citep{wuyts_etal11}.
The red shaded area highlights the main sequence of star-forming galaxies
\citep{duarte_etal17}. The green dashed curve marks the bottom of the
green valley that separates star-forming and passive galaxies.
}
\label{SFR_mass}
\end{figure}

The results of {\sc GalICS~2.1} depend on the metallicity of the IGM. Therefore, it is important to check that our predictions for $Z_{\rm IGM}$ are in agreement with observations.
The black curve in Fig.~\ref{Z_IGM} is computed by stacking the IGM of all the forests, so that $Z_{\rm IGM}(z)=\langle Z_{\rm accr}(z)\rangle$.
The data points show the redshift - metallicity relation for the damped Lyman-$\alpha$ absorbers (DLAs) observed by \citet[black symbols]{rafelski_etal12,rafelski_etal14} and \citet[red symbols]{berg_etal16}.
DLAs signal systems with a considerable column density of neutral hydrogen on the line of sight.
Most DLAs and definitely the most metal-rich ones are galaxies but some must be associated with {\sc Hi} clouds within filaments. They will be those with the lowest metallicities.
Therefore, the lower envelope of the redshift - metallicity relation
for DLAs may be interpreted as a measurement of how the metallicity of
the filaments evolves with $z$.
The black curve follows this lower envelope reasonably well. $Z_{\rm IGM}(z)$ is a genuine prediction of {\sc GalICS~2.1} because no free parameter was tuned to
reproduce it.

For a given halo at a given time, the shock-heated fraction can take only two values: $f_{\rm hot}=1$ or $f_{\rm hot}=0$. On a population average, however, $f_{\rm hot}$ can take all values between zero and one.
Fig.~\ref{fhotmean} shows $\langle f_{\rm hot}\rangle$ in bins of
$M_{\rm vir}$ at $z\simeq 0.1$ and $z\simeq 2.25$ (solid black curve
and dashed black curve, respectively).
{ The squares with arrows show the results of a cosmological hydrodynamic
  simulation \citep{nelson_etal13} at $z=2$. The comparison with
  simulations will be discussed in Section~4.

The black curves in Fig.~\ref{fhotmean}  show that shock-heating is
negligible at} $M_{\rm vir}<10^{10}\,{\rm M}_\odot$ ($f_{\rm
  hot}=0$ in nearly all haloes). { In contrast, at} $M_{\rm vir}>4\times
10^{12}\,{\rm M}_\odot$, shock-heating is complete ($f_{\rm hot}=1$ in all haloes).
The differences between $z\simeq 0.1$ and $z\simeq 2.25$ are
small. 

\begin{figure}
\includegraphics[width=1.0\hsize]{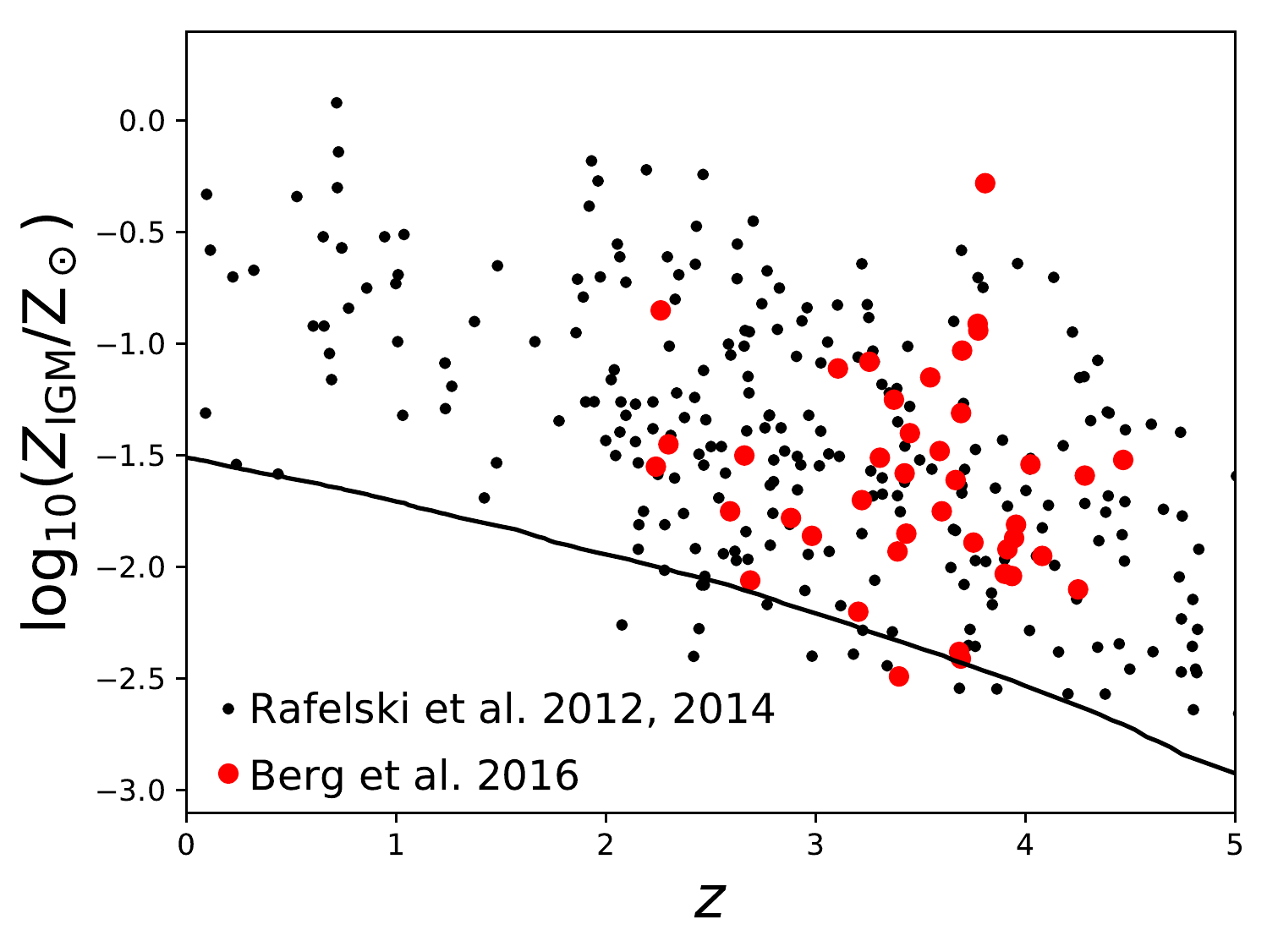} 
\caption{Metallicity of the IGM  as a function of redshift in the full {\sc GalICS~2.1} model (black curve).
The black/red symbols show the redshift - metallicity relation for the DLAs observed by \citet{rafelski_etal12,rafelski_etal14} and \citet{berg_etal16}, respectively.}
\label{Z_IGM}
\end{figure}

The $M_{\rm vir}^{\rm crit}$ determined from
Figs.~\ref{tcomp_over_tcoolSD}a and \ref{tcomp_over_tcoolSD}b
(vertical red solid and vertical red dashed line, respectively)
correspond to an average shock-heated fraction of $\gsim 0.9$ in both cases.
Shock-heating is complete at $M_{\rm vir}>3M_{\rm vir}^{\rm crit}$ at both $z=0$ and $z=2.25$.
However, at $M_{\rm vir}<3\times 10^{11}\,{\rm M}_\odot$ there is more cold accretion at low $z$, where $Z_{\rm IGM}$ is higher,
whereas at $M_{\rm vir}>3\times 10^{11}\,{\rm M}_\odot$ there is more cold accretion at high $z$, where the effect of narrower solid angles dominates.
Therefore, the statement that cold accretion is more important at high
$z$ is correct for massive systems but not in general.

\section{Discussion}

{ Our article covers two topics: the role of cold flows and cooling
  in the growth of galaxies, and SN feedback. We therefore split our
  discussion into two parts.}

\subsection{Cold flows and cooling in galaxy formation}

{ Disentangling the roles of cold- and hot-mode accretion has been
  the major goal of this article.
  The purpose of this discussion is: i) to compare our results with previous work,
  ii) to put them in a broader context and perspective, and iii) to
  identify potential issues that require further research.
We start by comparing our results to DB06's, since their work is the
closest to ours. We then consider cosmological hydrodynamic simulations,
where the comparison is less straightforward.}

As many of our assumptions are the same as in DB06, it is not surprising that many of our conclusions echo theirs. Below a critical mass $M_{\rm vir}^{\rm crit}$, cold accretion dominates. When the halo mass reaches the critical value a stable shock develops
at $r_{\rm s}\sim 0.1r_{\rm vir}$ and rapidly propagates out to the virial radius. $M_{\rm vir}^{\rm crit}$ increases with the metallicity $Z_{\rm accr}$ of the accreted gas. Denser, narrower high-$z$ filaments penetrate massive haloes more effectively
than their low-$z$ counterparts. 

The main differences with  DB06 derive from  the  density and the metallicity of the accreted gas. DB06 computed the gas density at the
shock radius by assuming that it scales as $r_{\rm s}^{-2}$. 
{\sc GalICS~2.1} uses the continuity equation, so that   $\rho_1\propto (u_1r_{\rm s}^2)^{-1}$.
The two assumptions would be identical if the infall speed $u_1$ were constant but the gas gains speed while it falls in.
Hence, $\rho_1$ increases less rapidly at small radii in our SAM than in DB06.
Using the continuity equation gives densities that are
$\sim 27$ times lower, but
 we have also shown that 
inhomogeneities within the filaments are likely to increase
the density of the cold gas by a factor of nine. Therefore, 
our final pre-shock densities are lower than DB06's by 
a factor of three 
and this lowers $M_{\rm vir}^{\rm crit}$ by a factor of $\sim 1.5$.
Another factor of about two comes from the metallicity of the
accreted gas, which is lower in {\sc GalICS~2.1} ($Z_{\rm accr}=0.03{\rm\,Z}_\odot$
at $z=0$) than in DB06 ($Z_{\rm accr}=0.1{\rm\,Z}_\odot$
at $z=0$). This is why 
we find $M_{\rm vir}^{\rm crit}\sim 2\times 10^{11}\,{\rm M}_\odot$
instead of $M_{\rm vir}^{\rm crit}\sim 6\times 10^{11}\,{\rm M}_\odot$ as in DB06
($M_{\rm vir}^{\rm crit}\sim 2\times 10^{11}\,{\rm M}_\odot$ is a typical
value at $z=0$ because in {\sc GalICS~2.1} each halo has its own
critical mass).

\begin{figure}
\includegraphics[width=1.0\hsize]{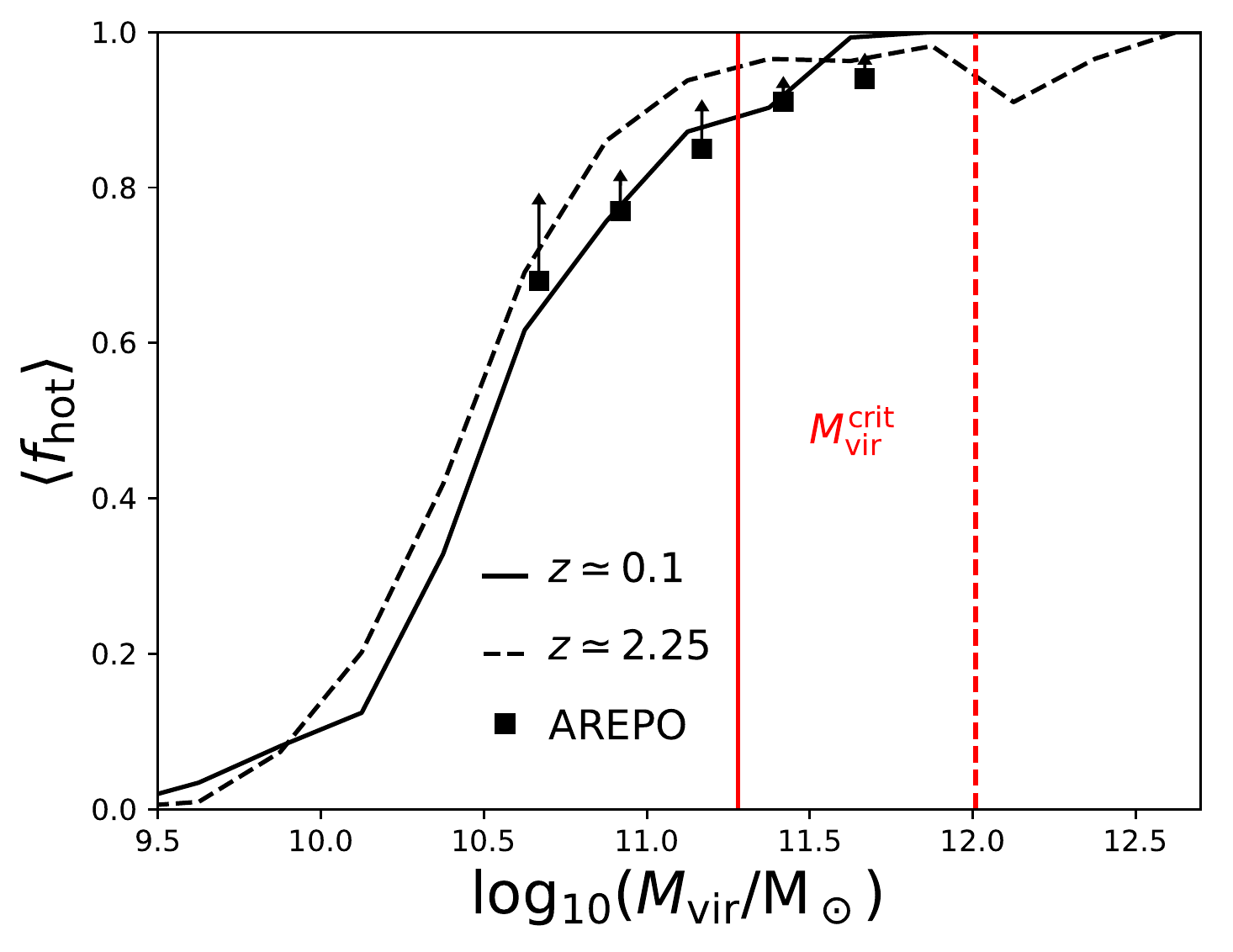} 
\caption{Average shock-heated mass fraction for the baryons that accrete onto a halo of mass $M_{\rm vir}$ at $z\sim 2.25$ (black dashed curve) and $z\sim 0.1$ (black solid curve) in the full {\sc GalICS~2.1} model.
The black symbols show the contribution of hot-mode accretion to the growth of central galaxies in a cosmological hydrodynamic simulation by \citet[$z=2$]{nelson_etal13}.
The vertical solid red line and the vertical dashed red line
correspond to $M_{\rm vir}^{\rm crit}=1.9\times 10^{11}\,{\rm M}_\odot$ and
$M_{\rm vir}^{\rm crit}=1.02\times 10^{12}\,{\rm M}_\odot$, i.e.,
the critical masses at $z=0$
(Fig.~\ref{tcomp_over_tcoolSD}) and $z=2.25$
(Fig.~\ref{tcomp_over_tcoolSD}b), respectively.
Comparing the black solid curve to the red solid line and the black dashed curve to the red dashed line shows that a model with $f_{\rm hot} = 0$ at $M_{\rm vir}<M_{\rm vir}^{\rm crit}$ and $f_{\rm hot} = 1$ at $M_{\rm vir}>M_{\rm vir}^{\rm crit}$ is too simplistic.
}\label{fhotmean}
\end{figure}

The comparison should consider that,
 in {\sc GalICS~2.1}, $r_{\rm s}$ and $Z_{\rm accr}$
are predictions of our SAM, while in DB06 they were tuned
to reproduce the critical halo mass that separates red and blue
galaxies in the SDSS. 
DB06 needed a relatively high $M_{\rm vir}^{\rm
  crit}$ because they assumed that galaxies formed through cold
accretion only.
C17 confirmed the necessity of a high $M_{\rm vir}^{\rm
  crit}$ if one makes that assumption.
Here, we can
reproduce the GSMF with a lower $M_{\rm vir}^{\rm
  crit}$ because cold accretion is not the only mechanism through
which galaxies are assembled.

Furthermore, DB06 studied $M_{\rm vir}^{\rm crit}(z)$ for different
values of $Z_{\rm accr}$ but they did not let $Z_{\rm accr}$ vary with
redshift.
In {\sc GalICS~2.1}, we find that the metallicity of the IGM grows by a factor of three
between $z=2.25$ and $z=0$, and this is the reason why, at low masses, we have less cold accretion at high $z$ (Fig.~\ref{fhotmean}).
Our predictions are consistent with data from DLAs \citep{rafelski_etal12,rafelski_etal14,berg_etal16}.
A caveat  is that $Z_{\rm accr}$ may be higher than $Z_{\rm IGM}$ if the KH instability stirs metals from the enriched intracluster or circumgalactic medium into the filaments
(\citealp{mcdonald_etal16}  find that the intracluster medium had
already attained its current metallicity $Z_{\rm hot}\sim 0.2\,{\rm Z}_\odot$ at $z=1.5$).

The comparison with simulations is not straightforward because the
relative contribution of cold- and hot-mode accretion is sensitive to
the way these two modes are defined, which cannot be exactly the same in a SAM
and a hydrodynamic simulation.
\citet{keres_etal05} used the fraction $f_{\rm cold}$ of the accreted
baryons that have never been hot as a measure of the importance of cold
accretion,
but:  a)$f_{\rm cold}$ depends on the criterion used to decide which particles are hot and b) ``never" depends on the
time resolution with which simulations are analysed.

{ Cosmological hydrodynamic simulations run by
  \citet{nelson_etal13} with the 
{\sc arepo} moving-mesh code} showed that $f_{\rm cold}$ varies considerably
depending on whether  accretion is onto the halo or onto the galaxy
 and on whether hot is
defined in absolute terms or relative to the virial temperature.
If $f_{\rm cold}$ is the fraction of the halo baryons 
that have never been hot, then $f_{\rm cold}$ includes gas that is
still cold at the time of measurement but will be shock heated when it
reaches smaller radii;
the real $f_{\rm cold}$ will be lower.
Defining $f_{\rm cold}$ as the fraction of  galaxy baryons
that have never been hot ($T<T_{\rm  vir}$ at all times)
gives a better estimate of the contribution of
cold accretion to the growth of galaxies. The black symbols in
Fig.~\ref{fhotmean}
shows the results that \citet{nelson_etal13} find at $z=2$ when they
apply this procedure.

However, comparing the dashed curve with the black symbols is not
straightforward either, because the dashed curve refers to all baryons that
have accreted onto a halo whereas the black symbols refer only to those
baryons that have accreted onto the central galaxy.
Baryons that avoid shock-heating have a higher chance of accreting
onto the galaxy. Hence, this method underestimates the
shock-heated fraction, too.
This is why  the black symbols in Fig.~\ref{fhotmean}  put a lower limit to the
$\langle f_{\rm
  hot}\rangle$ that we expect to find in our SAM.
Reassuringly, our predictions at $z=2.25$ (black dashed curve)
lie above the symbols. The arrows show how the symbols could move up
if the definition of hot gas was relaxed just a little bit
($T>0.8T_{\rm  vir}$).
The reader should also be aware of a difference in metallicity. In
{\sc GalICS~2.1}, the metallicity of the IGM at $z=2.25$ is small
($Z_{\rm IGM}=0.01$) but  not primordial.  \citet{nelson_etal13}
did not include metal-line cooling.

Previous work by \citet[Horizon-Mare Nostrum simulation]{ocvirk_etal08} had
included metal enrichment and found that the metallicity of the filaments
is indeed negligible ($\lsim 10^{-3}\,{\rm Z}_\odot$). However, the basis for
this conclusion deserves further scrutiny.
\citet{ocvirk_etal08} looked at the metallicity - temperature
distribution at $0.2r_{\rm vir}<r<r_{\rm vir}$
and identified the filaments with the strip of
low-temperature low-metallicity gas that stretches from $Z\sim
10^{-2}\,{\rm Z}_\odot$ all the way down to $Z\sim
10^{-8}\,{\rm Z}_\odot$ in the diagram at $z=4$.
Its disappearance in the diagram at $z=2.5$ was interpreted as
evidence that the filaments have been destroyed and accretion has
become isotropic.
\citet{ocvirk_etal08} supported this interpretation with images of the
gas-density distribution.
Our results agree well with their findings: our model predicts that accretion should still be fairly isotropic at $z=2.25$,
where $M_{\rm nl}\simeq 1.3\times 10^{11}\,{\rm M}_\odot$ and where we predict $\Omega\sim 2\pi$ for a halo with
$M_{\rm vir}\sim M_{\rm vir}^{\rm crit}\sim 10^{12}{\rm\,M}_\odot$, but not at
$z=4$, where $M_{\rm
  nl}\simeq 1.5\times 10^{10}\,{\rm M}_\odot$.
However, the disappearance of narrow filaments does not imply the end
of cold accretion, which \citet{ocvirk_etal08} keep finding at $z=2$.
The crucial point is how one interprets cold gas with  $Z\sim
10^{-2}\,{\rm Z}_\odot$ at $0.2r_{\rm vir}<r<r_{\rm vir}$.
\citet{ocvirk_etal08} considered that halo gas with $Z\sim 10^{-2}$--$10^{-1}\,{\rm Z}_\odot$ was part of the ISM of satellite galaxies.
To an extent, this is certainly true, at least for the
highest-metallicity gas.
However, if all this gas comes from satellites, then the logical conclusion
is that all the cold accretion that \citet{ocvirk_etal08} measure
at $z\lsim 2.5$ is in fact due to minor mergers.
The alternative interpretation is that $Z\sim 10^{-2}\,{\rm Z}_\odot$ is the
metallicity of cold flows at $z\sim 2.5$.
If this interpretation is correct, then the metallicity that
\citet{ocvirk_etal08} find for cold flows 
is comparable to ours.

\citet{ocvirk_etal08} used a Eulerian code ({\sc ramses};
\citealp{teyssier02}). Hence, they could not track particles and check
if they had always been cold. Instead, they followed the inflow of gas through a spherical
surface of radius $r=0.2r_{\rm vir}$. Very little gas had $10^5{\rm\,K}<T<2.5\times 10^5\,$K. Most
of the  accreted gas was either below or above this temperature range.
They therefore defined $\langle f_{\rm
  hot}\rangle$ as the fractional contribution of gas with
$T>10^5\,$K to the mass inflow rate at $0.2r_{\rm vir}$.  

\citet{ocvirk_etal08}'s results
for $\langle f_{\rm
  hot}(M_{\rm vir})\rangle$ at $z=2$ are qualitatively similar to the black
dashed curve in Fig.~\ref{fhotmean}. Quantitatively, they are shifted
to high masses by at least $0.5\,$dex. Three considerations explain
this discrepancy. \citet{ocvirk_etal08}'s cold flows have higher
metallicity than ours or include a contribution from minor mergers
that we do not consider. Gas may be shock heated at $r<0.2r_{\rm
  vir}$. Gas may be shock heated at $r>0.2r_{\rm
  vir}$ and cool before crossing the spherical surface $r=0.2r_{\rm
  vir}$. All three effects lower $\langle f_{\rm
  hot}\rangle$ and this explain why \citet{nelson_etal13} find
different results even when they adopt the same temperature threshold
to identify the shock-heated gas.

A less quantitative but more robust comparison has been obtained by
looking at the temperature - density diagrams for the NIHAO galaxies.
At $M_{\rm vir}\gsim 2\times 10^{11}\,{\rm M}_\odot$ (corresponding to
$M_{\rm stars}\gsim 2\times 10^9\,{\rm M}_\odot$),
the shock-heated gas stands out as a distinct
thermodynamic phase with high temperature and low density. Its
characteristic feature is having higher entropy than the IGM (T19).
This phase is absent at lower masses.
A critical mass of $M_{\rm vir}^{\rm crit}\sim 2\times
10^{11}\,{\rm M}_\odot$ at $z=0$ is in good agreement with our theoretical
predictions (Fig.~\ref{fhotmean}, red solid line).

Our discussion of hydrodynamic simulations has focussed on the methods
used to define and measure cold accretion. A more fundamental concern
is that all baryons may have passed through a hot phase before
accreting onto galaxies. Sometimes this phase was so short that it
could not be detected.
For \citet{croton_etal06}, who made this argument, cold accretion is  
just a misnomer for rapid cooling.
Our calculations do not exclude this possibility.
In fact, the post-shock temperatures used to compute the
cooling time are always larger than the virial temperature
(Fig.~\ref{Trs}), even in the so-called cold mode.
What cannot happen in the cold mode is the propagation of a {\it
  stable} shock.

A stable shock inevitably results in the formation of a hot atmosphere
because the condition 
$r_{\rm s}/u_2 < t_{\rm cool}$ means that shock-heated gas travelling at the post-shock
speed $u_2$ will fill a sphere of radius $r_{\rm s}$ before
cooling becomes important.
This point is essential  for a correct appreciation of the role of
shock stability in the formation and evolution of galaxies and leads
us into part (ii) of our discussion (see the plan at the beginning of
Section~4.1).

The maximum temperature that gas reaches before it accretes onto
galaxies is ultimately a detail and hydrodynamic simulations are
better equipped to deal with it than SAMs.
Fig.~\ref{GSMFz} demonstrates that re-routing gas from the cold mode
to the hot mode makes very little difference to the final galaxy
masses:  the KH instability is a mechanism for re-routing accretion from the cold mode to the hot mode
but its effect on the GSMF is
  small (the red shaded area is narrow).

\citet{benson_bower11} were the first to present a SAM with both cold
accretion and cooling.
Our model is more accurate because the shock-heating condition is
evaluated at the shock radius rather than the virial radius.
Hence, our critical masses are lower (it is the difference between the
green curve and the yellow curve in Fig.~\ref{tcomp_over_tcoolSD}a).
However, the main conclusion is the same: the results of models with
two accretion modes are not substantially different from those of
simpler SAMs where all the gas is assumed to be at the virial temperature initially.
This is probably why nobody else has pursued this line
of research besides us. However, the conclusion that shock stability
does not make any difference ignores two possibilities. Feedback may affect
different thermodynamic phases differently.
The accretion mode may have an impact on the morphologies of galaxies.
{ We shall now elaborate on these possibilities.}

The SAM presented in this article has no AGN feedback.
SN feedback is included but it acts only on gas that has already
accreted onto galaxies (or, at the opposite end, gas that not accreted
onto haloes yet, but pre-emptive feedback is effective only at very
low masses). The absence of any mechanisms that mitigate
cooling in massive systems is the reason why the number density of
massive galaxies is overpredicted by more than an order of magnitude
(Fig.~\ref{GSMFz}).  \citet{bower_etal06},
\citet{cattaneo_etal06} and \citet{croton_etal06} have shown
that cooling must be heavily suppressed above $M_{\rm shutdown}\sim
10^{12}\,{\rm M}_\odot$
to solve this problem.

Considerable evidence suggests  that mechanical feedback from radio
galaxies (radio-mode feedback)
maintains the thermal equilibrium of the hot gas in X-ray groups and
clusters (\citealp{birzan_etal04, forman_etal05,forman_etal17,fabian_etal06,rafferty_etal06}; also see \citealp{cattaneo_etal09} for a review).
The detailed physics (shock waves, sound waves, cosmic-ray heating)
are heavily debated \citep{mcnamara_nulsen12,fabian_etal17}
but they all assume an atmosphere that
absorbs, thermalises and distributes the mechanical energy of the
jets.
The mere existence of isolated regions with $T\sim T_{\rm vir}$ is not
enough.
The transition from the cold mode to the hot mode could thus be the switch that
activates feedback from radio galaxies if a supermassive black hole is
present \citep{cattaneo_etal06}.

Fig.~\ref{GSMFz} shows that cold accretion alone can account for: a) the
bulk of the galaxy population at $z>1.5$, as well as b) dwarf ($M_{\rm
  stars}< 3\times 10^9\,{\rm M}_\odot$) and giant ($M_{\rm
  stars}> 3\times 10^{11}\,{\rm M}_\odot$) galaxies at all $z$. Only for
intermediate-mass galaxies at low $z$ is cooling unavoidable.

The fraction of irregular morphologies increases at high $z$ (e.g., \citealp{huertas_etal15}) and at
low masses in the local Universe (e.g., \citealp{delapparent_etal11}). The most massive galaxies are
elliptical but their progenitors were most likely clumpy discs, such as those
observed by \citet{foersterschreiber_etal09}.
Grand-design spirals correspond to the population of low-$z$
intermediate-mass galaxies that acquired most of their mass through
cooling in {\sc GalICS~2.1}. The EAGLE simulation, too, finds
prevalently disc-dominated galaxies at intermediate but not low or high
masses \citep{clauwens_etal18}.
We speculate that this finding is not a
coincidence. Cold flows come in at high speed.
Only cooling provides the gentle smooth accretion that is needed to
assemble dynamically fragile systems such as thin discs.
{ Bulge-to-total mass ratios are the only morphological predictions that {\sc GalICS~2.1} makes at the moment
(Devergne et al., sub.),
but we plan to address the distinction between grand-design and clumpy
discs in a future study.}

This picture for the formation of spiral galaxies implies
that the halo mass $M_{\rm shutdown}$ above which cooling is suppressed 
is larger than the halo mass $M_{\rm
  vir}^{\rm crit}$ above which gas is shock-heated but cooling is
still possible.
\citet{cattaneo_etal06} found the best fit to the joint colour--magnitude distribution of SDSS galaxies \citep{baldry_etal04} for
$M_{\rm shutdown}= 2.3\times 10^{12}\,{\rm M}_\odot$.
Our explanation for $M_{\rm shutdown}>M_{\rm
  vir}^{\rm crit}$ is that a hot atmosphere of shock-heated gas is a
necessary but not sufficient condition for effective AGN feedback.
Quiescence is strongly related to the mass of the central black hole \citep{bluck_etal14, bluck_etal16,bluck_etal19,terrazas_etal16},
which is much higher in elliptical  than in spiral galaxies even for a same
stellar mass \citep{davis_etal18,sahu_etal19}. A plausible scenario is
that cooling flows keep feeding spiral galaxies until
mergers transform them into ellipticals.
Mergers fuel the growth of supermassive
black holes \citep{hutchings_campbell83,sanders_etal88,cattaneo_etal99,springel_etal05,hopkins_etal06}.
Only after these black holes are in place does radio-mode feedback become important
(\citealp{croton_etal06, bower_etal06}; and \citealp{cattaneo_etal09}
for a review).

Fig.~\ref{GSMFz} shows that the GSMF at $M_{\rm stars}\sim 10^{12}{\rm\,M}_\odot$ is overpredicted even when cooling is completely shut down (blue curves).
The above finding suggests that
maintenance feedback alone may not be enough to bring our predictions in agreement with the observations and that additional physics may be needed.
They may be ejective quasar feedback \citep{springel_etal05,hopkins_etal07}, or tidal and ram-pressure stripping \citep{conroy_etal07,cattaneo_etal11,steinhauser_etal16,tollet_etal17}. None of these processes is included in the version of {\sc GalICS~2.1} used for this article.

\subsection{SN feedback and the evolution of the GSMF}

Feedback is a complex multi-scale problem. Many aspects would
  have been very difficult to understand without the invaluable help
  from hydrodynamic simulations. However, the exchange between hydrodynamic simulations and 
SAMs is fruitful both ways. Studies that resolve
the internal structure of individual galaxies are limited by the
number of galaxies they can simulate\footnote{
  We refer to high-resolution simulations such as FIRE
  \citep{hopkins_etal14,angles_etal17} and NIHAO
  (\citealp{wang_etal15}; T19). Cosmological
  hydrodynamic simulations with $\sim 1\,$kpc resolution now routinely
  reach volumes comparable to ours (\citealp{schaye_etal15}: EAGLE, \citealp{dolag15}:
  Magneticum, \citealp{dubois_etal16}:
  HorizonAGN, \citealp{pillepich_etal18}:
  IllustrisTNG; \citealp{dave_etal19}: Simba)}. By plugging their results into SAMs,
we can determine whether the simulations' findings are consistent with
the evolution of galaxies across the Hubble time.
The feedback prescriptions that we have derived from the NIHAO simulations give results in good agreement with the GSMF and its evolution with redshift.

Most SAMs reproduce the GSMF at $z=0$ by construction but fail at higher redshifts because they overpredict the number density of low-mass high-$z$ galaxies \citep{knebe_etal18,asquith_etal18}.
{\sc L-Galaxies} \citep{henriques_etal15}, { {\sc shark}
\citep{lagos_etal18}} and {\sc GalICS~2.1} are the only SAMs
that reproduce the low-mass end of the GSMF over the entire redshift
range $0<z<2.5$, { to the best of our knowledge}.
 { {\sc GalICS~2.1} is based on a different reaccretion scenario
   from that of
 {\sc L-Galaxies} and {\sc shark}. In {\sc L-Galaxies} and {\sc shark},
gas is expelled from haloes and then slowly reincorporated into them}. In the NIHAO simulations and {\sc GalICS~2.1}, the fountain never leaves the inner halo.
However, { the common feature of all three SAMs is 
  a longer reaccretion timescale in low-mass systems.
  In the conventional semianalytic picture, it is the opposite. The
  reaccretion time is determined by the cooling time, which is longer at
  high masses.
  The fact that only SAMs
with a longer reaccretion time at low masses are able to reproduce
the evolution of the GSMF is a strong argument to conclude that this was the missing ingredient.}}

In T19 and our article, the physical justification for longer $t_{\rm reaccr}$ at
low $M_{\rm vir}$ is that most of the outflow mass is in
a cold component (the fountain), the motion of which is essentially though not purely ballistic (e.g., \citealp{collins_etal02,fraternali_binney06,fraternali_binney08}; also see \citealp{concas_etal19} for an observational perspective).
For a same initial speed (set by the physics of SNe and the physical conditions in the interstellar medium), the ejected gas reaches greater distances 
and thus has a longer reaccretion time in the shallower potential wells of low-mass systems.
This picture differs from the traditional one in which feedback is a
synonym of reheating, although $30$ per cent of the ejected gas is reheated
in {\sc GalICS~2.1}, too.

\section{Conclusion}

We summarise the main innovations and results of the article.
\begin{enumerate}
\item We have implemented a new SAM that uses a shock-stability
  criterion (Eq.~\ref{tcomp_over_tcool2}) to decide when cold flows give way to the formation of a hot
  quasi-hydrostatic atmosphere.
  The originality is that the shock radius is computed
self-consistently. The critical shock-heating mass is computed
individually for each halo.

\item
 For a typical halo at $z=0$, a shock appears for the first time at
 one tenth of the virial radius when $M_{\rm vir}\sim M_{\rm vir}^{\rm
   crit}\sim 2\times 10^{11}\,{\rm M}_\odot$. The shock reaches the virial
 radius by the time $M_{\rm vir}\sim 5\times 10^{11}\,{\rm M}_\odot$ (Fig.~\ref{tcomp_over_tcoolSD}a).

\item
  $M_{\rm vir}^{\rm
   crit}$ is sensitive to the metallicity of the accreted gas. We
 assume that the filaments have the same metallicity as the IGM and
 find results consistent with the metallicity of the IGM that we infer
 from DLAs (Fig.~\ref{Z_IGM}).

\item In massive haloes ($M_{\rm vir}>3\times 10^{11}\,{\rm M}_\odot$), cold
  accretion is more important at high redshift because the filaments
  are narrower. Hence, they radiate more effectively. This is not true
  at lower masses, where the lower metallicity of the IGM at high $z$
  is the dominant factor and produces an opposite effect (Fig.~\ref{fhotmean}).

\item
 { Cold accretion forms dwarf galaxies. It can also explain the
   formation of elliptical
 galaxies at high redshift.} Cooling is necessary to explain the
 intermediate-mass population around the knee of the GSMF, which is
 mainly composed of spiral galaxies
such as Andromeda and the Milky Way (Fig.~\ref{GSMFz}).

\item
  The contribution of cold-mode accretion to the masses of galaxies is
  sensitive
  to the ability of filaments to penetrate hot atmospheres without
  being disrupted (e.g., by the KH instability), but,
  without feedback that couples preferentially to the hot gas,
  the total mass that accretes onto galaxies is
insensitive to these uncertainties  (Fig.~\ref{GSMFz}).

\item
  We have implemented a new model for SN-driven winds based on
  cosmological hydrodynamic simulations (Section~2.5).
  Winds have a multiphase structure. The hot phase carries
  most of the energy. The cold phase carries most of the mass.
  Putting most of the mass in the slower cold phase allows us to
  achieve
  high-mass loading factors without violating the constraint of the
  energy available from SNe.

\item
  Cold winds do not reach the escape speed but form a galactic
  fountain. Contrary to reheated gas, which has a longer cooling
  time at high masses, the dominant cold component has a longer
  reaccretion time at low masses (Eq.~\ref{treaccr}).
  This behaviour is essential to delay the formation of low-mass
  galaxies and reproduce the evolution of the GSMF (Fig.~\ref{GSMFz}).

\item
The shallow slope of the GSMF at intermediate masses is linked to a
progressive but steep
decrease of the energetic efficiency of SN feedback at $v_{\rm
  vir}>75{\rm\,km\,s}^{-1}$, that is, $M_{\rm vir}>2\times
10^{11}\,{\rm M}_\odot$
(Eqs.~\ref{eta} and \ref{eta_cold}).
In high-mass haloes, most of the energy from SNe is wasted accelerating
gas that is not  able to escape from the galaxy and cycle through the
fountain (let alone escape from the halo).
The  mass $M_{\rm stars}\sim 10^9\,{\rm M}_\odot$ below which the GSMF
becomes steeper
corresponds to the mass-scale at which the efficiency of SN feedback
reaches its maximum saturation value.

 \end{enumerate}

\section*{Acknowledgements}

AC thanks Imene Belahcene, Yuval Birnboim, Andreas Faltenbacher and Patrick Petitjean for useful conversation.

Part of the numerical simulations used in this work were performed using the DiRAC Data Intensive service at Leicester, 
operated by the University of Leicester IT Services, which forms part of the STFC DiRAC HPC Facility (www.dirac.ac.uk). 
The equipment was funded by BEIS capital funding via STFC capital grants ST/K000373/1 and ST/R002363/1 and STFC DiRAC Operations grant ST/R001014/1. DiRAC is part of the UK National e-Infrastructure.

\bibliographystyle{mn2e} 

\bibliography{ref_av}

\appendix 

\section{The solid angle of accretion}

{ In this appendix, we compute the solid angle $\Omega$ covered by
  the filaments. Its value is important because if affects their density and thus their ability to
  radiate (the cold gas is denser if it is accreted from a narrow solid angle).
  
 $\Omega$ is the product of two quantities: the solid angle
  $\Omega_{\rm fil}$ covered by one filament and the number of
  filaments $n_{\rm fil}$.
This appendix is therefore subdivided into two parts, the calculation of
$\Omega_{\rm fil}$ and the calculation of $n_{\rm fil}$, followed by a
comparison with previous work.

Let us consider a model where filaments are cylindrical with radius
$r_{\rm fil}$ outside haloes and conical inside them. The part of a
filament inside a halo is assumed to be a cone with base radius
$r_{\rm fil}$ and height $r_{\rm vir}$. The cone aperture is
$\theta=\arctan(r_{\rm fil}/r_{\rm vir})$ and the corresponding solid angle is:
\begin{equation}
  \Omega_{\rm fil}=\int_0^{2\pi}{\rm
    d}\phi\int_0^\theta\sin\theta{\rm\,d}\theta=2\pi(1-\cos\theta)=
  \label{solidangle_onefilament}
  \end{equation}
$$=2\pi\left(1-\cos\arctan{r_{\rm fil}\over r_{\rm vir}}\right)=
2\pi\left(1-{ 1\over\sqrt{1+{r_{\rm fil}^2\over r_{\rm vir}^2}} }\right).$$

We assume that $r_{\rm fil}(z)$ is the virial radius of a halo with mass
equal to the non-linear mass $M_{\rm nl}$ at redshift $z$ (DB06), so that:
\begin{equation}
{r_{\rm fil}\over r_{\rm vir}}=\left[{M_{\rm nl}(z)\over M_{\rm vir}}\right]^{1\over 3}
\label{rvir}
\end{equation}
($r_{\rm vir}\propto M_{\rm vir}^{1/3}$).}

$M_{\rm nl}(z)$ is computed by solving the equation:
\begin{equation}
\sigma(M_{\rm nl})={\delta_{\rm c}\over D(z)},
\label{Mnl}
\end{equation}
where $\sigma(M)$ is the mass variance at $z=0$,
$\delta_{\rm c}=1.67572$ is the critical amplitude for collapse at $z=0$ \citep{eke_etal96},
and $D(z)$ is the linear growth factor of density fluctuations
in our cosmology with
$\Omega_{\rm M}=0.308$, $\Omega_\Lambda=0.692$, $\Omega_{\rm b} = 0.0481$ and  $\sigma_8=0.807$
(\citealp{planck14}, {\it Planck} + WP + BAO). $D$ is normalised so that $D=1$ for $z=0$.
Cosmological hydrodynamic simulations confirm that our model provides a reasonable estimate for the characteristic sizes of filaments \footnote{The fitting formula:
  $$r_{\rm fil} = {0.56{\rm\,Mpc}\over 0.6+0.4(1+z)^{2.8}}$$
approximates { our results to a  precision of a few per cent. It is based on
  Eq.~(\ref{Mnl}) and uses the virial density at redshift $z$ to pass
  from $M_{\rm nl}$ to $r_{\rm fil}$}.} both at low redshift \citep{gheller_etal15}
and at $z\sim 3$ (Ramsoy et al., in prep.)

{ For the number of filaments, we assume:
\begin{equation}
n_{\rm fil}\simeq {20\over 3}+{10\over 3}{\rm\,log}_{10}{M_{\rm vir}\over M_{\rm nl}}
\label{nfil}
\end{equation}
with a floor at $n_{\rm fil}=4$ and a ceiling at $n_{\rm fil}=12$
based on analytic work and numerical
simulations of Gaussian random fields by
\citet{codis_etal18}\footnote{\citet{codis_etal18} determined $n_{\rm
    fil}$ 
  as a function of $M_{\rm vir}/{\rm M}_\odot$ at $z=0$. We have extended
  their result to $z>0$ by using the
  self-similarity of the hierarchical evolution of the cold DM and
  replacing $M_{\rm vir}$ with $M_{\rm vir}/M_{\rm nl}(z)
  \times M_{\rm nl}(0)$.}.

The total solid angle:
\begin{equation}
  \Omega=\min(n_{\rm fil}\Omega_{\rm fil},\,4\pi)
  \label{anisotropy}
\end{equation}
equals $4\pi$ for $M_{\rm vir}\lsim 0.87 M_{\rm nl}$ and decreases
with $M_{\rm vir}$ at higher masses.

Anisotropy boosts the mean density $\langle\rho\rangle$ of the cold gas
by a factor of $4\pi/\Omega$, but the cooling rate is proportional to
$\langle\rho^2\rangle$ and $C=\langle\rho^2\rangle/\langle\rho\rangle^2>1$
if the cold gas is inhomogeneous. Hence, the cooling rate increases not
by a factor $4\pi/\Omega$ but by a
factor of $4\pi/\Omega_{\rm eff}$, where $\Omega_{\rm eff}=\Omega/C$.
We  estimate $C\simeq 9$ based on the density distribution within 
filaments measured in cosmological hydrodynamical simulations\footnote{Ramsoy et al. (in prep.) find that the density within a filament
scales as $(1+x^2)^{-2}$, where $x$ is the distance from the centre of
the filament in filament's core radii, and that the truncation radius
at which gas from regions of lower density (walls, voids) shocks onto
the filament
corresponds to $x\sim 5$}}.
This calculation neglects the clumpiness of the gas, which \citet{cantalupo_etal14} and \citet{wisotzki_etal18} find to be significant, even though theoretical arguments based on the Jeans length suggest the contrary.

{ We conclude this appendix by comparing our findings to those by
\citet{danovich_etal12}, who used the Horizon Mare Nostrum cosmological
hydrodynamic simulation} to study the feeding of massive galaxies
through cold streams at high redshift\footnote{ The Mare Nostrum
  simulation was run with the adaptive-refinement code {\sc ramses}
  \citep{teyssier02}. See appendix~A of \citet{danovich_etal12} for
  details about the numerical resolution and the assumptions for
  radiative cooling, ultraviolet background radiation, star formation
  and SN feedback.}. They considered a sample of 350 haloes with
$M_{\rm vir}\simeq 10^{12}\,{\rm M}_\odot$ at $z=2.5$
and found that $70\%$ of the influx into the virial radius came from narrow streams covering $10\%$ of the virial shell; 
$64\%$ of the stream influx came from one stream and $95\%$ from the three dominant ones, for which $\Omega_{\rm fil}\sim 0.4$ on average.

Differently from the Mare Nostrum simulation,
our SAM cannot distinguish between larger and smaller filaments. For a
halo of mass $M_{\rm vir}= 10^{12}\,{\rm M}_\odot\simeq 10\,M_{\rm nl}(z=2.5)$,
{\sc GalICS~2.1} finds $n_{\rm fil}\simeq 10$ (Eq.~\ref{nfil}) because it counts several small filaments that contribute to barely $5\%$ of the baryonic accretion rate onto the halo,
{ but it also finds a lower mean solid angle per filament.
For  $r_{\rm fil}=40{\rm\,kpc}$ and $r_{\rm vir}=70{\rm\,kpc}$ (the
typical values in our SAM for a halo with $M_{\rm vir}=
10^{12}\,{\rm M}_\odot$ at $z=2.5$), Eq.~(\ref{solidangle_onefilament})
gives $\Omega_{\rm fil,\,eff}=\Omega_{\rm fil}/C\simeq 0.092$.
However, the total effective solid angle covered by all filaments, $\Omega_{\rm eff}\simeq 0.92$, is very similar to the one found by \citet{danovich_etal12}, 
$\Omega\simeq 0.4\times 3=1.2$.}

\section{The Kelvin-Helmholtz instability timescale}

{ Filaments are disrupted by the KH instability when KH waves
  develop on scale comparable to their radii.
This happens on a timescale:}
\begin{equation}
t_{\rm KH}={r_{\rm fil}\over u_1}\sqrt{\rho_{\rm fil}\over\rho_{\rm
    hot}},
\label{tKH}
\end{equation}
(see, e.g., \citealp{agertz_etal07}), where $r_{\rm fil}$ is the radius of the filaments 
and $u_1$ is the relative speed of the cold gas in the filaments with respect to the hot gas,
while $\rho_{\rm fil}$ and $\rho_{\rm hot}$ are the characteristic
densities of the filaments and the hot gas, respectively.
If $t_{\rm KH}<t_{\rm ff}$, the filaments will be disrupted faster
than they accrete onto the galaxy.

{ Our evaluation of $t_{\rm KH}$ is based on virial quantities. For a conical geometry, filaments become narrower as one moves to
  smaller radii.
  However,  the freefall time is $t_{\rm ff}\sim r/u_1$. Therefore, the ratio
  $t_{\rm KH}/t_{\rm ff}$ is not sensitive to the radius at which it is
  evaluated.
  We also note that the way $\rho_{\rm fil}$ and $\rho_{\rm hot}$ increase
  at small radii may not be the same but we neglect this complication.

 To estimate
  $\rho_{\rm fil}/\rho_{\rm hot}$,
let $V$ be the volume filled by the hot gas
(the virial volume in the approximation of narrow filaments), so that
 $\rho_{\rm hot}={M_{\rm hot}/V}.$
    If the cold gas is accreted from a solid angle $\Omega\ll 4\pi$,
    it will fill a volume $V\Omega/(4\pi)$, so that
  $\rho_{\rm
        fil}=(4\pi/\Omega)(M_{\rm fil}/V)$ and:
 \begin{equation}
  {\rho_{\rm fil}\over\rho_{\rm hot}}=
 {1\over\epsilon_{\rm KH}}{4\pi\over\Omega}{M_{\rm fil}\over M_{\rm hot}},
\label{rrhofil}
\end{equation}
where $\epsilon_{\rm KH}$ is a parameter that sets the efficiency of
the KH instability ($\epsilon_{\rm KH}=1$ in our default model).

For $\Omega=4\pi$,
the volume $V$ occupied by the cold gas is the same volume
that is occupied by the hot gas. Both occupy the totality of the
volume within the halo. { This naive prediction of Eq.~(\ref{rrhofil})
cannot be correct and signifies
that our geometric approximation has broken down.}
Physically, we no longer have cold filaments penetrating a hot atmosphere but a multiphase medium.
In this case, we assume that the largest scale on which perturbations can
grow is the virial radius and we replace $r_{\rm fil}$
with $r_{\rm vir}$. { The general formula is:
\begin{equation}
t_{\rm KH}={{\rm\,min}(r_{\rm fil},r_{\rm vir})\over u_1(r_{\rm
    vir})}\sqrt{{1\over\epsilon_{\rm KH}}{4\pi\over\Omega}{M_{\rm fil}\over M_{\rm hot}}},
\label{tKH3}
\end{equation}
where $4\pi/\Omega$ is never allowed to be smaller than unity.
For $\Omega=4\pi$, the term in front of the square root is the
freefall time. Therefore, the interpretation of Eq.~(\ref{tKH3}) is
very simple: filaments start to be disrupted when $M_{\rm hot}>M_{\rm
  fil}$.
}

The reader should notice that our analysis is based on the standard KH theory for two homogeneous media with a contact discontinuity.
The shear instability could develop differently if the dense gas has an inhomogeneous density profile, as it the case for cold filamentary flows.
Therefore, $\epsilon_{\rm KH}$ parametrizes not only the density profile of the cold gas but also our ignorance of the way to deal with such a configuration,
for which there is no analytic theory to the best of our knowledge.

{ We can, nevertheless, present an argument why we expect that our results should be robust. In  a cylindrical or conical geometry, $\rho_{\rm
    fil}\propto r_{\rm fil}^{-2}$.
 Therefore,  $r_{\rm fil}\sqrt{\rho_{\rm
      fil}}$ (Eq.~\ref{tKH}) is insensitive to errors in our estimate of $r_{\rm fil}$.

  We can also derive a lower limit for $t_{\rm KH}$ by considering the
  lowest possible $\rho_{\rm fil}$, i.e., the density of the cold gas
  at the discontinuity surface that separates it from the hot gas.
  Ramsoy et al. (in prep.) find that this density is  $\epsilon_{\rm
    KH}\sim 26$ times lower than the mean density inside the filaments
  (footnote~14).
  We use this extreme model to estimate the maximum effect that the KH
  instability could have.}

\section{Numerical tests of the cooling scheme}

We have used hydrodynamic simulations with {\sc ramses}
\citep{teyssier02} in three dimensions to test \citet{white_frenk91}'s
cooling scheme  { and to determine the value of the fudge factor
  $\epsilon_{\rm cool}$.
 The compared quantity is the mass $M_{\rm cooled}$ that cools within time $t$.
 Our goal is to demonstrate that, for $\epsilon_{\rm cool}=0.8$, our SAM.
 reproduces $M_{\rm cooled}(t)$ as measured in hydrodynamic simulations
 when the initial conditions for $\rho$
 and $T$, the gravitational potential, the metallicity of the gas and
 the cooling function are the same.

The density profile assumed in {\sc GalICS~2.1} is not a suitable
 initial condition for hydrodynamic simulations 
because it is not an exact solution of the equation of hydrostatic
equilibrium. The violent relaxation of the hot gas to a quasi-equilibrium profile would propagate disturbances
that may artificially affect its cooling rate.
We have therefore preferred to use the \citet[KS]{komatsu_seljak01}
solution, which is the exact hydrostatic equilibrium solution for a
polytropic gas ($T\propto\rho^{\gamma-1}$) in the gravitational
potential of an NFW halo.

The KS contains three free parameters (the central density, the
central temperature and the polytropic index $\gamma$), which KS
constrained by requiring that $\rho/\rho_{\rm NFW}\simeq\Omega_{\rm
  b}/\Omega_{\rm M}$ at $0.5r_{\rm vir}<r<2r_{\rm vir}$. Here $\rho$ is the
density of the hot gas and $\rho_{\rm NFW}$ is the density profile that
generates the gravitational potential.
We have verified that the density profile used in {\sc GalICS~2.0} and the KS
are very similar at $r\lsim 2r_{\rm vir}$, where cooling takes place.}

The computational set-up of our hydrodynamic simulations and
all numerical parameters are the same as in \citet{cattaneo_teyssier07}'s control simulation without AGN feedback.
We therefore refer to that article for all technical details.

We have considered a halo with mass $M_{\rm vir}$, concentration
$c(M_{\rm vir}) = 11(M_{\rm vir}/10^{12}{\rm\,M}_\odot)^{-0.097}$
\citep{dutton_maccio14} and metallicity $Z$, { and we have explored
  four parameter combinations corresponding to two halo masses and two
  metallicities (Fig.~\ref{mod_vs_sim}).}
We have used the KS solution to set up hydrostatic initial conditions and we have let them evolve for several gigayears to verify that they are indeed hydrostatic.
Then, we have let the gas cool.
As our simulations do not contain star formation, $M_{\rm cooled}$ is simply the mass of cold gas (gas with $T<2\times 10^5\,$K) within
our computational volume (a cube of side-length $2.6\,$Mpc);
$t=0$ is the time at which the gas starts cooling.

The comparison with hydrodynamic simulations has shown that \citet{white_frenk91}'s
cooling scheme  tends to overestimate the cooling time for a given
halo mass
because SAMs assume static density and temperature profiles,
while in the hydrodynamic simulations
the central density increases and the central temperature decreases as time passes (Fig.~2 of \citealp{cattaneo_teyssier07}).
In Eq.~(\ref{ttcool}), we partially correct this effect by introducing the fudge factor $\epsilon_{\rm cool}=0.8$.
With this correction, the results of
{\sc GalICS~2.1} are  in reasonably good agreement with hydrodynamic simulations over a broad range of halo masses and metallicities 
(compare the solid and the dashed curves in Fig.~\ref{mod_vs_sim}).

\begin{figure}
\includegraphics[width=1.00\hsize]{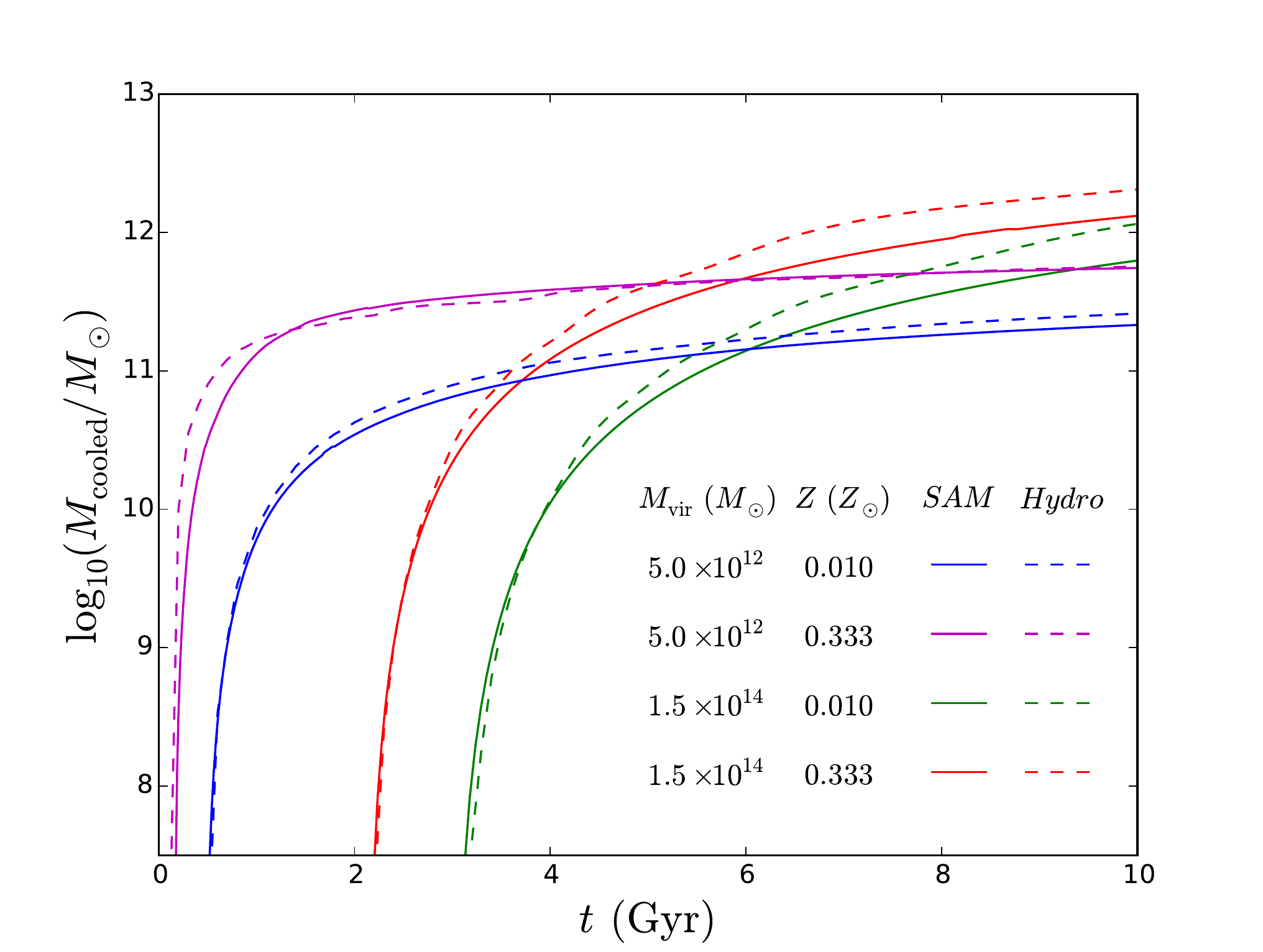} 
\caption{The cooled mass $M_{\rm cooled}=\int_0^t\dot{M}_{\rm
    cool}{\rm\,d}t$ computed with our SAM for $\epsilon_{\rm cool}=0.8$ (solid curves)
compared to the mass of cold gas ($T<2\times 10^5\,$K) 
that has accumulated by time $t$ in control simulations with the adaptive-mesh-refinement code {\sc ramses} (dashed curves).
The SAM and the simulations assume the same halo properties, the same
initial conditions, the same metallicity of the gas and the same
cooling function \citep{sutherland_dopita93}.}
\label{mod_vs_sim}
\end{figure}


\label{lastpage}
\end{document}